\documentclass[%
aip,
jcp,
amsmath,amssymb,
reprint,%
]{revtex4-2}

\usepackage{graphicx}
\usepackage{dcolumn}
\usepackage{bm}

\usepackage{xr-hyper}
\usepackage{hyperref}

\usepackage[utf8]{inputenc}
\usepackage[T1]{fontenc}
\usepackage{mathptmx}
\usepackage{etoolbox}
\usepackage{xcolor}

\makeatletter
\def\@email#1#2{%
	\endgroup
	\patchcmd{\titleblock@produce}
	{\frontmatter@RRAPformat}
	{\frontmatter@RRAPformat{\produce@RRAP{*#1\href{mailto:#2}{#2}}}\frontmatter@RRAPformat}
	{}{}
}%
\makeatother
\begin{document}

\preprint{AIP/123-QED}

\title{Analysis and Simulation of Generalized Langevin Equations with Non-Gaussian Orthogonal Forces}
\author{Henrik Kiefer}
 \author{Benjamin J. A. Héry} 
 \author{Lucas Tepper}
 \author{Benjamin A. Dalton}
\author{Cihan Ayaz}
 \author{Roland R. Netz}
 \altaffiliation{Author to whom correspondence should be addressed: rnetz@physik.fu-berlin.de}%
 \affiliation{%
Department of Physics,  Freie  Universität  Berlin,  Arnimallee  14,  14195  Berlin,  Germany.
}%

\date{\today}

\begin{abstract}
 \noindent The generalized Langevin equation (GLE) is a useful framework for analyzing and modeling the dynamics of many-body systems in terms of low-dimensional reaction coordinates, with its specific form determined by the choice of projection formalism.  
We compare parameters derived from different GLE formulations using molecular dynamics simulations of butane's dihedral angle dynamics.
Our analysis reveals non-Gaussian contributions of the orthogonal force in different GLEs, being most enhanced for the Mori-GLE, where all non-linearities are relegated to the orthogonal force.
We establish a simulation technique that correctly accounts for non-Gaussian orthogonal forces, which is critical for accurately predicting dihedral-angle mean first-passage times. We find that the accuracy of GLE simulations depends significantly on the chosen GLE formalism; the Mori-GLE offers the most numerically robust framework for capturing the statistical observables of the dihedral angle dynamics, provided the correct non-Gaussian orthogonal force distribution is used.

\end{abstract}
\maketitle

\section{Introduction}
Coarse-graining of complex systems in terms of low-dimensional observables provides useful models for the description of many-body kinetics; examples are protein folding \cite{plotkin1998non,monticelli2008martini,clementi2008coarse,hills2009insights} or chemical reactions between individual molecules \cite{minichino1997potential,kamerlin2011coarse,zhou2019variational, brunig2021proton}.
The resulting reduced equations of motion, typically stochastic, not only give the advantage of numerical efficiency and speed over all-atomistic simulations of a complex system, but also provide the possibility of analysis and interpretation of atomistic simulations and experiments.\\
\indent The generalized Langevin equation (GLE), comprising a deterministic force, non-Markovian friction, and an orthogonal force, serves as a powerful framework for constructing coarse-grained models of many-body systems \cite{berne1966calculation,adelman1980generalized,berne1990dynamic,chorin2000optimal,chorin2002optimal,darve2006numerical, lange2006collective, carof2014two, ma2016derivation, lee2019multi,loos2019heat, klippenstein2021introducing,doerries2021correlation}. By reducing the system's full degrees of freedom to a few relevant observables, the GLE enables accurate modeling and kinetic predictions of conformational transitions \cite{bagchi1983effect,straub1987calculation,canales1998generalized,satija2019generalized,ayaz2021non,dalton2022protein, dalton2024role,milster2024tracer}, reaction dynamics \cite{brunig2022pair, vroylandt2022likelihood}, and the vibrational spectra of molecules in the presence of non-linear bond-length and bond-angle potentials \cite{brunig2021proton, brunig2022timedependent}.\\
\indent First introduced by R. Zwanzig \cite{zwanzig1961memory} and H. Mori \cite{mori1965transport}, different forms of the GLE arise from different projection operator formalisms. It has been argued that the Mori-projection technique is less suitable for modeling non-linear observables because the resulting GLE includes a deterministic force linear in the coordinate and a frictional force linear in the velocity. However, since the Mori-GLE is an exact decomposition of the total force acting on an observable, non-linearities in the observable are incorporated into the orthogonal force, which becomes non-Gaussian and thereby challenging to parameterize \cite{mazur1991and, vroylandt2022position}. By contrast, GLEs derived using the Zwanzig-projection method are considered more suitable, as they explicitly include the term associated with the potential of mean force (PMF) \cite{li2017computing,di2022derivation, jung2023dynamic}.
\\\indent To explicitly account for non-linearities in the deterministic forces and to simplify the friction kernel compared to the Zwanzig formulation, a hybrid projection scheme was recently introduced \cite{ayaz2022generalized}. This approach combines the linear Mori and conditional Zwanzig projection operators to derive a hybrid GLE, which comprises a memory friction term that splits into two components: one that is linear in the past velocity and another that is generally non-linear in the coordinate but independent of velocities. Numerical methods for extracting all parameters of this GLE from trajectories have been proposed \cite{ayaz2022generalized}, however methods to simulate the hybrid GLE are currently not known.\\ 
\indent Until recently, it was widely assumed that a GLE incorporating the PMF and a friction force only linear in velocity \cite{lange2006collective}, which is suitable for various applications \cite{daldrop2018butane, lee2019multi,ayaz2021non, brunig2022timedependent, dalton2022protein, dalton2024role,dalton2024memory}, could only be derived approximately by neglecting position-dependent friction terms. However, it was demonstrated \cite{vroylandt2022gle} that such a GLE can be rigorously derived by sequentially applying the Mori and hybrid projection techniques during different steps of the derivation.\\ 
\indent With the various forms of GLEs discussed, a key question arises: which GLE formulation is most robust for simulating non-linear systems? Here, we evaluate robustness based on the ability to accurately reproduce commonly used statistical observables of the modeled coordinates, such as two-point correlation functions and mean first-passage times. \\
\indent Previous work on coarse-grained simulations relied on replacing the GLE by coupled Markovian Langevin equations with Gaussian noise, a method commonly referred to as Markovian embedding \cite{ceriotti2010colored, li2017computing, klippenstein2021introducing, vroylandt2022likelihood}. These studies concluded that faithfully reproducing statistical observables requires involving the correct non-linear PMF \cite{ayaz2021non, brunig2022timedependent, jung2023dynamic}. Furthermore, although the orthogonal force resulting from projection is, in principle, a deterministic function with well-defined initial conditions \cite{mori1965transport}, practical simulations assume that it follows a stationary Gaussian distribution, leveraging a relationship between the orthogonal force autocorrelation and the memory kernel. However, experimental studies \cite{wiesenfeld1994stochastic,wio1999stochastic,wang2009anomalous, wang2012brownian} and simulations across various fields \cite{shin2010brownian, carof2014two, kanazawa2015minimal,jung2023dynamic} frequently reveal deviations from Gaussian behavior of the orthogonal force. The modeling of the system’s kinetics can be strongly influenced by the probability distribution and higher-order correlations of the orthogonal force \cite{zwanzig2001chemical,wio2004effect,majee2005colored,chechkin2017brownian,mutothya2021first,baule2023exponential,caspers2024nonlinear}. Accurate modeling of the orthogonal force in GLE simulations is, therefore, key for capturing the underlying physics of soft-matter systems, particularly when rare events dominate the long-time dynamics. This is the topic of the present paper.\\
\indent After reviewing the derivations and relationships between the various GLEs discussed, using dihedral angle trajectories from molecular dynamics (MD) simulations, we compare the parameters extracted for the different GLEs. In particular, we explore the behavior of the orthogonal force in the different GLE formulations. Our extraction methods reveal significant non-Gaussian features in the orthogonal force for all GLEs. \\
\indent Ultimately, we present a GLE simulation algorithm for arbitrary orthogonal force distributions. This non-Gaussian force (NGF) technique samples the orthogonal force from MD simulations and operates without additional approximations when simulating the GLE, in contrast to methods used in previous work \cite{ceriotti2010colored, ayaz2021non, widder2022generalized, jung2023dynamic}. Contrary to common Markovian embedding techniques, we can precisely reproduce the mean first-passage times of the butane dihedral angle when using an orthogonal force with the correct distribution. 
By comparing simulations of different GLEs, we show that the Mori-GLE, which relegates all non-linearities in the deterministic and friction forces into the orthogonal force, is the numerically most accurate approach for simulating dihedral angle dynamics. Consequently, accurately modeling the statistics of the orthogonal force is more critical than including the appropriate PMF, contrary to previous suggestions \cite{ayaz2022generalized, glatzel2022interplay, jung2023dynamic}.

\section{Theoretical Background}
\subsection{Different generalized Langevin equations}
We consider a dynamical system of $N$ interacting particles in three-dimensional space. The time evolution of a microstate $\omega(t) $, which is a $6N$ vector of the Cartesian positions $\vec{q}_i$ and the conjugate momenta $\vec{p}_i$ of the $i=1, 2, ..., N$ particles in the system, is described by Hamilton’s equation of motion 
\begin{equation}
\label{eq:phase_space}
    \Dot{\omega}(t) = \mathcal{L} \omega(t),
\end{equation}
where $\mathcal{L}$ is the Liouville operator, depending on the system's Hamiltonian $H(\omega)$. We assume that the system is initially at phase-space position $\omega(0) = \omega_0$. Hence, Eq.~\eqref{eq:phase_space} has the solution $\omega(t) = e^{t\mathcal{L}}\omega_0$. 
\\ \indent Since we are not interested in the time evolution of all degrees of freedom in $\omega$, we coarse-grain the dynamics. We restrict ourselves to one-dimensional observables, but note that we can proceed analogously for higher dimensions. A versatile starting point for coarse-graining is the projection operator technique introduced by Zwanzig \cite{zwanzig1961memory} and Mori \cite{mori1965transport}. For an arbitrary scalar observable $A$, which is a real-valued function of phase-space coordinates and depends on time only implicitly via the time dependence of $\omega(t)$, i.e. $A(t) \equiv A(\omega,t)$, the Liouville equation for its acceleration, i.e. $\Ddot{A}(t) = e^{t\mathcal{L}}\mathcal{L}^2 A_0$ \cite{zwanzig2001nonequilibrium}, can be decomposed into a relevant subspace by using the projection operator $\mathcal{P}_0$, and an orthogonal subspace by using the operator $Q_0 = \textit{1} - \mathcal{P}_0$, with $\textit{1}$ being the identity operator,
\begin{equation}
\label{eq:projection_liouville}
    \Ddot{A}(t) = e^{t\mathcal{L}}(\mathcal{P}_0+Q_0)\mathcal{L}^2 A_0,
\end{equation}
where $A_0 = A(\omega,0)$. 
The coarse-grained equation of motion for $A(t)$, the generalized Langevin equation (GLE), reads \cite{zwanzig1961memory,mori1965transport,zwanzig2001nonequilibrium,ayaz2022generalized}
\begin{equation}
\label{eq:basic_gle}
    \Ddot{A}(t) = e^{tL}\mathcal{P}_0\mathcal{L}^2 A_0 + \int_0^t ds\:e^{(t-s)\mathcal{L}}\mathcal{P}_1\mathcal{L}F_\text{Q}(s) + F_\text{Q}(t).
\end{equation}
Note that we use different projection operators for the decomposition in Eq.~\eqref{eq:projection_liouville} ($\mathcal{P}_0 = 1- Q_0)$ and to derive the memory friction in Eq.~\eqref{eq:basic_gle} ($\mathcal{P}_1=1- Q_1$) \cite{vroylandt2022gle}.  
Eq.~\eqref{eq:basic_gle} is an exact decomposition of the Liouville equation into three terms. The first describes the evolution of the relevant subspace and reflects the deterministic mean force due to a potential. The second term accounts for friction between the relevant and orthogonal subspace and includes a memory kernel. The last term $F_\text{Q}(t) = e^{tQ_1\mathcal{L}}Q_0\mathcal{L}\dot{A}_0$ is the orthogonal force. Since neither the Liouville operator $\mathcal{L}$ nor the initial conditions $\omega_0$ are known, $F_\text{Q}(t)$ is often described as a stochastic process and therefore often referred to as the random force.
\\ \indent The GLE’s form depends on the projection operators $\mathcal{P}_0$ and $\mathcal{P}_1$.
Using the Mori-projection operator ($\mathcal{P}_0 = \mathcal{P}_1 = \mathcal{P}_\text{M}$, see supplementary material Sec.~\ref{app:proof_equiv_kernels}) leads to a GLE with a deterministic force that is linear in the position and a frictional force independent on $A$ and linear in the velocity \cite{mori1965transport}
\begin{equation}
    \label{eq:gle_mori}
    \Ddot{A}(t) = - \frac{k}{M_0} A(t) - \int_0^t ds\:\Gamma^\text{M}(t-s) \Dot{A}(s) + F^\text{M}_\text{Q}(t),
\end{equation}
where $k=k_BT$/$\langle A_0^2\rangle$ is the potential curvature and $M_0 = k_BT/\langle \Dot{A}_0^2 \rangle$ is the effective mass, both following from the equipartition theorem. $k_BT$ is the product of Boltzmann constant and temperature.
The Mori memory kernel $\Gamma^\text{M}(t)$ and orthogonal force $F^\text{M}_\text{Q}(t)$ are related via \cite{mori1965transport}
\begin{equation}
\label{eq:FDT_Mori}
\langle F^\text{M}_\text{Q}(0), F^\text{M}_\text{Q}(t) \rangle = \langle \dot{A}_0^2 \rangle  \Gamma^\text{M}(t).
\end{equation}
$\langle A(t), B(s) \rangle$ denotes an equilibrium ensemble average 
\begin{equation}
    \label{eq:dot_product}
    \langle A(t),B(s) \rangle = \int d\omega \:\rho_\text{eq}(\omega)A(\omega,t)B(\omega,s),
\end{equation}
including the canonical Boltzmann distribution $\rho_\text{eq}(\omega)=Z^{-1}e^{-H(\omega)/k_BT}$ with $Z$ being the partition function $Z = \int d\omega\:e^{-H(\omega)/k_BT}$.
Note that $A(t)$, its derivatives and the orthogonal force are phase-space dependent, $A(t) = A(\omega,t)$, but $\Gamma^\text{M}(t)$, $k$ and $M_0$ are not.\\
\indent The Mori-GLE contains a deterministic force and friction linear in $A$ and $\dot{A}$. Hence, a non-linearity in the observable is absorbed into the orthogonal force, making it impossible to model the orthogonal force as a stationary Gaussian process for non-linear observables. Contrary, the Zwanzig projection introduces the non-linear PMF into the GLE \cite{zwanzig1961memory}. However, for this projection, the memory kernel depends explicitly on position and velocity, which is difficult to handle practically \cite{ayaz2022generalized}. 
\\\indent The hybrid projection ($\mathcal{P}_0 = \mathcal{P}_1 = \mathcal{P}_\text{H}$, see supplementary material Sec.~\ref{app:proof_equiv_kernels}), developed by Ayaz \textit{et al.} \cite{ayaz2022generalized}, combines the Mori and Zwanzig projection, resulting in a GLE that includes the PMF, a memory kernel that is independent of $A$ and $\dot{A}$ and couples to the velocity, and a memory kernel that depends non-linearly on the position. 
The hybrid GLE is given by
\begin{align}
    \nonumber
     \Ddot{A}(t) =& - F_{\text{eff}}\bigl(A(t)\bigr) - \int_0^t ds\:\Gamma^\text{H}_\text{L}(t-s) \Dot{A}(s) \\ \label{eq:hybrid_gle}
     &+\int_0^t ds\:\Gamma^\text{H}_\text{NL}\bigl(A(s),t-s\bigr) + F^\text{H}_\text{Q}(t),
\end{align}
where the effective non-linear force $F_{\text{eff}}$ is given by
\begin{equation}
    \label{eq:eff_force}
    F_{\text{eff}}\bigl(A(t)\bigr) = \frac{1}{M\bigl(A(t)\bigr)}\frac{d}{dA}\bigl[U_{\text{PMF}}\bigl(A(t)\bigr) + k_BT\:\ln\:M\bigl(A(t)\bigr)\bigr].
\end{equation}
$M(A) = k_BT/\langle \Dot{A}_0^2 \rangle_{A}$ is the position-dependent mass, and $U_{\text{PMF}}(A) = - k_BT\ln \rho(A)$ is the PMF, with $\rho(A)$ being the equilibrium probability distribution of $A$. $\langle..\rangle_{A}$ denotes a conditional average (supplementary material Sec.~\ref{app:proof_equiv_kernels}). 
The linear memory kernel $\Gamma^\text{H}_\text{L}(t)$ is related to the orthogonal force $F^\text{H}_\text{Q}(t)$ via 
\begin{equation}
    \label{eq:FDT_hybrid}
    \Gamma^\text{H}_\text{L}(t) = \frac{\langle F^\text{H}_\text{Q}(0),F^\text{H}_\text{Q}(t)\rangle}{\langle \Dot{A}_0^2 \rangle}.
\end{equation}
The non-linear memory kernel $\Gamma^\text{H}_\text{NL}$ is a general function of $A(t)$ and accounts for position-dependent friction effects. It is defined as
\begin{align}
\nonumber
    \Gamma^\text{H}_\text{NL}\bigl(A(t),s\bigr) =& \frac{d}{dA}D\bigl(A(t),s\bigr) \\     \label{eq:kernel_nl}
 &- \frac{1}{k_BT}D\bigl(A(t),s\bigr)\frac{d}{dA}U_{\text{PMF}}\bigl(A(t)\bigr),
\end{align}
where $D$ is the conditional correlation function between the velocity at the initial time and the orthogonal force 
\begin{equation}
    \label{eq:dfunc}
    D(A(t),s) = \langle \Dot{A}_0,F^\text{H}_\text{Q}(s)\rangle_{A}.
\end{equation}
All parameters in the hybrid GLE can be calculated from a trajectory $A(t)$ by the algorithm introduced in Ref.~\onlinecite{ayaz2022generalized}, utilizing a forward propagation method inspired by Carof \textit{et al.} \cite{carof2014two}. Volterra methods do not apply for the hybrid GLE, which severely complicates the practical applicability of the hybrid GLE \cite{ayaz2022generalized}.
\\
\indent The position-dependent friction term in Eq.~\eqref{eq:kernel_nl} vanishes if the conditional correlation function between the orthogonal force and the time derivative of the observable, $D\bigl(A(t),s\bigr)$, is zero. A simpler form of the GLE, including the PMF and only a linear friction term, has been widely used in previous work \cite{daldrop2018butane, ayaz2021non, brunig2022timedependent, brunig2022pair, dalton2022protein, dalton2024role}, and has been derived by applying the hybrid ($\mathcal{P}_0 = \mathcal{P}_\text{H}$) and Mori ($\mathcal{P}_1 = \mathcal{P}_\text{M}$) projection operators at different stages of the derivation \cite{vroylandt2022gle}.
 We give details in supplementary material Sec.~\ref{app:proof_equiv_kernels}. This dual-projection (DP-)GLE follows as
\begin{equation}
\label{eq:dp_gle}
    \Ddot{A}(t) = - F_{\text{eff}}\bigl(A(t)\bigr) - \int_0^t ds\:\Gamma^\text{DP}(t-s) \Dot{A}(s) + F^\text{DP}_\text{Q}(t),
\end{equation}
with memory kernel $\Gamma^\text{DP}(t)$ and orthogonal force $F^\text{DP}_\text{Q}(t)$.  The effective force $F_{\text{eff}}\bigl(A(t)\bigr)$ remains the same as in Eq.~\eqref{eq:eff_force}. 
\begin{figure*}[hbt!]
    \centering
    \includegraphics[width=\textwidth]{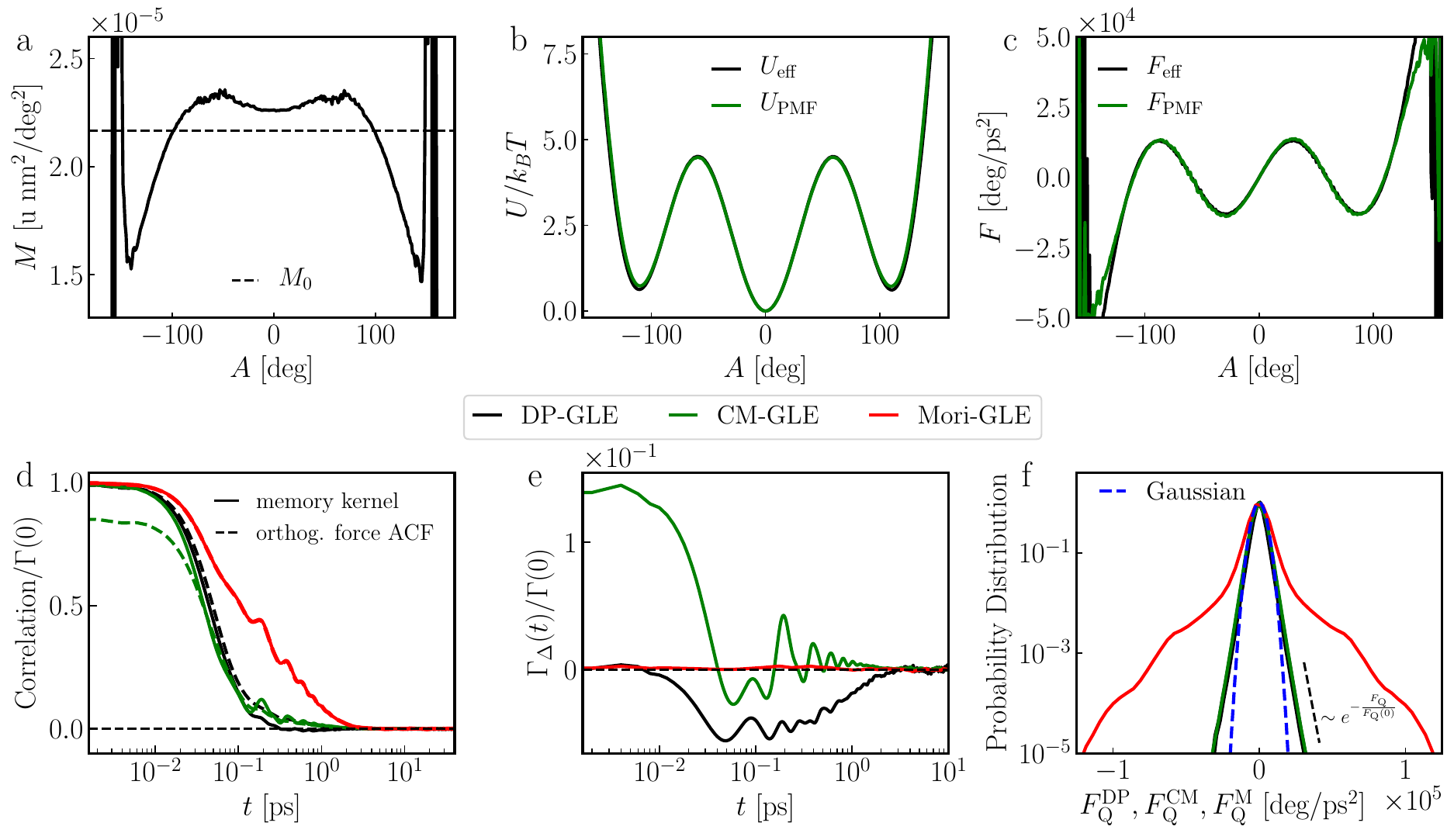}
    \caption{GLE parameter extraction for butane dihedral angle dynamics.
    (a) Position-dependent mass profile $M(A)$ = $k_BT/\langle \Dot{A}_0^2 \rangle_{A}$ extracted for the butane dihedral angle from an MD simulation (explained in Appendix \ref{sec:mds}). 
    The broken line denotes the constant mass, i.e. $M_0 = k_BT / \langle \Dot{A}_0^2 \rangle $. (b) Comparison between the potential governing the force in Eq.~\eqref{eq:eff_force} without, i.e. $U(A) = U_{\text{PMF}}(A)$ (green), and with the mass term, i.e. $U_{\text{eff}}(A) = U_{\text{PMF}}(A) + k_BT \ln M(A)$ (black). (c) Comparison between the deterministic force without, i.e. $F_{\text{PMF}}(A) = \left(\nabla_A U_{\text{PMF}}(A)\right)/M_0$, and with the mass term, $F_{\text{eff}}(A)$, as given in Eq.~\eqref{eq:eff_force}. 
 (d) Extracted memory kernel and orthogonal force ACF (divided by $\langle \dot{A}_0^2 \rangle$) from the DP-GLE ($\Gamma^\text{DP}(t)$ and $F_\text{Q}^\text{DP}(t)$) in Eq.~\eqref{eq:dp_gle} (black), CM-GLE ($\Gamma^\text{CM}(t)$ and $F_\text{Q}^\text{CM}(t)$) in Eq.~\eqref{eq:const_mass_gle} (green) and the Mori-GLE ($\Gamma^\text{M}(t)$ and $F_\text{Q}^\text{M}(t)$) in Eq.~\eqref{eq:gle_mori} (red). Solid lines denote the memory kernels from the Volterra extraction scheme (see supplementary material Sec.~\ref{app:MemKernExtrac} for details) and broken lines denote the corresponding orthogonal force ACFs, computed via Eq.~\eqref{eq:random_force}, where the parameters are calculated using Eq.~\eqref{eq:dp_gle} (black), Eq.~\eqref{eq:const_mass_gle} (green) and Eq.~\eqref{eq:gle_mori} (red). 
    (e) Residual memory kernels $\Gamma_{\Delta}(t)$ computed from the difference between the memory kernel and orthogonal force ACF data in (d). (f) Distribution of the orthogonal forces $F^\text{DP}_\text{Q}(t)$, $F^\text{CM}_\text{Q}(t)$ and $F^\text{M}_\text{Q}(t)$, normalized by their maximal values in the y-axis, compared with a Gaussian distribution with zero mean and variance $\langle \dot{A}_0^2 \rangle\Gamma^\text{DP}(0)$ (blue broken line). The black broken line represents an exponential function ($\sim e^{-F_\text{Q}/F_\text{Q}(0)}$), where $F_\text{Q}(0)$ is a typical force constant.}
    \label{fig:butane_check_fdt}
\end{figure*}
The memory kernel in Eq.~\eqref{eq:dp_gle} is a sum of a memory kernel $\Gamma^\text{DP}_\text{Q}(t)$ connected to the orthogonal force $F^\text{DP}_\text{Q}(t)$ via
\begin{equation}
    \label{eq:FDT_dp}
    \Gamma^\text{DP}_\text{Q}(t) = \frac{\langle F^\text{DP}_\text{Q}(0),F^\text{DP}_\text{Q}(t)\rangle}{\langle \Dot{A}_0^2 \rangle},
\end{equation}
and a residual part \cite{vroylandt2022gle, wolf2025cross}
\begin{equation}
    \label{eq:kernel_dp_res}
\Gamma^\text{DP}_{\Delta}(t) = \Gamma^\text{DP}(t) -\Gamma^\text{DP}_\text{Q}(t).
\end{equation}
For further discussion, see supplementary material Sec.~\ref{app:proof_equiv_kernels}.\\
\indent In this work, we compare the extracted parameters from the DP-GLE with a simplified DP-GLE ~\cite{plotkin1998non,medina2018transition,kappler2019cyclization,ayaz2021non,satija2019generalized,brunig2022timedependent,brunig2022pair,dalton2022protein}.
If the velocity distribution does not depend on $A$, the mass $M(A)$ is configuration-independent, i.e. $M(A) = M_0$, and the DP-GLE in Eq.~\eqref{eq:dp_gle} simplifies to
\begin{align}
    \nonumber
    \Ddot{A}(t) = & - M_0^{-1}\frac{d}{dA} U_{\text{PMF}}\bigl(A(t)\bigr) + F^\text{CM}_\text{Q}(t) \\ \label{eq:const_mass_gle} & - \int_0^t ds\:\Gamma^\text{CM}(t-s) \Dot{A}(s).
\end{align}
In the following, we refer to the GLE in Eq.~\eqref{eq:const_mass_gle} as the constant-mass (CM-)GLE. The approximation of a constant mass has been shown to be accurate for several systems \cite{lee2019multi,ayaz2022generalized, dalton2022protein,brunig2022timedependent, dalton2024memory,wolf2025cross}, and has been used for Markovian embedding simulations \cite{ceriotti2010colored, li2017computing, klippenstein2021introducing, ayaz_embedding_nl}. 

\subsection{Obtaining GLE parameters from time-series data}
From the Mori-GLE in Eq.~\eqref{eq:gle_mori}, all parameters can be extracted from time-series trajectories $A(t)$ by well-established Volterra extraction methods \cite{berne1970calculation,shin2010brownian,kowalik2019memory}. The same is true for the DP-GLE in Eq.~\eqref{eq:dp_gle} and the CM-GLE in Eq.~\eqref{eq:const_mass_gle} \cite{darve2009computing, daldrop2018butane,kowalik2019memory,ayaz2021non}; see supplementary material Sec.~\ref{app:MemKernExtrac} for details.
The difference memory kernel $\Gamma^\text{DP}_{\Delta}(t)$ in Eq.~\eqref{eq:kernel_dp_res} is directly accessible from data via the Volterra extraction method: From a given trajectory $A(t)$, we can extract the memory kernel $\Gamma^\text{DP}(t)$ from the DP-GLE in Eq.~\eqref{eq:dp_gle} by the Volterra scheme. 
Using the acceleration $\ddot{A}(t)$, the effective force $F_{\text{eff}}\bigl(A(t)\bigr) $ in Eq.~\eqref{eq:eff_force} and the memory kernel, we can then compute the orthogonal force from the trajectory $A(t)$ by rearranging Eq.~\eqref{eq:dp_gle} according to 
\begin{equation}
    \label{eq:random_force}
    F^\text{DP}_\text{Q}(t) = \Ddot{A}(t) + F_{\text{eff}}\bigl(A(t)\bigr) + \int_0^t ds\:\Gamma^\text{DP}(t-s) \Dot{A}(s).
\end{equation}
If the autocorrelation function (ACF) of $F^\text{DP}_\text{Q}(t)$ does not equal the memory kernel $\Gamma^\text{DP}(t)$, i.e. $\langle F^\text{DP}_\text{Q}(0), F^\text{DP}_\text{Q}(t) \rangle \neq \langle \Dot{A}_0^2 \rangle \Gamma^\text{DP}(t)$, then $\Gamma^\text{DP}_{\Delta}(t) \neq 0$ as given in Eq.~\eqref{eq:kernel_dp_res}, which, in terms of the hybrid GLE in Eq.~\eqref{eq:hybrid_gle}, means that position-dependent friction effects are present in the trajectory \cite{vroylandt2022gle, wolf2025cross} (see supplementary material Sec.~\ref{app:proof_equiv_kernels} for further discussion). 
\\\indent Accordingly, the orthogonal forces $F_\text{Q}^\text{M}(t)$ and $F_\text{Q}^\text{CM}(t)$ can be calculated from Eq.~\eqref{eq:random_force} adapted for the Mori-GLE in Eq.~\eqref{eq:gle_mori} and the CM-GLE in Eq.~\eqref{eq:const_mass_gle}, respectively.


\section{GLE Parameters and Orthogonal Forces From Butane Dihedral Dynamics}
We compare the parameters obtained for the different GLE types for an observable that is a non-linear function of atomic distances, namely the dihedral angle dynamics of a butane molecule in water, obtained from
MD simulations (see Appendix \ref{sec:mds} for details). It has been shown in a previous study \cite{ayaz2022generalized}, that the butane dihedral angle $A$ exhibits a configuration-dependent mass $M(A)$ and significant position-dependent frictional effects within the hybrid GLE in Eq.~\eqref{eq:hybrid_gle}, making this observable ideal for the analyses with different GLEs.
\\ \indent In Fig.~\ref{fig:butane_check_fdt}(a), we show the computed position-dependent mass $M(A)$ (see supplementary material Sec.~\ref{app:MemKernExtrac} for details) together with the constant mass $M_0$, obtained via $M_0 = k_BT/\langle \Dot{A}_0^2 \rangle$. The mass coordinate-dependence only slightly influences the potential $U(A)$ in Fig.~\ref{fig:butane_check_fdt}(b) (see definition in caption) and the effective force $F_{\text{eff}}\bigl(A(t)\bigr)$ (Eq.~\eqref{eq:eff_force}) in Fig.~\ref{fig:butane_check_fdt}(c). \\
\indent The memory kernel $\Gamma^\text{DP}(t)$ from the DP-GLE in Eq.~\eqref{eq:dp_gle} is displayed as a black solid line in Fig.~\ref{fig:butane_check_fdt}(d), extracted using the Volterra extraction scheme \cite{ayaz2021non} as explained in supplementary material Sec.~\ref{app:MemKernExtrac}. Comparing the data with the corresponding orthogonal force ACF (black broken line), i.e. $\Gamma^\text{DP}_\text{Q} = \langle F^\text{DP}_\text{Q}(0), F^\text{DP}_\text{Q}(t)\rangle / \langle \Dot{A}_0^2\rangle$, computed via Eq.~\eqref{eq:random_force}, in Fig.~\ref{fig:butane_check_fdt}(e) we observe small but distinct differences  $\Gamma^\text{DP}_{\Delta}(t) = \Gamma^\text{DP}(t) - \Gamma^\text{DP}_\text{Q}(t)$ in the time regime between 0.01 and 1 ps. This difference is related to the position-dependent memory kernel $\Gamma^\text{H}_\text{NL}$ in the hybrid GLE (see supplementary material Sec.~\ref{app:hybrid_dp_butane} for details). Thus, we demonstrate that the DP-GLE violates the equality between the orthogonal force ACF and the memory kernel, given in Eq.~\eqref{eq:FDT_Mori} for the Mori-GLE, which means that position-dependent friction is present \cite{wolf2025cross}. \\
\indent We suggest that these position-dependent friction effects, quantified by $\Gamma^\text{H}_\text{NL}$, arise from the coupling of internal degrees of freedom in the freely moving butane molecule in water, denoted as internal friction \cite{daldrop2018butane}. This follows from the fact that, as shown in supplementary material Sec.~\ref{app:results_2constr}, we find only weak position-dependent friction for a system with less internal friction, namely MD simulations of butane where the two inner carbon atoms are fixed in space.\\ 
\indent The memory kernel $\Gamma^\text{CM}(t)$ from the CM-GLE for the butane dihedral angle, displayed in Fig.~\ref{fig:butane_check_fdt}(d) (green line), slightly differs from the result for the DP-GLE (black line), signifying the influence of the configuration-dependent mass $M(A)$ in Fig.~\ref{fig:butane_check_fdt}(a) on the memory kernel. The difference $\Gamma_{\Delta}(t)$ between the memory kernel $\Gamma^\text{CM}(t)$ and the corresponding orthogonal force ACF (green broken line, computed by Eq.~\eqref{eq:random_force} with the CM-GLE), visible as green line in Fig.~\ref{fig:butane_check_fdt}(e), is higher than for the DP-GLE.
\\\indent The memory kernel $\Gamma^\text{M}(t)$ and the orthogonal force ACF from the Mori-GLE in Eq.~\eqref{eq:gle_mori} (red lines in Fig.~\ref{fig:butane_check_fdt}(d)) perfectly overlap as given by Eq.~\eqref{eq:FDT_Mori}, demonstrating the accuracy of the orthogonal force computation via Eq.~\eqref{eq:random_force}. 
The kernel differences $\Gamma_{\Delta}(t)$ from the DP-GLE and from the CM-GLE shown in Fig.~\ref{fig:butane_check_fdt}(e) are significantly larger than the Mori-GLE result, which demonstrates that the former are not due to numerical artifacts.\\ 
\begin{figure*}
	\centering
	\includegraphics[width=0.7\textwidth]{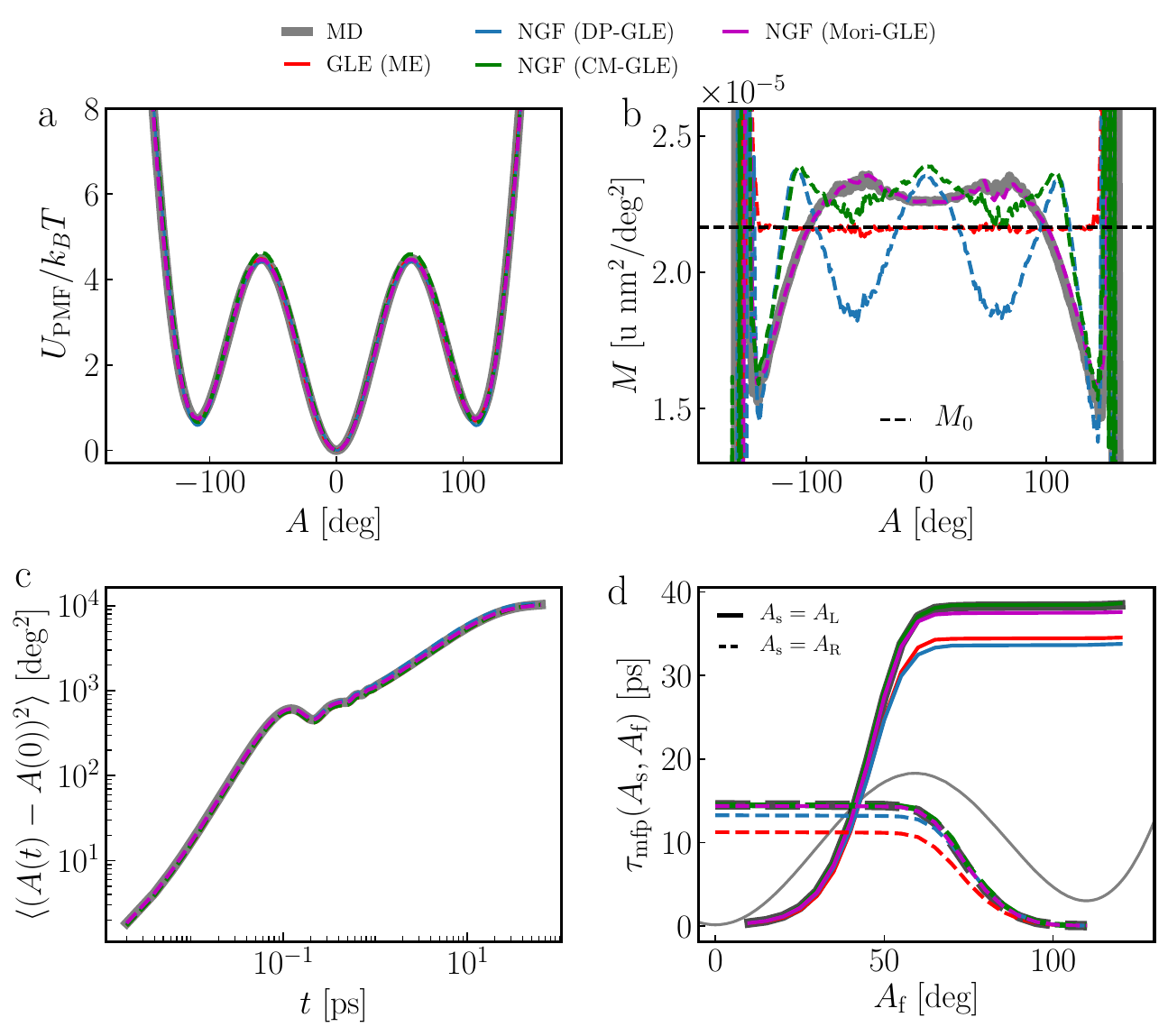}
	\caption{Results for GLE simulations of butane dihedral dynamics. 
		(a - c) Comparison of the PMF $U_{\text{PMF}}(A)$ (a), shown in Fig.~\ref{fig:butane_check_fdt}(b), the position-dependent mass $M(A)$ (b), shown in Fig.~\ref{fig:butane_check_fdt}(a), and the mean-squared displacement $\langle \left(A(t)-A(0)\right)^2 \rangle$ (c) between the MD data (black) and different GLE simulation approaches (colored).
		NGF simulations are performed using the procedure described in Appendix \ref{sec:ngf}, for the GLEs in Eq.~\eqref{eq:dp_gle} (blue), in Eq.~\eqref{eq:const_mass_gle} (green) and in Eq.~\eqref{eq:gle_mori} (purple).
		The data in red visualizes results from Markovian embedding simulations (ME) of the CM-GLE, which assumes the orthogonal force distribution to be Gaussian (see supplementary material Sec.~\ref{app:const_mass_ME} for details). Exemplary trajectories for the NGF simulations, where the initial conditions for the orthogonal force computation do no match with those from the NGF simulations, can be found in Fig.~\ref{fig:butane_brute_vf_samples_diff_init_cond} of supplementary material Sec.~\ref{app:acc_brute}.
		(d) Comparison between the mean first-passage time $\tau_{\text{mfp}}$ of the butane dihedral angle from MD simulations (black) and predicted by Markovian embedding (ME) GLE (red) and NGF GLE simulations, as a function of the final position $A_\text{f}$ for different starting positions $A_\text{s}$ ($A_\text{L} = 0$ deg (trans-state) and $A_\text{R} = 110$ deg (cis-state)). The statistical errors are calculated but are smaller than the data linewidth. The gray curve displays the PMF landscape from Fig.~\ref{fig:butane_check_fdt}(b).} 
	\label{fig:butane4}
\end{figure*}
\indent In Fig.~\ref{fig:butane_check_fdt}(f), the orthogonal force distributions from all GLEs reveal deviations from a Gaussian distribution with zero mean and variance $\langle \Dot{A}_0^2 \rangle\Gamma^\text{DP}(0)$ (blue broken line).
This non-Gaussian behavior presumably originates from the interaction between butane and the solvent, as discussed in supplementary material Sec.~\ref{app:expotails}. Non-Gaussianity indicates that the two-point autocorrelation cannot fully describe the orthogonal force, and higher-order correlation functions are needed (supplementary material Sec.~\ref{app:higher_orders_fr}). 
The orthogonal force distributions from DP-GLE and CM-GLE are rather similar, and decay exponentially for high forces (black broken line in Fig.~\ref{fig:butane_check_fdt}(f)). The non-Gaussianity is enhanced for $F_\text{Q}^\text{M}$ (red data); we remark distinct shoulders, presumably due to the relegation of the non-linear PMF component in the deterministic force and of position-dependent friction into $F_\text{Q}^\text{M}$.

\section{Simulating GLEs with non-Gaussian orthogonal forces}

The central question of this work is which GLE discussed so far yields the most accurate comparison with the MD simulations, and, particularly, how non-Gaussian contributions in the orthogonal force can be incorporated into a GLE simulation to accurately reproduce the dynamics of butane isomerization. Since the hybrid and DP-GLE are equivalent, as demonstrated in supplementary material Sec.~\ref{app:proof_equiv_kernels}, simulating the hybrid GLE in Eq.~\eqref{eq:hybrid_gle} is unnecessary. Observing in Fig.~\ref{fig:butane_check_fdt}(e) that $\Gamma^\text{DP}_\Delta \neq 0$ implies that the memory kernel $\Gamma^\text{DP}(t)$ in the DP-GLE in Eq.~\eqref{eq:dp_gle} is not proportional to the orthogonal force $F^\text{DP}_\text{Q}(t)$; performing a Markovian embedding simulation \cite{ayaz2021non, brunig2022timedependent, brunig2022pair} with purely Gaussian $F_\text{Q}^\text{DP}(t)$ would thus lead to inaccurate predictions of the system’s dynamics, especially since the orthogonal force is non-Gaussian distributed (see Fig.~\ref{fig:butane_check_fdt}(f)). Therefore, we implement an alternative approach by directly solving the GLE with extracted parameters for the memory kernel, deterministic force, and an accurate realization of the non-Gaussian orthogonal force. \\
\indent In this technique, which we refer to as the non-Gaussian force (NGF) method, realizations of the orthogonal force are computed from the MD trajectories using Eq.~\eqref{eq:random_force} and then used for the numerical solution of the GLE. 
The orthogonal-force realizations are derived from position trajectories and initial conditions different from those used as input for the NGF simulation. We apply the NGF method to different GLEs; more details are given in Appendix \ref{sec:ngf}. \\
\begin{figure*}
	\centering
	\includegraphics[width=1
	\linewidth]{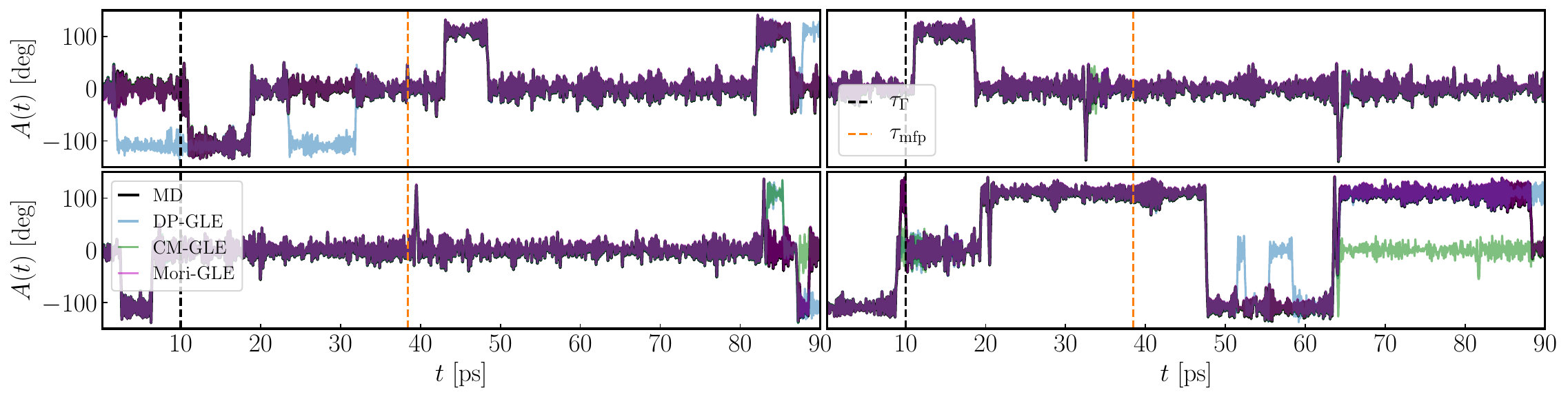}
	\caption{Investigation of the NGF simulation numerical accuracy for butane dihedral trajectories. We compare three different dihedral trajectories $A(t)$ from GLE simulations with the MD data (black). Contrary to the results shown in Fig.~\ref{fig:butane4}, we here simulate the GLE by computing the orthogonal force $F_\text{Q}(t)$ according to Eq.~\eqref{eq:random_force} starting at an arbitrarily chosen time and inserting them into an NGF simulation according to Eq.~\eqref{eq:ODE_discret} in Appendix \ref{sec:ngf} starting at the same time, with the initial conditions $A_0$ and $\dot{A}_0$ matching the MD simulations. Different colors denote results for different GLEs: The blue trajectories are generated from the DP-GLE in Eq.~\eqref{eq:dp_gle}, the green trajectories from the CM-GLE in Eq.~\eqref{eq:const_mass_gle}, and the purple data is from the Mori-GLE in Eq.~\eqref{eq:gle_mori}. The black broken line illustrates the time $\tau_\Gamma \approx$ 10 ps where the extracted memory kernel $\Gamma^\text{DP}(t)$ in Fig.~\ref{fig:butane_check_fdt}(d) has approximately decayed to zero. The orange broken line represents the mean first-passage time $\tau_{\text{mfp}}$ shown in Fig.~\ref{fig:butane4}(d) going from the trans- to the cis-state. }
	\label{fig:butane_brute_vf_samples}
\end{figure*} 
\indent Fig.~\ref{fig:butane4} depicts NGF simulation results for the Mori-GLE in Eq.~\eqref{eq:gle_mori}, for the DP-GLE in Eq.~\eqref{eq:dp_gle}, and for the CM-GLE in Eq.~\eqref{eq:const_mass_gle}. Additionally, we show simulation results from a Markovian embedding simulation, where we approximate the extracted memory kernel $\Gamma^\text{CM}$ in the CM-GLE by a sum of oscillatory-exponential functions, and model the orthogonal force as a stationary Gaussian process similar to the relation in Eq.~\eqref{eq:FDT_Mori} for the Mori-GLE (see supplementary material Sec.~\ref{app:const_mass_ME} for details).\\
\indent  To test the accuracy of these different GLE simulations, we compare the PMF $U_\text{PMF}(A)$ (a), the position-dependent mass $M(A)$ (b) and the mean-squared displacement $\langle \left(A(t)-A(0)\right)^2 \rangle$ (c) from GLE simulatons with those from the MD data in black. All simulations reproduce the PMF and the mean-squared displacement very well. Further discussion about the self-consistency of the simulations is presented in supplementary material Sec.~\ref{app:ngf_poc_butane}. The position-dependent mass in Fig.~\ref{fig:butane4})(b), on the other hand, is only accurately simulated with the Mori-GLE. This is at first sight surprising since the Mori-GLE does not involve a position-dependent mass and reflects the numerical robustness of the Mori-GLE.\\
\indent To evaluate the accuracy of the GLE models, we compare computed mean first-passage time profiles ($\tau_\text{mfp}$) \cite{ayaz2021non} of the simulated GLE trajectories with the MD simulation (see Appendix \ref{sec:mfpts} for details).  The $\tau_\text{mfp}$-profiles from the CM-GLE (green) and the Mori-GLE (purple) are in very good agreement with the MD data (black), and are significantly improved compared to the Markovian embedding scheme (red), strongly suggesting that non-Gaussian behavior of the orthogonal force has an influence on the dihedral dynamics. In supplementary material Sec.~\ref{app:butane_params_higher_visc}, NGF simulations of butane in water with higher viscosities $\eta/\eta_0$ underline our findings.
\\ \indent Remarkably, the Mori-GLE provides an accurate prediction of all statistical observables such as the mean-squared displacement in Fig.~\ref{fig:butane4}(c) or mean first-passage times in Fig.~\ref{fig:butane4}(d). Although it includes only a harmonic potential, the orthogonal force alone contains all necessary information to guide the system to the correct equilibrium non-Gaussian distribution in the simulation, see Fig.~\ref{fig:butane4}(a). 
\\ \indent The optimal representation of non-Markovian systems can only be achieved with a precise description of the full first-passage time distribution \cite{zhou2024rapid}. In supplementary material Sec.~\ref{app:transition_times_butane}, we discuss the prediction of transition-path and first-passage time distributions, demonstrating that the Mori-GLE accurately reproduces both short- and long-term transitions. In contrast, the Markovian embedding (ME) technique disproportionately enhances state-recrossing events, which accounts for its inferior performance.
\\ \indent We note that the NGF simulations of the Mori- and DP-GLEs are in principle exact, except numerical inaccuracies. Although being approximate, a CM-GLE simulation for the dihedral angle (green) outperforms the DP-GLE (blue), as visible in Fig.~\ref{fig:butane4}(b, d). The NGF technique based on Eq.~\eqref{eq:ODE_discret} in Appendix \ref{sec:ngf} without numerical errors should, hypothetically, perfectly reconstruct a trajectory $A(t)$ starting from initial conditions $A_0$ and $\dot{A}_0$, if using matching past velocities $\dot{A}(t)$ and the appropriate orthogonal force computed from Eq.~\eqref{eq:random_force}. In Fig.~\ref{fig:butane_brute_vf_samples}, we investigate the numerical accuracy of the NGF simulation, where we present samples of reconstructed dihedral trajectories $A(t)$ from GLE simulations compared with the MD trajectories of the dihedral angle. We reconstruct a trajectory by computing the orthogonal force $F_\text{Q}(t)$ according to Eq.~\eqref{eq:random_force} starting at an arbitrary time and, in contrast to the results shown in Fig.~\ref{fig:butane4}, inserting them into an NGF simulation starting at the same time of $A(t)$, using the initial position $A_0$ and initial velocity $\dot{A}_0$ matching the MD simulation. The blue trajectories are generated from the DP-GLE in Eq.~\eqref{eq:dp_gle}, the green trajectories from the CM-GLE in Eq.~\eqref{eq:const_mass_gle}, and the purple data stems from the Mori-GLE in Eq.~\eqref{eq:gle_mori}. \\
\indent For the CM- and Mori-GLEs, even at longer simulation times than $\tau_\Gamma \approx$ 10 ps, being the time the extracted memory kernel $\Gamma^\text{DP}(t)$ in Fig.~\ref{fig:butane_check_fdt}(d) has decayed to zero, the reconstruction looks fairly well. Numerical inaccuracies are mainly seen for the DP-GLE, which involves a position-dependent mass $M(A)$. Deviations in the CM-GLE simulation become evident at times longer than the mean first-passage time from the trans- to the cis-state, i.e. $\tau_{\text{mfp}} \approx 38$ ps (see orange broken line in Fig.~\ref{fig:butane_brute_vf_samples}). More discussion on the accuracy of the GLE simulations is given in supplementary material Sec.~\ref{app:acc_brute}. We hypothesize that the dephasing effect visible in Fig.~\ref{fig:butane_brute_vf_samples} arises from a desynchronization between the orthogonal force and the non-linear PMF, which leads to numerical inaccuracies in both the DP-GLE and CM-GLE simulations (for details, see supplementary material Sec.~\ref{app:pos_dep_fr}), being visible in Fig.~\ref{fig:butane4}(b, d). The numerical inaccuracy of DP-GLE and CM-GLE simulations mirrors the fact that they cannot reproduce the position-dependent mass $M(A)$ of the dihedral angle in Fig.~\ref{fig:butane4}(b).\\ 
\indent In essence, we find that the Mori-GLE most accurately reproduces the MD simulation results, provided the orthogonal force is drawn from the correct non-Gaussian distribution. We conclude that the correct modeling of the orthogonal force appears to be more relevant than including the non-linear force from the PMF.

\section{Conclusions}
By analyzing MD trajectories of the butane dihedral angle, we find that all GLEs considered in this work feature a non-Gaussian orthogonal force, with the non-Gaussianity being most pronounced for the Mori-GLE. The choice of GLE depends on the observable onto which the complex system’s dynamics is projected. For harmonic confinement, the Mori-GLE is a natural choice. For observables with a non-Gaussian distribution, the hybrid or DP-GLEs seem better suited; however, we note that the hybrid GLE cannot be simulated by an Markovian embedding scheme \cite{ayaz_embedding_nl}. In fact, we have demonstrated that the Mori-GLE can  very accurately reproduce the dynamics of even non-linear observables, provided that the non-Gaussianity of the orthogonal force is properly accounted for. We find that a CM-GLE model is also rather accurate for simulations of the butane dihedral angle. \\
\indent The Mori-GLE as an exact model offers considerable robustness as a simulation framework, with the orthogonal force being the sole non-linear term in the equation of motion, minimizing the influence of numerical dephasing. Therefore, accurately modeling the orthogonal-force statistics is more crucial than including the non-linear PMF or position-dependent memory kernels. Additionally, the Mori-GLE provides practical advantages over GLEs with a non-harmonic PMF as the parameter extraction is less sensitive to discretization effects, a key consideration when analyzing experimental data \cite{mitterwallner2020non, klimek2023data, tepper2024accurate, kiefer2024predictability}. \\
\indent Our investigation of NGF GLE simulations suggests that methods for sampling non-Gaussian orthogonal forces, which go beyond Markovian embedding, are crucially important. Current Markovian embedding methods cannot accurately reproduce key statistical observables, such as the mean first-passage time, when non-Gaussianity effects are prominent. For NGF GLE simulations, we used orthogonal-force trajectories derived from MD simulations; an independent sampling method for non-Gaussian, higher-order correlated trajectories would be desirable.

\section*{Supplementary Material}
The supplementary material includes supporting theory and additional results and is attached as a separate file.\\

\section*{Acknowledgements}
We gratefully acknowledge funding from the Deutsche Forschungsgemeinschaft (DFG) via Grant No. SFB 1114 Project Id 235221301, Project C02 and No. SFB 1449 Project Id 431232613, Project A02, and
funding from the European Research Council (ERC) under the European Union’s Horizon 2020 Research and Innovation Program under Grant Agreement No. 835117 (NoMaMemo). We acknowledge computing time on the HPC clusters at the Physics department and ZEDAT, FU Berlin.

\section*{Author Declarations}
\subsection*{Conflict of Interests}
The authors declare no competing interests.

\subsection*{Author Contributions}

\noindent \textbf{Henrik Kiefer}: Investigation (equal);
Methodology (equal); Validation (equal); Visualization (equal); Writing – original draft (equal). 
\textbf{Benjamin J. A. Héry}: Investigation (supporting);
Methodology (supporting); Writing – original draft (supporting).
\textbf{Lucas Tepper}: Methodology (supporting). 
\textbf{Benjamin A. Dalton}: Investigation (supporting);
Methodology (supporting). 
\textbf{Cihan Ayaz}: Conceptualization
(supporting); Methodology (supporting). 
\textbf{Roland R. Netz}: Conceptualization
(equal); Funding acquisition (equal); Methodology (equal); Supervision (equal); Writing - original draft (equal).


\section*{Data Availability}
Input files of the MD simulations and custom computer codes for GLE analyses and simulations are available from the corresponding author upon reasonable request.

\appendix
\section{MD Simulations}
\label{sec:mds}
We perform MD simulations using the GROMACS software package (version 2021.3) \cite{pronk2013gromacs}, where we place a single butane molecule in a cubic water box with a length of 4.1 nm. We use
a GROMOS united-atom force field for butane \cite{Oostenbrink_2004} that neglects the hydrogens and approximates butane by four Lennard–Jones beads subject to fixed bond lengths and fixed bond angles (SHAKE algorithm \cite{Ryckaert_1977}).
For the 1000 water molecules, we use the extended simple point charge (SPC/E)
model \cite{berendsen1987missing}. We pre-equilibrate in an NPT ensemble for 1 ns, using a Berendsen barostat \cite{Berendsen_1984} set to 1 atm. We perform constant particle number, volume, and
temperature (NVT) MD simulations with a total length of 1$\:\mu$s and an MD time step of 2 fs. The temperature $T$ = 300 K is controlled by the velocity rescaling thermostat \cite{bussi2007canonical}, coupled only to the solvent with a time constant of $\eta/\eta_0\:\cdot$ 1 ps. For electrostatics, we use the particle-mesh Ewald method \cite{Darden_1993}, with a cut-off length of 1 nm. The center-of-mass
motion of the simulation system was removed every 0.2 ps.\\
\indent Moreover, we perform MD simulations with varying water molecule mass $m_\text{w}$ to change the water viscosity while keeping the butane mass fixed \cite{daldrop2018butane} (supplementary material Sec.~\ref{app:butane_params_higher_visc}). This modifies all intrinsic water time scales, particularly the water viscosity, according to $\eta \propto \sqrt{m_\text{w}}$. For the results we show in supplementary material Sec.~\ref{app:results_2constr}, we run MD simulations where the two inner carbon atoms of the butane molecule are frozen \cite{daldrop2018butane}.
\section{NGF Simulations Using the GLE}
\label{sec:ngf}
We discretize the GLE in Eq.~\eqref{eq:dp_gle} (and analogous for the GLEs in Eq.~\eqref{eq:gle_mori} and Eq.~\eqref{eq:const_mass_gle}) using a discrete time step $\Delta t$ = 2 fs, i.e. $A(i\Delta t) = A_i$, according to
\begin{equation}
	\Ddot{A}_i= - F_{\text{eff}}(A_i) - \Delta t \sum_{j=0}^i \omega_{i,j}\:\Gamma^\text{DP}_j \Dot{A}_{i-j} + F^{\text{DP},i}_\text{Q},
	\label{eq:ODE_discret}
\end{equation}
and solve this equation for $A$ utilizing a 4th-order Runge-Kutta integrator.
The term $\omega_{ij} = 1 - \delta_{ij}/2 - \delta_{0j}/2$ follows from the trapezoidal rule for the memory-integral in Eq.~\eqref{eq:dp_gle}. Starting with the last known position $A_l$ at time $\tilde{t}=t-t_{l}$, we insert a realization of the orthogonal force $F^{\text{DP},i}_\text{Q}$, the memory kernel $\Gamma^\text{DP}_i$, as well as the past time steps $\Dot{A}_{i-j}$ in each iteration step to obtain the next trajectory point $A_{i+1}$ consecutively. 
For multi-step simulations, we solve the GLE one step ahead and treat the simulated $A_{i+1}$ and $\dot{A}_{i+1}$ as past values for the next simulation step $A_{i+2}$, and so forth. To accelerate the numerical solution of Eq.~\eqref{eq:ODE_discret}, we truncate the memory kernel $\Gamma^\text{DP}_i$ after a length of $\tau_\Gamma$, which is the time after which the memory kernel has decayed to zero. For all memory kernels ($\Gamma^\text{DP}(t)$, $\Gamma^\text{CM}(t)$ and $\Gamma^\text{M}(t)$), we used $\tau_\Gamma$ = 10 ps.\\
\indent We compute realizations of $F^\text{DP}_\text{Q}(t)$ from the given MD trajectories using Eq.~\eqref{eq:random_force} and insert them as realizations to solve Eq.~\eqref{eq:ODE_discret} numerically. In detail, for an NGF simulation, we generate 500 trajectories with a discrete time step of 2 fs and a length of $\:10^6$ time steps, which is in total 1 $\mu$s, the same length as for the MD simulations. The initial conditions of $A$ and $\dot{A}$ of the 500 trajectories are randomly selected from the MD simulations. For the results we show in Fig.~\ref{fig:butane4}, we ensure that the orthogonal-force realizations do not originate from the same position trajectory and initial conditions we use as input for the NGF simulations. Contrary, for the simulations in Fig.~\ref{fig:butane_brute_vf_samples}, we use orthogonal forces from initial conditions that match those from the NGF simulations, to reconstruct the MD trajectory.
Note that Eq.~\eqref{eq:random_force} and Eq.~\eqref{eq:ODE_discret} are modified for the GLEs in Eq.~\eqref{eq:gle_mori} and Eq.~\eqref{eq:const_mass_gle}.\\
\indent In supplementary material Sec.~\ref{app:ngf_poc_butane}, we demonstrate the self-consistency of the NGF simulation technique for the butane dihedral angle. 
In supplementary material Sec.~\ref{app:ngf_model_system}, we discuss the self-consistency of the NGF simulation technique for a model system. 
\section{Calculation of Mean First-Passage Time Profiles}
\label{sec:mfpts}
We compute the mean first-passage time $\tau_\text{{mfp}}$($A_\text{s}$,$A_\text{f}$) from trajectories $A(t)$ as the mean time to go from one position $A_\text{s}$ to some final position $A_\text{f}$. Here, our starting positions are chosen as the minima in the PMF in Fig.~\ref{fig:butane_check_fdt}(b), where the butane molecule is in the trans-state at a dihedral angle of $A = 0$ deg and in the cis-state at $A = \pm 110$ deg. From the simulations, we obtain the
first-passage time (FPT) distribution, defined as the distribution of times necessary for the particle to go from
$A_\text{s}$ to $A_\text{f}$ for the first time. By averaging over the FPT distribution, thereby considering all FPT events including recrossings of the initial state $A_\text{s}$ \cite{dalton2024role}, we obtain the mean first-passage time $\tau_\text{mfp}$. The $\tau_\text{mfp}$-values are computed for different combinations of $A_\text{s}$ and $A_\text{f}$.

\bibliographystyle{unsrt}
\bibliography{refs}

\begin{thebibliography}{10}

\bibitem{plotkin1998non}
S.~S. Plotkin and P.~G. Wolynes.
\newblock {Non-Markovian Configurational Diffusion and Reaction Coordinates for
  Protein Folding}.
\newblock {\em Physical Review Letters}, 80(22):5015, 1998.

\bibitem{monticelli2008martini}
L.~Monticelli, S.~K. Kandasamy, X.~Periole, R.~G. Larson, D.~P. Tieleman, and
  S.-J. Marrink.
\newblock {The MARTINI Coarse-Grained Force Field: Extension to Proteins}.
\newblock {\em Journal of Chemical Theory and Computation}, 4(5):819--834,
  2008.

\bibitem{clementi2008coarse}
C.~Clementi.
\newblock {Coarse-Grained Models of Protein Folding: Toy Models or Predictive
  Tools?}
\newblock {\em Current Opinion in Structural Biology}, 18(1):10--15, 2008.

\bibitem{hills2009insights}
R.~D. Hills~Jr and C.~L. Brooks~III.
\newblock {Insights from Coarse-Grained G{\=o} Models for Protein Folding and
  Dynamics}.
\newblock {\em International Journal of Molecular Sciences}, 10(3):889--905,
  2009.

\bibitem{minichino1997potential}
C.~Minichino and G.~A. Voth.
\newblock {Potential Energy Surfaces for Chemical Reactions: An Analytical
  Representation From Coarse Grained Data with an Application to Proton
  Transfer in Water}.
\newblock {\em The Journal of Physical Chemistry B}, 101(23):4544--4552, 1997.

\bibitem{kamerlin2011coarse}
S.~C.~L. Kamerlin, S.~Vicatos, A.~Dryga, and A.~Warshel.
\newblock {Coarse-Grained (Multiscale) Simulations in Studies of Biophysical
  and Chemical Systems}.
\newblock {\em Annual Review of Physical Chemistry}, 62(1):41--64, 2011.

\bibitem{zhou2019variational}
S.~Zhou, R.~G. Wei{\ss}, L.-T. Cheng, J.~Dzubiella, J.~A. McCammon, and B.~Li.
\newblock {Variational Implicit-Solvent Predictions of the Dry--Wet Transition
  Pathways for Ligand--Receptor Binding and Unbinding Kinetics}.
\newblock {\em Proceedings of the National Academy of Sciences},
  116(30):14989--14994, 2019.

\bibitem{brunig2021proton}
F.~N. Br{\"u}nig, P.~Hillmann, W.~K. Kim, J.~O. Daldrop, and R.~R. Netz.
\newblock {Proton-Transfer Spectroscopy Beyond the Normal-Mode Scenario}.
\newblock {\em The Journal of Chemical Physics}, 157(17):174116, 2022.

\bibitem{berne1966calculation}
B.~J. Berne, J.~P. Boon, and S.~A. Rice.
\newblock {On the Calculation of Autocorrelation Functions of Dynamical
  Variables}.
\newblock {\em The Journal of Chemical Physics}, 45(4):1086--1096, 1966.

\bibitem{adelman1980generalized}
S.~A. Adelman.
\newblock {Generalized Langevin Theory for Many-Body Problems in Chemical
  Dynamics: Reactions in Liquids}.
\newblock {\em The Journal of Chemical Physics}, 73(7):3145--3158, 1980.

\bibitem{berne1990dynamic}
B.~J. Berne, M.~E. Tuckerman, J.~E. Straub, and A.~L.~R. Bug.
\newblock {Dynamic Friction on Rigid and Flexible Bonds}.
\newblock {\em The Journal of Chemical Physics}, 93(7):5084--5095, 1990.

\bibitem{chorin2000optimal}
A.~J. Chorin, O.~H. Hald, and R.~Kupferman.
\newblock {Optimal Prediction and the Mori-Zwanzig Representation of
  Irreversible Processes}.
\newblock {\em Proceedings of the National Academy of Sciences},
  97(7):2968--2973, 2000.

\bibitem{chorin2002optimal}
A.~J. Chorin, O.~H. Hald, and R.~Kupferman.
\newblock {Optimal Prediction with Memory}.
\newblock {\em Physica D: Nonlinear Phenomena}, 166(3-4):239--257, 2002.

\bibitem{darve2006numerical}
E.~Darve.
\newblock {Numerical Methods for Calculating the Potential of Mean Force}.
\newblock {\em New Algorithms for Macromolecular Simulation}, pages 213--249,
  2006.

\bibitem{lange2006collective}
O.~F. Lange and H.~Grubm{\"u}ller.
\newblock {Collective Langevin Dynamics of Conformational Motions in Proteins}.
\newblock {\em The Journal of Chemical Physics}, 124(21):214903, 2006.

\bibitem{carof2014two}
A.~Carof, R.~Vuilleumier, and B.~Rotenberg.
\newblock {Two Algorithms to Compute Projected Correlation Functions in
  Molecular Dynamics Simulations}.
\newblock {\em The Journal of Chemical Physics}, 140(12):124103, 2014.

\bibitem{ma2016derivation}
L.~Ma, X.~Li, and C.~Liu.
\newblock {The Derivation and Approximation of Coarse-Grained Dynamics from
  Langevin Dynamics}.
\newblock {\em The Journal of Chemical Physics}, 145(20):204117, 2016.

\bibitem{lee2019multi}
H.~S. Lee, S.-H. Ahn, and E.~F. Darve.
\newblock {The Multi-Dimensional Generalized Langevin Equation for
  Conformational Motion of Proteins}.
\newblock {\em The Journal of Chemical Physics}, 150(17):174113, 2019.

\bibitem{loos2019heat}
S.~A.~M. Loos and S.~H.~L. Klapp.
\newblock {Heat Flow Due to Time-Delayed Feedback}.
\newblock {\em Scientific reports}, 9(1):2491, 2019.

\bibitem{klippenstein2021introducing}
V.~Klippenstein, M.~Tripathy, G.~Jung, F.~Schmid, and N.~F.~A. van~der Vegt.
\newblock {Introducing Memory in Coarse-grained Molecular Simulations}.
\newblock {\em The Journal of Physical Chemistry B}, 125(19):4931--4954, 2021.

\bibitem{doerries2021correlation}
T.~J. Doerries, S.~A.~M. Loos, and S.~H.~L. Klapp.
\newblock {Correlation Functions of Non-Markovian Systems out of Equilibrium:
  Analytical Expressions Beyond Single-Exponential Memory}.
\newblock {\em Journal of Statistical Mechanics: Theory and Experiment},
  2021(3):033202, 2021.

\bibitem{bagchi1983effect}
B.~Bagchi and D.~W. Oxtoby.
\newblock {The Effect of Frequency Dependent Friction on Isomerization Dynamics
  in Solution}.
\newblock {\em The Journal of Chemical Physics}, 78(5):2735--2741, 1983.

\bibitem{straub1987calculation}
J.~E. Straub, M.~Borkovec, and B.~J. Berne.
\newblock {Calculation of Dynamic Friction on Intramolecular Degrees of
  Freedom}.
\newblock {\em Journal of Physical Chemistry}, 91(19):4995--4998, 1987.

\bibitem{canales1998generalized}
M.~Canales and G.~Sese.
\newblock {Generalized Langevin Dynamics Simulations of NaCl Electrolyte
  Solutions}.
\newblock {\em The Journal of ChemicalPphysics}, 109(14):6004--6011, 1998.

\bibitem{satija2019generalized}
R.~Satija and D.~E. Makarov.
\newblock {Generalized Langevin equation as a Model for Barrier Crossing
  Dynamics in Biomolecular Folding}.
\newblock {\em The Journal of Physical Chemistry B}, 123(4):802--810, 2019.

\bibitem{ayaz2021non}
C.~Ayaz, L.~Tepper, F.~N. Br{\"u}nig, J.~Kappler, J.~O. Daldrop, and R.~R.
  Netz.
\newblock {Non-Markovian Modeling of Protein Folding}.
\newblock 118(31):e2023856118, 2021.

\bibitem{dalton2022protein}
B.~A. Dalton, C.~Ayaz, H.~Kiefer, A.~Klimek, L.~Tepper, and R.~R. Netz.
\newblock {Fast Protein Folding is Governed by Memory-Dependent Friction}.
\newblock {\em Proceedings of the National Academy of Sciences},
  120(31):e2220068120, 2023.

\bibitem{dalton2024role}
B.~A. Dalton, H.~Kiefer, and R.~R. Netz.
\newblock {The Role of Memory-Dependent Friction and Solvent Viscosity in
  Isomerization Kinetics in Viscogenic Media}.
\newblock {\em Nature Communications}, 15(1):3761, 2024.

\bibitem{milster2024tracer}
S.~Milster, F.~Koch, C.~Widder, T.~Schilling, and J.~Dzubiella.
\newblock {Tracer Dynamics in Polymer Networks: Generalized Langevin
  dDscription}.
\newblock {\em The Journal of Chemical Physics}, 160(9), 2024.

\bibitem{brunig2022pair}
F.~N. Br{\"u}nig, J.~O. Daldrop, and R.~R. Netz.
\newblock {Pair-Reaction Dynamics in Water: Competition of Memory, Potential
  Shape, and Inertial Effects}.
\newblock {\em The Journal of Physical Chemistry B}, 126(49):10295--10304,
  2022.

\bibitem{vroylandt2022likelihood}
H.~Vroylandt, L.~Gouden{\`e}ge, P.~Monmarch{\'e}, F.~Pietrucci, and
  B.~Rotenberg.
\newblock {Likelihood-Based Non-Markovian Models from Molecular Dynamics}.
\newblock {\em Proceedings of the National Academy of Sciences},
  119(13):e2117586119, 2022.

\bibitem{brunig2022timedependent}
F.~N. Br{\"u}nig, O.~Geburtig, A.~von Canal, J.~Kappler, and R.~R. Netz.
\newblock {Time-Dependent Friction Effects on Vibrational Infrared Frequencies
  and Line Shapes of Liquid Water}.
\newblock {\em The Journal of Physical Chemistry B}, 126(7):1579--1589, 2022.

\bibitem{zwanzig1961memory}
R.~Zwanzig.
\newblock {Memory Effects in Irreversible Thermodynamics}.
\newblock {\em Physical Review}, 124(4):983, 1961.

\bibitem{mori1965transport}
H.~Mori.
\newblock {Transport, Collective Motion, and Brownian Motion}.
\newblock {\em Progress of Theoretical Physics}, 33(3):423--455, 1965.

\bibitem{mazur1991and}
P.~Mazur and D.~Bedeaux.
\newblock {When and Why is the Random Force in Brownian Motion a Gaussian
  Process}.
\newblock {\em Biophysical Chemistry}, 41(1):41--49, 1991.

\bibitem{vroylandt2022position}
H.~Vroylandt and P.~Monmarch{\'e}.
\newblock {Position-Dependent Memory Kernel in Generalized Langevin Equations:
  Theory and Numerical Estimation}.
\newblock {\em The Journal of Chemical Physics}, 156(24):244105, 2022.

\bibitem{li2017computing}
Z.~Li, H.~S. Lee, E.~Darve, and G.~E. Karniadakis.
\newblock {Computing the Non-Markovian Coarse-Grained Interactions Derived from
  the Mori-Zwanzig Formalism in Molecular Systems: Application to Polymer
  Melts}.
\newblock {\em The Journal of Chemical Physics}, 146(1):014104, 2017.

\bibitem{di2022derivation}
L.~Di~Cairano.
\newblock {On the Derivation of a Nonlinear Generalized Langevin Equation}.
\newblock {\em Journal of Physics Communications}, 6(1):015002, 2022.

\bibitem{jung2023dynamic}
Bernd Jung and Gerhard Jung.
\newblock {Dynamic Coarse-Graining of Linear and Non-Linear Systems:
  Mori-Zwanzig Formalism and Beyond}.
\newblock {\em The Journal of Chemical Physics}, 159(8), 2023.

\bibitem{ayaz2022generalized}
C.~Ayaz, L.~Scalfi, B.~A. Dalton, and R.~R. Netz.
\newblock {Generalized Langevin Equation with a Nonlinear Potential of Mean
  Force and Nonlinear Memory Friction From a Hybrid Projection Scheme}.
\newblock {\em Physical Review E}, 105:054138, 2022.

\bibitem{daldrop2018butane}
J.~O. Daldrop, J.~Kappler, F.~N. Br{\"u}nig, and R.~R. Netz.
\newblock {Butane Dihedral Angle Dynamics in Water is Dominated by Internal
  Friction}.
\newblock {\em Proceedings of the National Academy of Sciences},
  115(20):5169--5174, 2018.

\bibitem{dalton2024memory}
B.~A. Dalton, A.~Klimek, H.~Kiefer, F.~N. Br{\"u}nig, H.~Colinet, L.~Tepper,
  A.~Abbasi, and R.~R. Netz.
\newblock {Memory and Friction: From the Nanoscale to the Macroscale}.
\newblock {\em Annual Review of Physical Chemistry}, 76, 2024.

\bibitem{vroylandt2022gle}
H.~Vroylandt.
\newblock {On the Derivation of the Generalized Langevin Equation and the
  Fluctuation-Dissipation Theorem}.
\newblock {\em Europhysics Letters}, 140(6):62003, 2022.

\bibitem{ceriotti2010colored}
M.~Ceriotti, G.~Bussi, and M.~Parrinello.
\newblock {Colored-noise Thermostats {\`a} la Carte}.
\newblock {\em Journal of Chemical Theory and Computation}, 6(4):1170--1180,
  2010.

\bibitem{wiesenfeld1994stochastic}
K.~Wiesenfeld, D.~Pierson, E.~Pantazelou, C.~Dames, and F.~Moss.
\newblock {Stochastic Resonance on a Circle}.
\newblock {\em Physical Review Letters}, 72(14):2125, 1994.

\bibitem{wio1999stochastic}
H.~S. Wio and S.~Bouzat.
\newblock {Stochastic Resonance: The Role of Potential Asymmetry and Non
  Gaussian Noises}.
\newblock {\em Brazilian Journal of Physics}, 29:136--143, 1999.

\bibitem{wang2009anomalous}
B.~Wang, S.~M Anthony, S.~C. Bae, and S.~Granick.
\newblock {Anomalous Yet Brownian}.
\newblock {\em Proceedings of the National Academy of Sciences},
  106(36):15160--15164, 2009.

\bibitem{wang2012brownian}
B.~Wang, J.~Kuo, S.~C. Bae, and S.~Granick.
\newblock {When Brownian Diffusion is Not Gaussian}.
\newblock {\em Nature Materials}, 11(6):481--485, 2012.

\bibitem{shin2010brownian}
H.~K. Shin, C.~Kim, P.~Talkner, and E.~K. Lee.
\newblock {Brownian Motion from Molecular Dynamics}.
\newblock {\em Chemical Physics}, 375(2-3):316--326, 2010.

\bibitem{kanazawa2015minimal}
K.~Kanazawa, T.~G. Sano, T.~Sagawa, and H.~Hayakawa.
\newblock {Minimal Model of Stochastic Athermal Systems: Origin of Non-Gaussian
  Noise}.
\newblock {\em Physical Review Letters}, 114(9):090601, 2015.

\bibitem{zwanzig2001chemical}
R.~Zwanzig.
\newblock {A Chemical Langevin Equation with Non-Gaussian Noise}.
\newblock {\em The Journal of Physical Chemistry B}, 105(28):6472--6473, 2001.

\bibitem{wio2004effect}
H.~S. Wio and R.~Toral.
\newblock {Effect of Non-Gaussian Noise Sources in a Noise-Induced Transition}.
\newblock {\em Physica D: Nonlinear Phenomena}, 193(1-4):161--168, 2004.

\bibitem{majee2005colored}
P.~Majee, G.~Goswami, and B.~C. Bag.
\newblock {Colored Non-Gaussian Noise Induced Resonant Activation}.
\newblock {\em Chemical Physics Letters}, 416(4-6):256--260, 2005.

\bibitem{chechkin2017brownian}
A.~V. Chechkin, F.~Seno, R.~Metzler, and I.~M. Sokolov.
\newblock {Brownian Yet Non-Gaussian Diffusion: from Superstatistics to
  Subordination of Diffusing Diffusivities}.
\newblock {\em Physical Review X}, 7(2):021002, 2017.

\bibitem{mutothya2021first}
N.~M. Mutothya, Y.~Xu, Y.~Li, R.~Metzler, and N.~M. Mutua.
\newblock {First Passage Dynamics of Stochastic Motion in Heterogeneous Media
  Driven by Correlated White Gaussian and Coloured Non-Gaussian Noises}.
\newblock {\em Journal of Physics: Complexity}, 2(4):045012, 2021.

\bibitem{baule2023exponential}
A.~Baule and P.~Sollich.
\newblock {Exponential Increase of Transition Rates in Metastable Systems
  Driven by Non-Gaussian Noise}.
\newblock {\em Scientific Reports}, 13(1):3853, 2023.

\bibitem{caspers2024nonlinear}
J.~Caspers and M.~Kr{\"u}ger.
\newblock {Nonlinear Langevin Functionals for a Driven Probe}.
\newblock {\em The Journal of Chemical Physics}, 161(12), 2024.

\bibitem{widder2022generalized}
C.~Widder, F.~Koch, and T.~Schilling.
\newblock {Generalized Langevin Dynamics Simulation with Non-Stationary Memory
  Kernels: How to Make Noise}.
\newblock {\em The Journal of Chemical Physics}, 157(19), 2022.

\bibitem{glatzel2022interplay}
F.~Glatzel and T.~Schilling.
\newblock {The Interplay Between Memory and Potentials of Mean Force: A
  Discussion on the Structure of Equations of Motion for Coarse-Grained
  Observables}.
\newblock {\em Europhysics Letters}, 136(3):36001, 2022.

\bibitem{zwanzig2001nonequilibrium}
R.~Zwanzig.
\newblock {\em {Nonequilibrium Statistical Mechanics}}.
\newblock Oxford university press, 2001.

\bibitem{wolf2025cross}
N.~Wolf, V.~Klippenstein, and N.~F.~A. Van~der Vegt.
\newblock {Cross-Correlations in the Fluctuation--Dissipation Relation
  Influence Barrier-Crossing Dynamics}.
\newblock {\em The Journal of Chemical Physics}, 162(5), 2025.

\bibitem{medina2018transition}
E.~Medina, R.~Satija, and D.~E. Makarov.
\newblock {Transition Path Times in Non-Markovian Activated Rate Processes}.
\newblock {\em The Journal of Physical Chemistry B}, 122(49):11400--11413,
  2018.

\bibitem{kappler2019cyclization}
J.~Kappler, F.~No{\'e}, and R.~R. Netz.
\newblock {Cyclization and Relaxation Dynamics of Finite-Length Collapsed
  Self-Avoiding Polymers}.
\newblock {\em Physical Review Letters}, 122(6):067801, 2019.

\bibitem{ayaz_embedding_nl}
C.~Ayaz, L.~Tepper, and R.~R. Netz.
\newblock {Self-Consistent Markovian Embedding of Generalized Langevin
  Equations with Configuration-Dependent Mass and a Nonlinear Friction Kernel}.
\newblock {\em Turkish Journal of Physics}, 46(6):194--205, 2022.

\bibitem{berne1970calculation}
B.~J. Berne and G.~D. Harp.
\newblock {On the Calculation of Time Correlation Functions}.
\newblock {\em Adv. Chem. Phys}, 17:63--227, 1970.

\bibitem{kowalik2019memory}
B.~Kowalik, J.~O. Daldrop, J.~Kappler, J.~C.~F. Schulz, A.~Schlaich, and R.~R.
  Netz.
\newblock {Memory-Kernel Extraction for Different Molecular Solutes in Solvents
  of Varying Viscosity in Confinement}.
\newblock {\em Physical Review E}, 100(1):012126, 2019.

\bibitem{darve2009computing}
E.~Darve, J.~Solomon, and A.~Kia.
\newblock {Computing Generalized Langevin Equations and Generalized
  Fokker--Planck Equations}.
\newblock {\em Proceedings of the National Academy of Sciences},
  106(27):10884--10889, 2009.

\bibitem{zhou2024rapid}
Q.~Zhou, R.~R. Netz, and B.~A. Dalton.
\newblock {Rapid State-Recrossing Kinetics in Non-Markovian Systems}.
\newblock {\em arXiv preprint arXiv:2403.06604}, 2024.

\bibitem{mitterwallner2020non}
B.~G. Mitterwallner, C.~Schreiber, J.~O. Daldrop, J.~O. R{\"a}dler, and R.~R.
  Netz.
\newblock {Non-Markovian Data-Driven Modeling of Single-Cell Motility}.
\newblock {\em Physical Review E}, 101(3):032408, 2020.

\bibitem{klimek2023data}
A.~Klimek, D.~Mondal, S.~Block, P.~Sharma, and R.~R. Netz.
\newblock {Data-driven Classification of Individual Cells by their
  Non-Markovian Motion}.
\newblock {\em Biophysical Journal}, 123(10):1173--1183, 2024.

\bibitem{tepper2024accurate}
L.~Tepper, B.~A. Dalton, and R.~R. Netz.
\newblock {Accurate Memory Kernel Extraction from Discretized Time-Series
  Data}.
\newblock {\em Journal of Chemical Theory and Computation}, 2024.

\bibitem{kiefer2024predictability}
H.~Kiefer, D.~Furtel, C.~Ayaz, A.~Klimek, J.~O. Daldrop, and R.~R. Netz.
\newblock {Predictability Analysis and Prediction of Discrete Weather and
  Financial Time-Series Data with a Hamiltonian-Based Filter-Projection
  Approach}.
\newblock {\em arXiv preprint arXiv:2409.15026}, 2024.

\bibitem{pronk2013gromacs}
S.~Pronk, S.~P{\'a}ll, R.~Schulz, P.~Larsson, P.~Bjelkmar, R.~Apostolov, M.~R.
  Shirts, J.~C. Smith, P.~M. Kasson, D.~van~der Spoel, et~al.
\newblock {GROMACS 4.5: a High-Throughput and Highly Parallel Open Source
  Molecular Simulation Toolkit}.
\newblock {\em Bioinformatics}, 29(7):845--854, 2013.

\bibitem{Oostenbrink_2004}
C.~Oostenbrink, A.~Villa, A.~E. Mark, and W.~F. {Van Gunsteren}.
\newblock {A Biomolecular Force Field based on the Free Enthalpy of Hydration
  and Solvation: The GROMOS Force-Field Parameter Sets 53A5 and 53A6}.
\newblock {\em Journal of Computational Chemistry}, 25(13):1656--1676, oct
  2004.

\bibitem{Ryckaert_1977}
J.-P. Ryckaert, G.~Ciccotti, and H.~J.~C. Berendsen.
\newblock {Numerical Integration of the Cartesian Equations of Motion of a
  System with Constraints: Molecular Dynamics of n-Alkanes}.
\newblock {\em Journal of Computational Physics}, 23(3):327--341, 1977.

\bibitem{berendsen1987missing}
H.~J.~C. Berendsen, J.~R. Grigera, and T.~P. Straatsma.
\newblock {The Missing Term in Effective Pair Potentials}.
\newblock {\em Journal of Physical Chemistry}, 91(24):6269--6271, 1987.

\bibitem{Berendsen_1984}
H.~J.~C. Berendsen, J.~P.~M. Postma, W.~F. van Gunsteren, A.~DiNola, and J.~R.
  Haak.
\newblock {Molecular Dynamics with Coupling to an External Bath}.
\newblock {\em The Journal of Chemical Physics}, 81(8):3684--3690, 1984.

\bibitem{bussi2007canonical}
G.~Bussi, D.~Donadio, and M.~Parrinello.
\newblock {Canonical Sampling Through Velocity Rescaling}.
\newblock {\em The Journal of Chemical Physics}, 126(1):014101, 2007.

\bibitem{Darden_1993}
T.~Darden, D.~York, and L.~Pedersen.
\newblock {Particle Mesh Ewald: An N log(N) Method for Ewald Sums in Large
  Systems}.
\newblock {\em The Journal of Chemical Physics}, 98(12):10089--10092, jun 1993.

\bibitem{dyson1949radiation}
F.~J. Dyson.
\newblock {The Radiation Theories of Tomonaga, Schwinger, and Feynman}.
\newblock {\em Physical Review}, 75(3):486, 1949.

\bibitem{grabert1980microdynamics}
H.~Grabert, P.~H{\"a}nggi, and P.~Talkner.
\newblock {Microdynamics and Nonlinear Stochastic Processes of Gross
  Variables}.
\newblock {\em Journal of Statistical Physics}, 22(5):537--552, 1980.

\bibitem{samanta2021dielectric}
T.~Samanta and D.~V. Matyushov.
\newblock {Dielectric Friction, Violation of the Stokes-Einstein-Debye
  Relation, and Non-Gaussian Transport Dynamics of Dipolar Solutes in Water}.
\newblock {\em Physical Review Research}, 3(2):023025, 2021.

\bibitem{touchette2018introduction}
H.~Touchette.
\newblock {Introduction to Dynamical Large Deviations of Markov Processes}.
\newblock {\em Physica A: Statistical Mechanics and its Applications},
  504:5--19, 2018.

\bibitem{jack2020ergodicity}
R.~L. Jack.
\newblock {Ergodicity and Large Deviations in Physical Systems with Stochastic
  Dynamics}.
\newblock {\em The European Physical Journal B}, 93:1--22, 2020.

\bibitem{cherstvy2019non}
A.~G. Cherstvy, S.~Thapa, C.~E. Wagner, and R.~Metzler.
\newblock {Non-Gaussian, Non-Ergodic, and Non-Fickian Diffusion of Tracers in
  Mucin Hydrogels}.
\newblock {\em Soft Matter}, 15(12):2526--2551, 2019.

\bibitem{kappler2018memory}
J.~Kappler, J.~O. Daldrop, F.~N. Br{\"u}nig, M.~D. Boehle, and R.~R. Netz.
\newblock {Memory-Induced Acceleration and Slowdown of Barrier Crossing}.
\newblock {\em The Journal of Chemical Physics}, 148(1):014903, 2018.

\end{thebibliography}

\renewcommand{\thesection}{\arabic{section}} 
\setcounter{section}{0} 
\renewcommand{\appendixname}{} 

\newpage
\pagebreak
\newpage
\widetext
\begin{center}
	\textbf{\large Supplementary Material for: Analysis and Simulation of Generalized Langevin Equations with Non-Gaussian Orthogonal Forces}\\
	\vspace{5mm}
	\textbf{Henrik Kiefer$^1$, Benjamin J. A. Héry$^1$, Lucas Tepper$^1$, Benjamin A. Dalton$^1$, Cihan Ayaz$^1$, Roland R. Netz$^{1}$}\\
	\vspace{1mm}
	\textit{$^1$ Department of Physics,  Freie  Universität  Berlin,  Arnimallee  14,  14195  Berlin,  Germany}\\
	
\end{center}

\setcounter{equation}{0}
\setcounter{figure}{0}
\setcounter{table}{0}
\setcounter{page}{1}
\makeatletter
\renewcommand{\thepage}{S\arabic{page}}
\renewcommand{\thesection}{S\arabic{section}}
\renewcommand{\thefigure}{S\arabic{figure}}
\renewcommand{\theequation}{S\arabic{equation}}

\section{Relation Between DP-GLE and Hybrid GLE}
\label{app:proof_equiv_kernels}
The hybrid GLE in Eq.~\eqref{eq:hybrid_gle} and the DP-GLE in Eq.~\eqref{eq:dp_gle} in the main text origin from the same Eq.~\eqref{eq:basic_gle}, but are derived by using different projection operators for the decomposition in Eq.~\eqref{eq:projection_liouville} ($\mathcal{P}_0 = 1- Q_0)$ and in the Dyson decomposition to rewrite the second term on the r.h.s. in Eq.~\eqref{eq:projection_liouville} in the main text ($\mathcal{P}_1=1- Q_1$)  \cite{dyson1949radiation,vroylandt2022gle}, compare Eq.~\eqref{eq:basic_gle}.
To obtain the hybrid GLE in Eq.~\eqref{eq:hybrid_gle} in the main text, one chooses the same hybrid projection $\mathcal{P}_\text{H}$ for both operators, i.e. $\mathcal{P}_0 =  \mathcal{P}_1 =  \mathcal{P}_\text{H}$ \cite{ayaz2022generalized},
\begin{eqnarray}
	\label{eq:op_hybrid}
	\mathcal{P}_\text{H} A(t) = \langle A(t) \rangle_{A_0} + \frac{\langle A(t), \dot{A}_0 \rangle}{\langle  \dot{A}_0^2\rangle}  \dot{A}_0.
\end{eqnarray}
Note that this projection operator defines $A_0$ as a function of the phase-space positions $\vec{q}$ only, i.e., $A_0 = A(\vec{q},0)$, and $\dot{A}_0 = \dot{A}(\vec{q},\vec{p},0)$.
The orthogonal force $F^\text{H}_\text{Q}(t)$ is consequently given by $F^\text{H}_\text{Q}(t) = e^{tQ_\text{H}L}Q_\text{H}L\dot{A}_0$.
The conditional correlation function between two phase-space dependent observables $B(t)$ and $C(t)$ depending on the position $A_\text{s}$ is defined by \cite{grabert1980microdynamics, ayaz2022generalized}
\begin{equation}
	\label{eq:cond_corr}
	\langle B(t), C(t')\rangle_{A_\text{s}} =\frac{\langle \delta\bigl(A(\hat{\omega})-A(\omega_\text{s})\bigr),B(\hat{\omega},t)C(\hat{\omega},t') \rangle }{\langle\delta\bigl(A(\hat{\omega})-A(\omega_\text{s})\bigr)\rangle},
\end{equation}
where a phase-space variable with a hat denotes the variable that is integrated over. The memory kernel $\Gamma_\text{L}^\text{H}(t)$ in the hybrid GLE is defined via
\begin{align}
	\label{eq:hybrid-kernel}
	\Gamma_\text{L}^\text{H}(t) &\equiv \frac{\langle \Ddot{A}_{0}, F_\text{Q}^\text{H}(t) \rangle}{\langle \Dot{A}_{0}^{2} \rangle} = \frac{\langle F^\text{H}_\text{Q}(0), F_\text{Q}^\text{H}(t) \rangle}{\langle \Dot{A}_{0}^{2} \rangle},
\end{align}
where we used that $Q_\text{H}$ is self adjoint.
Contrary, to derive the DP-GLE in Eq.~\eqref{eq:dp_gle} from Eq.~\eqref{eq:basic_gle} in the main text, one uses $\mathcal{P}_0 = \mathcal{P}_\text{H}$, but the second operator in Eq.~\eqref{eq:basic_gle} is chosen to be the Mori projector, i.e. $\mathcal{P}_1 = \mathcal{P}_\text{M}$, defined as \cite{mori1965transport}
\begin{eqnarray}
	\label{eq:op_mori}
	\mathcal{P}_\text{M}  A(t) = \langle A(t) \rangle + \frac{\langle A(t), A_0 - \langle A_0 \rangle \rangle}{\langle  (A_0 - \langle A_0 \rangle)^2 \rangle}  (A_0 - \langle A_0 \rangle) + \frac{\langle A(t), \dot{A}_0 \rangle}{\langle  \dot{A}_0^2\rangle}  \dot{A}_0.
\end{eqnarray}
The orthogonal force $F^\text{DP}_\text{Q}(t)$ results as $F^\text{DP}_\text{Q}(t) = e^{tQ_\text{M}L}Q_\text{H}L\dot{A}_0$. The memory kernel $\Gamma^\text{DP}(t)$ in the DP-GLE is given by \cite{vroylandt2022gle}
\begin{align}
	\label{eq:DP-kernel}
	\Gamma^\text{DP}(t) &\equiv \frac{\langle \Ddot{A}_{0}, F_\text{Q}^\text{DP}(t) \rangle}{\langle \Dot{A}_{0}^{2} \rangle} = \frac{\langle F^\text{DP}_\text{Q}(0), F_\text{Q}^\text{DP}(t) \rangle}{\langle \Dot{A}_{0}^{2} \rangle}  + \frac{\langle F_\text{eff}(A_0), F_\text{Q}^\text{DP}(t) \rangle}{\langle \Dot{A}_{0}^{2} \rangle},\\
	&=  \Gamma_\text{Q}^\text{DP}(t) +  \Gamma_\Delta^\text{DP}(t).
\end{align}
The residual kernel $\Gamma^\text{DP}_{\Delta}(t)$ can be related to the position-dependent kernel $\Gamma^\text{H}_\text{NL}$ in the hybrid GLE (Eq.~\eqref{eq:hybrid_gle} in the main text) \cite{vroylandt2022gle}. Moreover, it was proposed that the orthogonal force ACFs of $\Gamma_\text{Q}^\text{DP}(t)$ and $\Gamma^\text{H}_\text{L}(t)$ should be equal \cite{vroylandt2022gle}. In order to relate the two orthogonal forces $F_\text{Q}^\text{DP}(t)$ and $F_\text{Q}^\text{H}(t)$, we first note that the l.h.s. of Eqs.~\eqref{eq:hybrid_gle} and \eqref{eq:dp_gle} in the main text are equal. 
Moreover, each of these equations of motion displays in their r.h.s. the same relevant force $F_\text{eff}\left(A(t)\right)$ (Eq.~\eqref{eq:eff_force} in the main text), which simplifies the equality to 
\begin{equation}
	- \int_{0}^{t} ds \: \Gamma_\text{L}^\text{H}(t-s) \Dot{A}(s) + \int_{0}^{t} ds \: \Gamma_\text{NL}^\text{H}\bigl(A(s),t-s\bigr) + F_\text{Q}^\text{H}(t) = - \int_{0}^{t} ds \: \Gamma^\text{DP}(t-s) \Dot{A}(s) + F_\text{Q}^\text{DP}(t).
	\label{eq:proof_fdts2}
\end{equation}
We define the difference between the orthogonal forces of the DP- and hybrid GLEs via
\begin{equation}
	\Delta F_\text{Q}(t) \equiv F_\text{Q}^\text{DP}(t) - F_\text{Q}^\text{H}(t),
	\label{eq:proof_fdts3}
\end{equation}
and use the definitions of $F_\text{Q}^\text{H}(t)$ and $F_\text{Q}^\text{DP}(t)$ from Eq.~\eqref{eq:FDT_hybrid} and Eq.~\eqref{eq:FDT_dp}, respectively, to rewrite Eq.~\eqref{eq:proof_fdts2} as 
\begin{equation}
	\Delta F_\text{Q}(t) - \int_{0}^{t} ds \: \frac{\langle \Ddot{A}_{0}, \Delta F_\text{Q}(s) \rangle}{\langle \Dot{A}_{0}^{2} \rangle} \Dot{A}(t-s) = \int_{0}^{t} ds \: \Gamma_\text{NL}^\text{H}\bigl(A( s),t-s \bigr),
	\label{eq:proof_fdts4}
\end{equation}
where we used Eqs.~\eqref{eq:hybrid-kernel} and ~\eqref{eq:DP-kernel}.
Eq.~\eqref{eq:proof_fdts4} is an integral equation for $\Delta F_\text{Q}(t)$, which has the explicit solution
\begin{align}
	\Delta F_\text{Q}(t)  = &   \int_{0}^{t} ds \: \Gamma_\text{NL}^\text{H}\bigl(A(s),t-s\bigr)  + \int_{0}^{t} dt_{1} \int_{0}^{t_{1}} dt_{2} \: \Dot{A}(t-t_{1}) \frac{\langle \Ddot{A}_{0}, \Gamma_\text{NL}^\text{H}\bigl(t_{1}-t_{2}, A(t_{2})\bigr) \rangle}{\langle \Dot{A}_{0}^{2} \rangle} \nonumber \\ & +
	\sum_{n \geq 2} \prod_{i = 1}^{n+1} \int_{0}^{t\delta_{i,1} + t_{i-1}(1-\delta_{i,1})} dt_{i} \Dot{A}(t-t_{1}) \prod_{j=2}^{n} \frac{\langle \Ddot{A}_{0}, \Dot{A}(t_{j-1} - t_{j}) \rangle}{\langle \Dot{A}_{0}^{2}\rangle} \frac{\langle \Ddot{A}_{0}, \Gamma_\text{NL}^\text{H}\bigl(t_{n}-t_{n+1}, A(t_{n+1})\bigr) \rangle}{\langle \Dot{A}_{0}^{2} \rangle},
	\label{eq:proof_fdts5}
\end{align}
which includes terms that are nested integrals, where
\begin{align}
	\prod_{i = 1}^{n} \int_{0}^{t\delta_{i,1} + t_{i-1}(1-\delta_{i,1})} dt_{i} \equiv \int_{0}^{t} dt_{1} \times ... \times \int_{0}^{t_{n-1}} dt_{n}.
	\label{eq:nested_integral}
\end{align}
The fact that Eq.~\eqref{eq:proof_fdts5} solves Eq.~\eqref{eq:proof_fdts4} can be easily seen by insertion.
Recalling Eq.~\eqref{eq:proof_fdts3}, we obtain an expression for $F_\text{Q}^\text{DP}(t)$ as a function of $F_\text{Q}^\text{H}(t)$
\begin{align}
	F_\text{Q}^\text{DP}(t)  = & \: F_\text{Q}^\text{H}(t) +  \int_{0}^{t} ds \: \Gamma_\text{NL}^\text{H}\bigl(t-s, A(s)\bigr)  + \int_{0}^{t} dt_{1} \int_{0}^{t_{1}} dt_{2} \: \Dot{A}(t-t_{1}) \frac{\langle \Ddot{A}_{0}, \Gamma_\text{NL}^\text{H}\bigl(t_{1}-t_{2}, A(t_{2})\bigr) \rangle}{\langle \Dot{A}_{0}^{2} \rangle} \nonumber \\
	& + \sum_{n \geq 2} \prod_{i = 1}^{n+1} \int_{0}^{t\delta_{i,1} + t_{i-1}(1-\delta_{i,1})} dt_{i} \Dot{A}(t-t_{1}) \prod_{j=2}^{n} \frac{\langle \Ddot{A}_{0}, \Dot{A}(t_{j-1} - t_{j}) \rangle}{\langle \Dot{A}_{0}^{2}\rangle} \frac{\langle \Ddot{A}_{0}, \Gamma_\text{NL}^\text{H}\bigl(t_{n}-t_{n+1}, A(t_{n+1})\bigr) \rangle}{\langle \Dot{A}_{0}^{2} \rangle},
	\label{eq:proof_fdts6}
\end{align}
which satisfies the initial condition $F_\text{Q}^\text{DP}(0) = F_\text{Q}^\text{H}(0)$.
We further compare the two autocorrelation memory kernels $\Gamma_\text{Q}^\text{DP}$ and $\Gamma_\text{L}^\text{H}$ by multiplying Eq.~\eqref{eq:proof_fdts6} with $F_\text{Q}^\text{DP}(0) = F_\text{Q}^\text{H}(0)$ and taking an ensemble average. Using Eqs.~\eqref{eq:FDT_hybrid} and \eqref{eq:FDT_dp}, we find that

\begin{align}
	\Delta\Gamma(t) \equiv	\Gamma_\text{Q}^\text{DP}(t) - \Gamma_\text{L}^\text{H}(t) = & \:  \int_{0}^{t} ds \: \frac{\langle F_\text{Q}^\text{H}(0), \Gamma_\text{NL}^\text{H}\bigl(t-s, A(s)\bigr) \rangle}{\langle \Dot{A}_{0}^{2}\rangle}  + \int_{0}^{t} dt_{1} \int_{0}^{t_{1}} dt_{2} \: \frac{\langle F_\text{Q}^\text{H}(0), \Dot{A}(t-t_{1}) \rangle}{\langle \Dot{A}_{0}^{2} \rangle} \frac{\langle \Ddot{A}_{0}, \Gamma_\text{NL}^\text{H}\bigl(t_{1}-t_{2}, A(t_{2})\bigr) \rangle}{\langle \Dot{A}_{0}^{2} \rangle}  \nonumber \\  +
	\sum_{n \geq 2} \prod_{i = 1}^{n+1}  \int_{0}^{t\delta_{i,1} + t_{i-1}(1-\delta_{i,1})} & dt_{i} \frac{\langle F_\text{Q}^\text{H}(0), \Dot{A}(t-t_{1}) \rangle}{\langle \Dot{A}_{0}^{2}\rangle} \prod_{j=2}^{n} \frac{\langle \Ddot{A}_{0}, \Dot{A}(t_{j-1} - t_{j}) \rangle}{\langle \Dot{A}_{0}^{2}\rangle} \frac{\langle \Ddot{A}_{0}, \Gamma_\text{NL}^\text{H}\bigl(t_{n}-t_{n+1}, A(t_{n+1})\bigr) \rangle}{\langle \Dot{A}_{0}^{2} \rangle}.
	\label{eq:proof_fdts7}
\end{align}
We see that the r.h.s. contains a countable infinite number of terms that may not vanish in general. Since $\mathcal{P}_\text{M}$ and $\mathcal{P}_\text{H}$ are not the same projection operators, the ACFs of the two orthogonal forces are not equal.
\\ \indent In the following, we compute a Taylor expansion of $\Delta\Gamma(t)$ around $t = 0$, and see how the difference between the orthogonal force ACFs in Eq.~\eqref{eq:proof_fdts7} behaves. Since the r.h.s. in Eq.~\eqref{eq:proof_fdts7} is a sum of integrals from $0$ to $t$, we obtain that 
\begin{eqnarray}
	\Delta\Gamma(0) = 0,
	\label{eq:proof_fdts8}
\end{eqnarray}
consistent with the fact that at time $t = 0$, the two complementary forces are  the same, i.e. $F_\text{Q}^\text{H}(0) = F_\text{Q}^\text{DP}(0)$.\\
\indent Computing the time derivatives of the integrals from $0$ to $t$, the first time derivative of the r.h.s. in Eq.~\eqref{eq:proof_fdts7} gives two terms for each integral

\begin{gather}
	\Delta\Gamma^{(1)}(t) = \frac{\langle F_\text{Q}^\text{H}(0), \Gamma_\text{NL}^\text{H}\bigl(0, A(t)\bigr) \rangle}{\langle \Dot{A}_{0}^{2}\rangle}  + \nonumber \\
	\int_{0}^{t} ds \: \frac{\langle F_\text{Q}^\text{H}(0), \partial_{t}\Gamma_\text{NL}^\text{H}\bigl(t-s, A(s)\bigr) \rangle}{\langle \Dot{A}_{0}^{2}\rangle}  + \frac{\langle F_\text{Q}^\text{H}(0), \Dot{A}_{0} \rangle}{\langle \Dot{A}_{0}^{2}\rangle} \left( \int_{0}^{t} ds \: \frac{\langle \Ddot{A}_{0}, \Gamma_\text{NL}^\text{H}\bigl(t-s, A(s)\bigr) \rangle}{\langle \Dot{A}_{0}^{2}\rangle} \right. \nonumber \\
	\left. + \sum_{n \geq 2} \prod_{i = 2}^{n+1} \int_{0}^{t\delta_{i,2} + t_{i-1}(1-\delta_{i,2})} dt_{i}  \frac{\langle \Ddot{A}_{0}, \Dot{A}(t - t_{2}) \rangle}{\langle \Dot{A}_{0}^{2}\rangle}  \prod_{j=3}^{n} \frac{\langle \Ddot{A}_{0}, \Dot{A}(t_{j-1} - t_{j}) \rangle}{\langle \Dot{A}_{0}^{2}\rangle} \frac{\langle \Ddot{A}_{0}, \Gamma_\text{NL}^\text{H}\bigl(t_{n}-t_{n+1}, A(t_{n+1})\bigr) \rangle}{\langle \Dot{A}_{0}^{2} \rangle} \right) \nonumber \\
	+ \int_{0}^{t} dt_{1} \int_{0}^{t_{1}} dt_{2} \: \frac{\langle F_\text{Q}^\text{H}(0), \Ddot{A}(t-t_{1}) \rangle}{\langle \Dot{A}_{0}^{2} \rangle} \frac{\langle \Ddot{A}_{0}, \Gamma_\text{NL}^\text{H}\bigl(t_{1}-t_{2}, A(t_{2})\bigr) \rangle}{\langle \Dot{A}_{0}^{2} \rangle}  \nonumber \\
	+ \sum_{n \geq 2} \prod_{i = 1}^{n+1} \int_{0}^{t\delta_{i,1} + t_{i-1}(1-\delta_{i,1})} dt_{i} \frac{\langle F_\text{Q}^\text{H}(0), \Ddot{A}(t-t_{1}) \rangle}{\langle \Dot{A}_{0}^{2}\rangle}  \prod_{j=2}^{n} \frac{\langle \Ddot{A}_{0}, \Dot{A}(t_{j-1} - t_{j}) \rangle}{\langle \Dot{A}_{0}^{2}\rangle} \frac{\langle \Ddot{A}_{0}, \Gamma_\text{NL}^\text{H}\bigl(t_{n}-t_{n+1}, A(t_{n+1})\bigr) \rangle}{\langle \Dot{A}_{0}^{2} \rangle}.
	\label{eq:proof_fdts9}
\end{gather}
Using $\langle F_\text{Q}^\text{H}(0), \Dot{A}_{0} \rangle = 0$ \cite{ayaz2022generalized}, Eq.~\eqref{eq:proof_fdts9} simplifies to

\begin{align}
	\Delta\Gamma^{(1)}(t) = \frac{\langle F_\text{Q}^\text{H}(0), \Gamma_\text{NL}^\text{H}\bigl(0, A(t)\bigr) \rangle}{\langle \Dot{A}_{0}^{2}\rangle} + \int_{0}^{t} ds \: \frac{\langle F_\text{Q}^\text{H}(0), \partial_{t}\Gamma_\text{NL}^\text{H}\bigl(t-s, A(s)\bigr) \rangle}{\langle \Dot{A}_{0}^{2}\rangle} + \nonumber \\
	+ \int_{0}^{t} dt_{1} \int_{0}^{t_{1}} dt_{2} \: \frac{\langle F_\text{Q}^\text{H}(0), \Ddot{A}(t-t_{1}) \rangle}{\langle \Dot{A}_{0}^{2} \rangle} \frac{\langle \Ddot{A}_{0}, \Gamma_\text{NL}^\text{H}\bigl(t_{1}-t_{2}, A(t_{2})\bigr) \rangle}{\langle \Dot{A}_{0}^{2} \rangle}  \nonumber \\
	\sum_{n \geq 2} \prod_{i = 1}^{n+1} \int_{0}^{t\delta_{i,1} + t_{i-1}(1-\delta_{i,1})} dt_{i} \frac{\langle F_\text{Q}^\text{H}(0), \Ddot{A}(t-t_{1}) \rangle}{\langle \Dot{A}_{0}^{2}\rangle} \prod_{j=2}^{n} \frac{\langle \Ddot{A}_{0}, \Dot{A}(t_{j-1} - t_{j}) \rangle}{\langle \Dot{A}_{0}^{2}\rangle} \frac{\langle \Ddot{A}_{0}, \Gamma_\text{NL}^\text{H}\bigl(t_{n}-t_{n+1}, A(t_{n+1})\bigr) \rangle}{\langle \Dot{A}_{0}^{2} \rangle}.
	\label{eq:proof_fdts10}
\end{align}
Evaluating this equation at $t = 0$, all integral terms on its r.h.s. vanish
\begin{eqnarray}
	\Delta\Gamma^{(1)}(0) = \frac{\langle F_\text{Q}^\text{H}(0), \Gamma_\text{NL}^\text{H}\bigl(0, A_{0}\bigr) \rangle}{\langle \Dot{A}_{0}^{2}\rangle}.
	\label{eq:proof_fdts11}
\end{eqnarray}
Next, we compute the correlation between $F_\text{Q}^\text{H}(0)$ and any function $f(A_{0})$ of $A_{0}$ (using the definition in Eq.~\eqref{eq:dot_product} in the main text)
\begin{eqnarray}
	\langle F_\text{Q}^\text{H}(0), f(A_{0}) \rangle = \int d\omega \:\rho_\text{eq}(\omega)  F_\text{Q}^\text{H}(\omega,0) f(A_{0}(\vec{q})).
	\label{eq:orthogonality_proof_1}
\end{eqnarray}
Introducing an integration over all possible values of $A_{0}$
\begin{eqnarray}
	\langle F_\text{Q}^\text{H}(0), f(A_{0}) \rangle = \int d\omega \: \rho_\text{eq}(\omega)  F_\text{Q}^\text{H}(\omega, 0) \int dA \: \delta (A_{0}(\vec{q}) - A) f(A),
	\label{eq:orthogonality_proof_2}
\end{eqnarray}
and switching  the integrals
\begin{eqnarray}
	\langle F_\text{Q}^\text{H}(0), f(A_{0}) \rangle = \int dA \: f(A) \int d\omega \: \rho_\text{eq}(\omega) \delta (A_{0}(\vec{q}) - A)  F_\text{Q}^\text{H}(\omega, 0) ,
	\label{eq:orthogonality_proof_3}
\end{eqnarray}
Eq.~\eqref{eq:orthogonality_proof_1} simplifies to
\begin{eqnarray}
	\langle F_\text{Q}^\text{H}(0), f(A_{0}) \rangle = \int dA \: f(A) \langle \delta ( A_{0} - A ) \rangle \langle F_\text{Q}^\text{H}(0) \rangle_{A}.
	\label{eq:orthogonality_proof_4}
\end{eqnarray}
We recall $\langle F_\text{Q}^\text{H}(0) \rangle_{A} = 0$ \cite{ayaz2022generalized}, which implies from Eq.~\ref{eq:orthogonality_proof_4} that the correlation between $F_\text{Q}^\text{H}(0)$ and any function of $A_{0}$ vanishes, i.e. $\langle F_\text{Q}^\text{H}(0), f(A_{0}) \rangle = 0$. Therefore, $\langle F_\text{Q}^\text{H}(0), \Gamma_\text{NL}^\text{H}(0, A_{0}) \rangle = 0$ and Eq.~\eqref{eq:proof_fdts11} reduces to
\begin{eqnarray}
	\Delta\Gamma^{(1)}(0) = 0.
	\label{eq:proof_fdts12}
\end{eqnarray}
This result is expected: since both $\Gamma_\text{Q}^\text{DP}(t)$ and $\Gamma_\text{L}^\text{H}(t)$ are invariant under the symmetry $t \mapsto - t$, $\Delta\Gamma(t)$ is symmetric as well, which implies that all odd powers in $t$ in its Taylor expansion around $t = 0$ vanish. We also find that $\Delta \Gamma^{(2)}(0) = 0$ for the second-order expansion term, as we demonstrate in supplementary material Sec.~\ref{app:second_order_expansion}.
Note that these findings do not exclude the possibility of non-vanishing higher-order terms in Eq.~\eqref{eq:proof_fdts7}. \\
\indent Assuming $\Gamma^\text{DP}_\text{Q}(t) \approx \Gamma^\text{H}_\text{L}(t)$, which is numerically true for the butane data shown in the main text (supplementary material Sec.~\ref{app:hybrid_dp_butane}), the DP-GLE and the hybrid GLE are related via
\begin{align}
	\label{eq:master_relation}   
	- \int_0^t ds\:\Gamma^\text{DP}_{\Delta}(t-s)&\langle \Dot{A}_0, \Dot{A}(s) \rangle  = \int_0^t ds \langle \Dot{A}_0,\Gamma^\text{H}_\text{NL}\bigl(A(s), t-s\bigr) \rangle,
\end{align}
which we derive in supplementary material Sec.~\ref{app:deriv_master_rel}. This relation means that a non-zero $\Gamma^\text{DP}_{\Delta}(t)$, defined in Eq.~\eqref{eq:kernel_dp_res} in the main text, translates into a non-zero $\Gamma^\text{H}_\text{NL}\bigl(A(s), t-s\bigr)$, and thereby allows for the detection of position-dependent friction utilizing the DP-GLE parameters extracted from MD simulations via Volterra extraction methods \cite{berne1970calculation, straub1987calculation, daldrop2018butane}. 

\section{Memory Kernel Extraction for an Arbitrary Potential}
\label{app:MemKernExtrac}
We compute the memory kernels $\Gamma^\text{DP}(t), \Gamma^\text{CM}(t)$ and $\Gamma^\text{M}(t)$ from a given one-dimensional time-series trajectory $A(t)$ by the Volterra extraction scheme \cite{ayaz2021non}, utilizing time-correlation functions \cite{berne1970calculation, straub1987calculation, daldrop2018butane}.
Multiplying Eq.~\eqref{eq:dp_gle} in the main text by the initial velocity $\dot{A}_0$, ensemble averaging and making use of the condition $\langle \dot{A}_0, F^\text{DP}_\text{Q}(t)\rangle=0$ \cite{mori1965transport} yields a Volterra equation of first kind
\begin{equation}
	\label{eq:Volterra}
	C^{\Dot{A}\Ddot{A}}(t) = - C^{\Dot{A}F_{\text{eff}}}(t) - \int_0^t ds\: \Gamma^\text{DP}(s)C^{\Dot{A}\Dot{A}}(t-s),
\end{equation}
where equilibrium time-correlations are denoted as $C^{AB}(t)=\langle A_0,B(t)\rangle$. Subsequently, Eq.~\eqref{eq:Volterra} can be integrated and hence becomes \cite{ayaz2021non}
\begin{equation}
	\begin{split}
		\bigl. C^{\Dot{A}\Dot{A}}(t)\bigr\vert_0^t = \bigl. C^{{A}F_{\text{eff}}}(t)\bigr\vert_0^t - \int_0^t ds \: G^\text{DP}(t-s)C^{\Dot{A}\Dot{A}}(s),
	\end{split}
	\label{eq:Volterrab}
\end{equation} 
where we defined the running integral $G^\text{DP}(t)=\int_{0}^{t} ds \: \Gamma^\text{DP}(s)$. Multiplying Eq.~\eqref{eq:dp_gle} in the main text by the initial position $A(0)$ and ensemble averaging gives \cite{ayaz2021non} 
\begin{equation}
	\label{eq:Volterra_pos}
	C^{A\Ddot{A}}(t) = - C^{AF_{\text{eff}}}(t) - \int_0^t ds\: \Gamma^\text{DP}(s)C^{A\Dot{A}}(t-s).
\end{equation}
Evaluating Eq.~\eqref{eq:Volterra_pos} at $t=0$, and using that $C^{A\Ddot{A}}(t) = - C^{\dot{A}\dot{A}}(t)$, we find that $C^{{A}F_{\text{eff}}}(0)=C^{\dot{A}\dot{A}}(0)$, which is used to simplify Eq.~\eqref{eq:Volterrab}. Discretizing Eq.~\eqref{eq:Volterrab} and using the trapezoidal rule for the integral, we obtain
\begin{equation}
	C_i^{\dot{A}\dot{A}} = C^{AF_{\text{eff}}}_{i} - \Delta t\sum_{j=0}^{i}\omega_{i,j} G^\text{DP}_{i-j} C^{\Dot{A}\Dot{A}}_{j},
\end{equation}
where $C^{AB}_i = \langle A_0,B_i\rangle$ and $\omega_{i,j} = 1 - \frac{\delta_{j,0}}{2} -\frac{\delta_{j,i}}{2}$ comes from the trapezoidal integration rule. Note that the subscript $i$ refers to the discrete trajectory time step $i\Delta t$, and the initial value $G^\text{DP}_0$ vanishes. Finally, after rearranging, we obtain the iterative formula for the running integral $G^\text{DP}_i$
\begin{equation}
	\label{eq:extraction}
	G^\text{DP}_i =  \frac{2}{\Delta t C^{\Dot{A}\Dot{A}}_0}\bigl(C^{{A}F_{\text{eff}}}_{i}-C_i^{\dot{A}\dot{A}} - \Delta t\sum_{j=1}^{i-1}G^\text{DP}_{i-j} C^{\Dot{A}\Dot{A}}_{j}\bigr),
\end{equation}
from which we can compute the discrete memory kernel $\Gamma^\text{DP}_i$ in the DP-GLE (Eq.~\eqref{eq:dp_gle}) by taking the numerical derivative of $G^\text{DP}_i$. The correlation $C^{AF_{\text{eff}}}_{i}$ is calculated by cubic spline interpolation of the discrete PMF, $U_{\text{PMF}}(A)$, and the position-dependent mass, $M(A)$, and by using Eq.~\eqref{eq:eff_force} in the main text. For this, the configuration space of $A$ is divided into $N_A$ bins with equal size $\Delta A$ = 1 deg, i.e. $A \in [A_\text{min}, A_\text{min} + N_A \Delta A]$, where $A_\text{min} = $ - 180 deg is the lower bound. We use $N_A$ = 360 bins for all extractions to discretize the configuration space. The PMF is computed from the probability distribution $\rho(A)$ according to $U_{\text{PMF}}(A) = -k_BT \ln \rho(A)$. The position-dependent mass is calculated via
\begin{equation}
	\label{eq:calc_mass}
	M^{-1}(A_l) = \frac{\sum_{i \leq N_\text{traj} -1,\:A_i \in \mathcal{I}_l} \dot{A}_i^2}{k_BT\sum_{i \leq N_\text{traj} - 1,\:A_i \in \mathcal{I}_l}1},
\end{equation}
where $N_\text{traj}$ is the trajectory length.
$\sum_{i \leq N_\text{traj} -1,\:A_i \in \mathcal{I}_l}$ means that we sum over the times $i$ where $A$ at time $t = i\Delta t$ is in the interval $\mathcal{I}_l$. We denote the intervals as $\mathcal{I}_l = [A_\text{min} + l \Delta A, A_\text{min} + (l + 1)\Delta A]$ with $l = 0, 1, ..., N_A - 1$. \\
\indent Using the same approach, the iterative equation for the CM-GLE in Eq.~\eqref{eq:const_mass_gle} in the main text reads \cite{ayaz2021non}
\begin{equation}
	\label{eq:extraction2}
	G^\text{CM}_i =  \frac{2}{\Delta t C^{\Dot{A}\Dot{A}}_0}\bigl(\frac{C_0^{\Dot{A}\Dot{A}}}{C_0^{A\nabla U_{\text{PMF}}}}C^{{A}\nabla U_{\text{PMF}}}_{i}-C_i^{\Dot{A}\Dot{A}} - \Delta t\sum_{j=1}^{i-1}G^\text{CM}_{i-j} C^{\Dot{A}\Dot{A}}_{j}\bigr),
\end{equation}
where $G^\text{CM}(t)=\int_{0}^{t} ds \: \Gamma^\text{CM}(s)$ and we employed $M_0=C^{{A}\nabla U_{\text{PMF}}}_0/C^{\dot{A}\dot{A}}_0$ \cite{ayaz2021non}. \\
\indent The extraction formula for the Mori-GLE in Eq.~\eqref{eq:gle_mori} in the main text follows as
\begin{equation}
	\label{eq:extraction3}
	G^\text{M}_i =  \frac{2}{\Delta t C^{\Dot{A}\Dot{A}}_0}\bigl(\frac{k}{M_0} C^{{A}A}_{i}-C_i^{\Dot{A}\Dot{A}} - \Delta t\sum_{j=1}^{i-1}G^\text{M}_{i-j} C^{\Dot{A}\Dot{A}}_{j}\bigr), 
	\end{equation}
	where $G^\text{M}(t)=\int_{0}^{t} ds \: \Gamma^\text{M}(s)$ and $k = k_BT/\langle A_0^2 \rangle $.
	
	\section{Comparison Between Hybrid and DP-GLE Parameters for Butane Dihedral Dynamics}
	\label{app:hybrid_dp_butane}
	\begin{figure*}
\centering
\includegraphics[width=\textwidth]{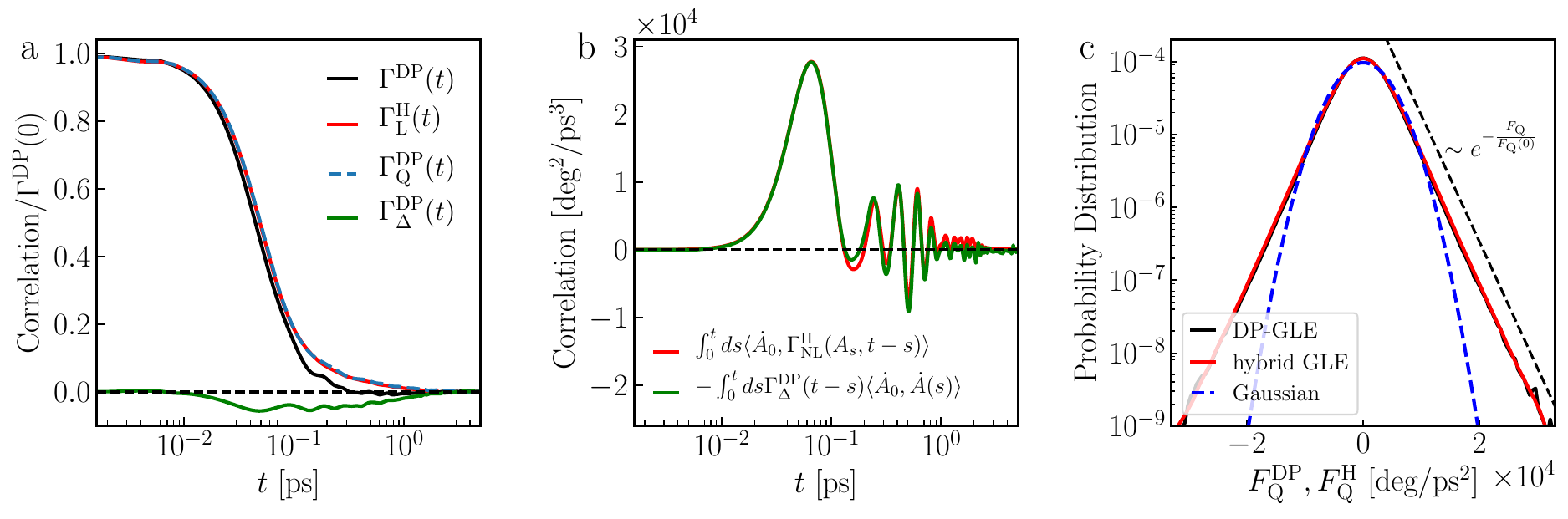}
\caption{Relation between the DP-GLE and hybrid GLE for butane dihedral angle dynamics.
	(a) ACF of the orthogonal force, i.e. $\Gamma^\text{DP}_\text{Q}(t) = \langle F^\text{DP}_\text{Q}(0), F^\text{DP}_\text{Q}(t)\rangle / \langle \dot{A}_0^2 \rangle$, denoted as blue broken line, compared with the extracted memory kernel $\Gamma^\text{DP}(t)$ from Eq.~\eqref{eq:dp_gle} in the main text using the Volterra extraction scheme (black line, see supplementary material Sec.~\ref{app:MemKernExtrac} for details), and the linear memory kernel $\Gamma^\text{H}_\text{L}(t)$ from Eq.~\eqref{eq:FDT_hybrid} in the main text (red line), which we compute using the forward propagation technique introduced in Ref.~\onlinecite{ayaz2022generalized}. The orthogonal force $F^\text{DP}_\text{Q}(t)$ follows from Eq.~\eqref{eq:random_force} in the main text. The green line denotes the residual memory kernel  $\Gamma^\text{DP}_{\Delta}(t) = \Gamma^\text{DP}(t) - \Gamma^\text{DP}_\text{Q}(t)$. (b) Comparison between the integrals including $\Gamma^\text{DP}_{\Delta}(t)$ and $\Gamma^\text{H}_\text{NL}\bigl(A(t),s\bigr)$ in the relation Eq.~\eqref{eq:master_relation}. The non-linear memory kernel data $\Gamma^\text{H}_\text{NL}\bigl(A(t),s\bigr)$ can be found in supplementary material Sec.~\ref{app:non_linear_kernels}. 
	(c) Distribution of the orthogonal force $F^\text{DP}_\text{Q}(t)$ (black) computed from Eq.~\eqref{eq:random_force}, compared with a Gaussian distribution with zero mean and variance $\langle \dot{A}_0^2 \rangle\Gamma^\text{DP}(0)$ (blue broken line), and the orthogonal force $F^\text{H}_\text{Q}(t)$ from the hybrid GLE in Eq.~\eqref{eq:hybrid_gle} (red).}
\label{fig:butane_check_fdt_hybrid}
\end{figure*}
\begin{figure*}[hbt!]
\centering
\includegraphics[width=1
\linewidth]{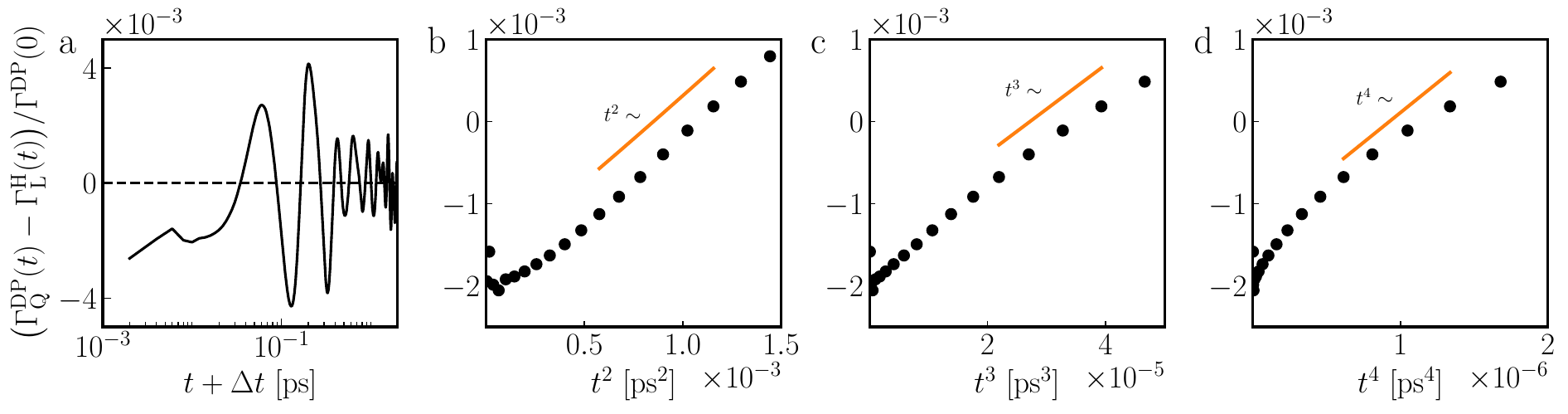}
\caption{(a) Difference between the orthogonal force ACFs from the DP-GLE and hybrid GLE (red solid and blue broken lines in Fig.~\ref{fig:butane_check_fdt_hybrid}(a)), i.e. $\bigl(\Gamma^\text{DP}_\text{Q}(t) - \Gamma^\text{H}_\text{L}(t)\bigr)/\Gamma^\text{DP}_0$, for the butane dihedral angle. The figures (b - d) show the result in (a) for different $t$-scalings. The small values of the difference suggest that possible non-vanishing terms of the r.h.s in Eq.~\eqref{eq:proof_fdts7} are masked by the numerical noise, and shows that $\Gamma^\text{H}_\text{L}(t) = \Gamma^\text{DP}_\text{Q}(t)$ is numerically true, which we use to obtain the relation in Eq.~\eqref{eq:master_relation} in supplementary material Sec.~\ref{app:proof_equiv_kernels}.}
\label{fig:difference_acf_kernels_butane}
\end{figure*}

In Fig.~\ref{fig:butane_check_fdt_hybrid}(a), we compare the GLE parameters from the hybrid GLE in Eq.~\eqref{eq:hybrid_gle} and the DP-GLE in Eq.~\eqref{eq:dp_gle} in the main text, for the butane dihedral angle (compare Fig. \ref{fig:butane_check_fdt} in the main text).
The orthogonal force ACF $\Gamma^\text{DP}_\text{Q}(t)$ is very similar to the linear memory kernel $\Gamma^\text{H}_\text{L}(t)$ from the hybrid GLE, which we calculate by using Eq.~\eqref{eq:FDT_hybrid} (red line) employing a forward propagation technique \cite{ayaz2022generalized}. \\
\indent In Fig.~\ref{fig:difference_acf_kernels_butane}, we see that the difference between the ACFs in Eq.~\eqref{eq:proof_fdts7} is very small, reflecting that $\Gamma^\text{H}_\text{L}$ and $\Gamma^\text{DP}_\text{Q}$ shown in Fig.~\ref{fig:butane_check_fdt_hybrid}(a) almost overlap. The values of $\Delta \Gamma(t) = \Gamma_\text{Q}^\text{DP}(t) - \Gamma_\text{L}^\text{H}(t)$ in Fig.~\ref{fig:butane_check_fdt_hybrid}(a) oscillate around zero and are less than 1$\%$ of $\Gamma^\text{DP}(0)$. In Fig.~\ref{fig:difference_acf_kernels_butane}(b - d), we observe finite quadratic and cubic terms in the short-time regime, which should vanish according to our calculations in supplementary material Sec.~\ref{app:proof_equiv_kernels}. Moreover, $\Delta \Gamma(0) \neq 0$ in Fig.~\ref{fig:difference_acf_kernels_butane}(a), which is presumably due to numerical noise. 
We conclude that higher-order even expansion terms of the difference $\Delta \Gamma(t) $ in Eq.~\eqref{eq:proof_fdts7}, if present, are below the numerical accuracy of the MD simulations. For this reason, we can rightfully assume, based on the numerical results, $\Gamma^\text{H}_\text{L}(t) = \Gamma^\text{DP}_\text{Q}(t)$.
It follows that the relation between $\Gamma^\text{DP}_{\Delta}(t)$ and $\Gamma^\text{H}_\text{NL}(A(t),s)$ in Eq.~\eqref{eq:master_relation} approximately holds for the butane dihedral angle, which we verify numerically in Fig.~\ref{fig:butane_check_fdt_hybrid}(b). In supplementary material Sec.~\ref{app:non_linear_kernels}, we provide the extracted non-linear memory kernel $\Gamma^\text{H}_\text{NL}$ data.
Note that, for computing the r.h.s. in Eq.~\eqref{eq:master_relation}, we express the data of $\Gamma^\text{H}_\text{NL}$ by a sum of scaled Legendre polynomials, as explained in supplementary material Sec.~\ref{app:legendre_non_linear_kernels}.\\
\indent Interestingly, the orthogonal force $F_\text{Q}^\text{DP}(t)$ has a similar distribution as $F_\text{Q}^\text{H}(t)$ extracted from the hybrid GLE, visualized Fig.~\ref{fig:butane_check_fdt_hybrid}(c). Both distributions are non-Gaussian with exponentially decreasing tails, denoted as a black broken line.

\section{Extraction Results for Butane with Two Frozen Carbon Atoms}
\label{app:results_2constr}
\begin{figure*}[hbt!]
\centering
\includegraphics[width=\textwidth]{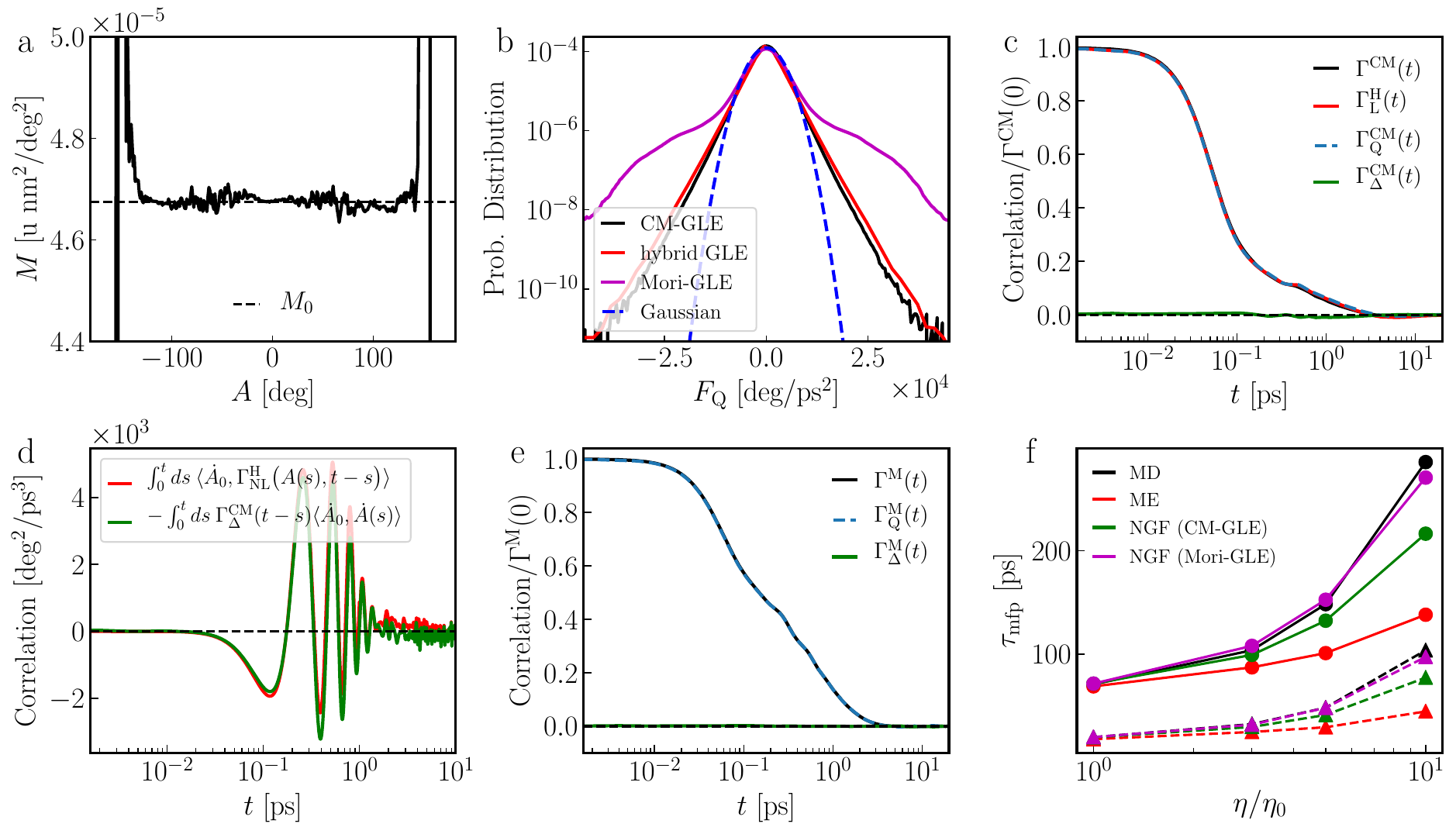}
\caption{Extraction results for the butane dihedral angle from MD simulations where the two inner carbon atoms are frozen in space. (a) Angle-dependent mass $M(A)$ = $k_BT/\langle \Dot{A}^2 \rangle_{A}$. Here, we show the standard viscosity results, i.e. $\eta/\eta_0=1$. The broken line denotes the constant mass, i.e. $M_0 = k_BT / \langle \Dot{A}_0^2 \rangle $. (b) Distribution of the orthogonal force from Eq.~\eqref{eq:random_force} in the main text (black), using the CM-GLE in Eq.~\eqref{eq:const_mass_gle}, compared with a Gaussian distribution with zero mean and variance $\langle \Dot{A}_0^2 \rangle\Gamma^\text{CM}(0)$ (blue broken line), and the orthogonal force extracted from the hybrid GLE in Eq.~\eqref{eq:hybrid_gle} (red), assuming a constant mass, i.e. $M(A) = M_0$. The purple line denotes the orthogonal force from the Mori-GLE in Eq.~\eqref{eq:gle_mori} in the main text.
	(c) Orthogonal force ACF $\Gamma^\text{CM}_\text{Q}(t) = \langle F^\text{CM}_\text{Q}(0), F^\text{CM}_\text{Q}(t)\rangle /\langle \Dot{A}_0^2 \rangle$ (blue broken line), compared with the extracted memory kernel $\Gamma^\text{CM}(t)$ from Eq.~\eqref{eq:const_mass_gle} using the Volterra extraction scheme (black line), and the linear memory kernel $\Gamma^\text{H}_\text{L}(t)$ which we calculate from the extraction scheme described in Ref.~\onlinecite{ayaz2022generalized} by using Eq.~\eqref{eq:FDT_hybrid} (red line). The green line shows the residual memory kernel $\Gamma^\text{CM}_{\Delta}(t) = \Gamma^\text{CM}(t) - \Gamma^\text{CM}_\text{Q}(t)$. (d) Comparison between the integrals including $\Gamma^\text{CM}_{\Delta}(t)$ and $\Gamma^\text{H}_\text{NL}(A(t),s)$ (Eq.~\eqref{eq:master_relation}), using the data shown in (c). The position-dependent memory kernel $\Gamma^\text{H}_\text{NL}\bigl(A(t),s\bigr)$ data can be found in supplementary material Sec.~\ref{app:non_linear_kernels}. (e) Orthogonal force ACF $\Gamma^\text{M}_\text{Q}(t) = \langle F^\text{M}_\text{Q}(0), F^\text{M}_\text{Q}(t)\rangle /\langle \Dot{A}_0^2 \rangle$, denoted as blue broken line, compared with the extracted memory kernel $\Gamma^\text{M}(t)$ (black line), both calculated from the Mori-GLE in Eq.~\eqref{eq:gle_mori} in the main text, together with the residual memory kernel $\Gamma^\text{M}_{\Delta}(t) = \Gamma^\text{M}(t) - \Gamma^\text{M}_\text{Q}(t)$ (green). (f) Comparison of the mean first-passage time $\tau_{\text{mfp}}$ of the butane dihedral angle from MD (black), ME GLE (red, supplementary material Sec.~\ref{app:const_mass_ME}), and NGF GLE simulations. NGF simulations are performed using the procedure described in Appendix \ref{sec:ngf} for the GLEs in Eq.~\eqref{eq:const_mass_gle} (green) and in Eq.~\eqref{eq:gle_mori} (purple). We show the mean first-passage times (Appendix \ref{sec:mfpts}) between the trans-state (0 deg) and the cis-state (110 deg) for MD simulations with varied water viscosity, i.e. $\eta/\eta_0$. Spheres denote the time from the trans-state to the cis-state, and triangles vice versa.}
\label{fig:butane_check_fdt_2constr}
\end{figure*}
We support our findings in the main text by analyzing trajectories of the dihedral angle dynamics of butane, where the two inner carbons are frozen in space \cite{daldrop2018butane}. 
This system significantly simplifies the analysis since the position-dependent mass of such a system is nearly constant, as numerically found in Fig.~\ref{fig:butane_check_fdt_2constr}(a), where we show the calculated position-dependent mass $M(A)$ together with the mass $M_0$. 
In the following, we do not consider the DP-GLE in Eq.~\eqref{eq:dp_gle} in the main text, but the CM-GLE in Eq.~\eqref{eq:const_mass_gle} in the main text, being a valid approximation, for the following analysis.\\
\indent The orthogonal force from the CM-GLE (black) and the hybrid GLE (red) in Fig.~\ref{fig:butane_check_fdt_2constr}(b) both exhibit non-Gaussian distributions and differ from the Mori-GLE result (purple), similar to the free scenario in Fig.~\ref{fig:butane_check_fdt}(f) in the main text. The orthogonal force ACF from the CM-GLE (blue) in Fig.~\ref{fig:butane_check_fdt_2constr}(c) reveals small differences from the extracted memory kernel $\Gamma^\text{CM}(t)$ (black line), shown as a green line, which are related to the position-dependent memory kernel from Eq.~\eqref{eq:hybrid_gle} via Eq.~\eqref{eq:master_relation} in the main text, as demonstrated in Fig.~\ref{fig:butane_check_fdt_2constr}(d). However, the amplitude of the residual memory kernel $\Gamma^\text{CM}_{\Delta}(t)$ is much smaller compared to the free butane result in Fig.~\ref{fig:butane_check_fdt}(e) in the main text. Internal friction in the butane molecule is reduced in this scenario \cite{daldrop2018butane}, suggesting a correlation of $\Gamma^\text{CM}_{\Delta}(t)$ with internal friction effects. This finding is supported by extracted parameters from the hybrid GLE shown in supplementary material Sec.~\ref{app:non_linear_kernels}.
Fig.~\ref{fig:butane_check_fdt_2constr}(e) illustrates the agreement between the orthogonal force ACF $\Gamma^\text{M}_\text{Q}(t)$ (blue broken line) and the extracted memory kernel $\Gamma^\text{M}(t)$ (black line) according to  Eq.~\eqref{eq:FDT_Mori}, both computed from the Mori-GLE in Eq.~\eqref{eq:gle_mori} in the main text.
\\ \indent In Fig.~\ref{fig:butane_check_fdt_2constr}(f), we verify the findings of Fig.~\ref{fig:butane4}(d) in the main text for the prediction of the mean first-passage times, here for the frozen scenario and as a function of the water viscosity, which is tuned by changing the water mass (Appendix \ref{sec:mds}). The best agreement with the MD data is reached with an NGF simulation using the Mori-GLE, followed by the CM-GLE. A Markovian embedding scheme employing a Gaussian orthogonal force performs the worst. Although the constant-mass assumption is accurate for this system (Fig.~\ref{fig:butane_check_fdt_2constr}(a)), the CM-GLE (green data) cannot reproduce the mean first-passage times of the MD data at high water viscosities. These deviations are presumably due to the numerical dephasing effects discussed in supplementary material Sec.~\ref{app:acc_brute}, which occur particularly for simulation times above 100 ps.

\section{Origin of Non-Gaussian Behavior in the Orthogonal Force}
\label{app:expotails}

\begin{figure*}[hbt!]
\centering
\includegraphics[width=0.7
\linewidth]{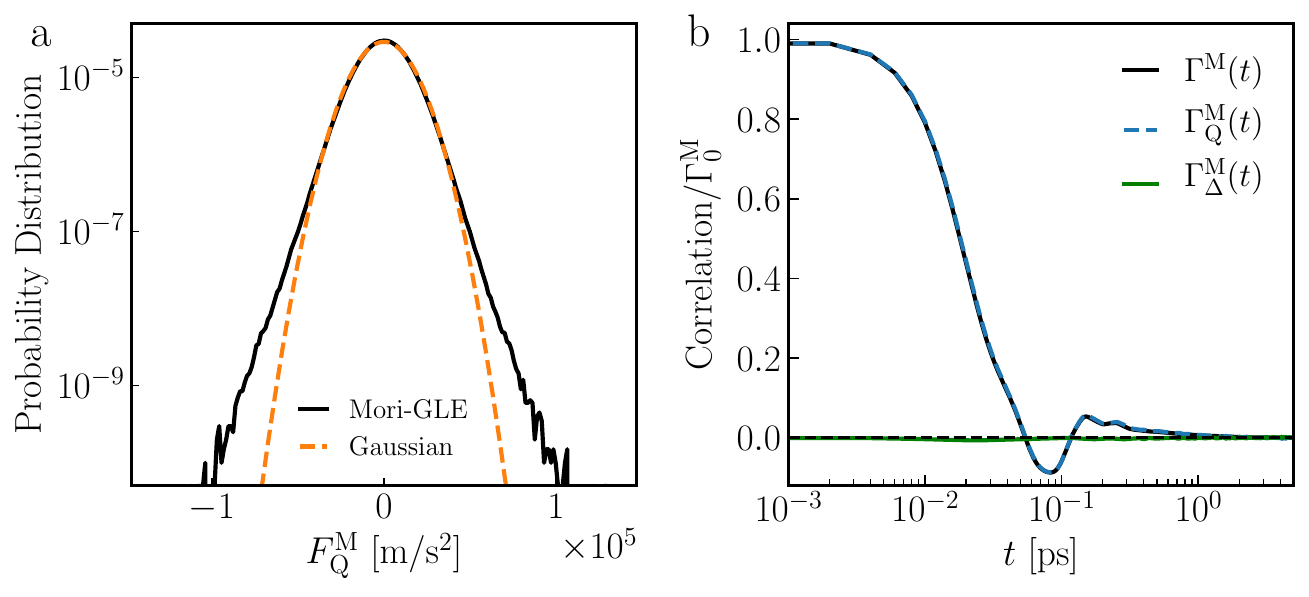}
\caption{Memory extraction results for the free diffusion of (SPC/E) water particles from MD simulations. 
	(a) Distribution of the orthogonal force $F^\text{M}_\text{Q}(t)$ from Eq.~\eqref{eq:random_force} in the main text (black), using the Mori-GLE in Eq.~\eqref{eq:gle_mori}, compared with a Gaussian distribution with zero mean and variance $\langle \Dot{A}_0^2 \rangle\Gamma^\text{M}(0)$ (orange broken line).
	(b) ACF of the orthogonal force (blue broken line), i.e. $\Gamma^\text{M}_\text{Q}(t) = \langle F^\text{M}_\text{Q}(0), F^\text{M}_\text{Q}(t)\rangle / \langle \Dot{A}_0^2 \rangle$, compared with the extracted memory kernel $\Gamma^\text{M}(t)$ from Eq.~\eqref{eq:gle_mori} in the main text using the Volterra extraction scheme (black line, see supplementary material Sec.~\ref{app:MemKernExtrac} for details), and the residual memory kernel $\Gamma^\text{M}_{\Delta}(t) = \Gamma^\text{M}(t) - \Gamma^\text{M}_\text{Q}(t)$ (green).}
\label{fig:check_fdt_spce}
\end{figure*}
The coupling between the butane molecule's translational and rotational dynamics influences its diffusive motion through the liquid environment \cite{daldrop2018butane}. According to the central limit theorem, the sum of many independent random variables, such as forces arising from random collisions with the environment, asymptotically exhibits a Gaussian distribution. However, diffusive motion involving solvent interactions, modeled by an additive Lennard-Jones potential combined with electrostatic interactions, does not fully satisfy the independence condition due to intricate correlations in the forces acting on the particles in the system \cite{samanta2021dielectric}. Consequently, the distribution of orthogonal forces can deviate from Gaussian behavior, displaying features such as non-zero skewness and kurtosis, which arise from rare events involving extremely low or high collision forces. These deviations align with predictions from large deviation theories \cite{touchette2018introduction, jack2020ergodicity}. Previous studies, such as that of Shin \textit{et al.} \cite{shin2010brownian}, have demonstrated that even in simple Brownian dynamics, where the solvent particles and the Brownian molecule are of comparable size and mass, the orthogonal forces can exhibit non-Gaussian distributions with exponential tails, particularly in systems characterized by low solvent densities.
\\
\indent
In Fig.~\ref{fig:check_fdt_spce}, we study the diffusion of the solvent particles in our MD simulations and the orthogonal force from trajectories of water particles. We perform MD simulations using the setup described in Appendix \ref{sec:mds}, but without the butane molecule. The GLE for the position of a freely diffusing SPC/E water particle follows the Mori-GLE in Eq.~\eqref{eq:gle_mori} in the main text for $k=0$ \cite{kowalik2019memory}, and we, therefore, use Eq.~\eqref{eq:extraction3} with $k=0$ for the memory kernel extraction. In Fig.~\ref{fig:check_fdt_spce}(a), we find that the orthogonal force of freely diffusing water particles exhibits non-Gaussian behavior with a pronounced exponential tail, similar to the butane dihedral angle in Fig.~\ref{fig:butane_check_fdt}(f) in the main text. Hence, we conclude that non-Gaussian fluctuating forces arise not only from internal interactions within the butane molecule but also from interactions between the butane molecule and water molecules.\\
\indent The statistics of the orthogonal force are strongly influenced by the number of solvent particles interacting with the solute particle, which in this case has a similar mass and size as the solvent particle \cite{shin2010brownian}. The number of interacting solvent particles fluctuates as fluid particles continuously enter and leave the interaction region surrounding the solute particle. These fluctuations lead to extreme events, such as those caused by many-body interactions or correlated collisions, which are likely influenced by the coupling between translational and rotational degrees of freedom \cite{samanta2021dielectric}. Such interactions explain the presence of pronounced shoulders in the orthogonal force distribution. Gaussian behavior is expected only in the limit of high solvent densities and large solute particles, such as proteins \cite{ayaz2021non,dalton2022protein}, where the likelihood of correlated collisions, as well as extreme events like exceptionally large forces from single-particle impacts, is significantly reduced. Under these conditions, the sum of all forces converges to a Gaussian distribution.\\
\indent The orthogonal force ACF shown in Fig.~\ref{fig:check_fdt_spce}(b) is in perfect agreement with the memory kernel for the Mori-GLE in Eq.~\eqref{eq:gle_mori}, as follows from Eq.~\eqref{eq:FDT_Mori} in the main text.
\section{Higher-Order Correlation Functions of the Orthogonal Force}
\label{app:higher_orders_fr}
\begin{figure*}[hbt!]
\centering
\includegraphics[width=\textwidth]{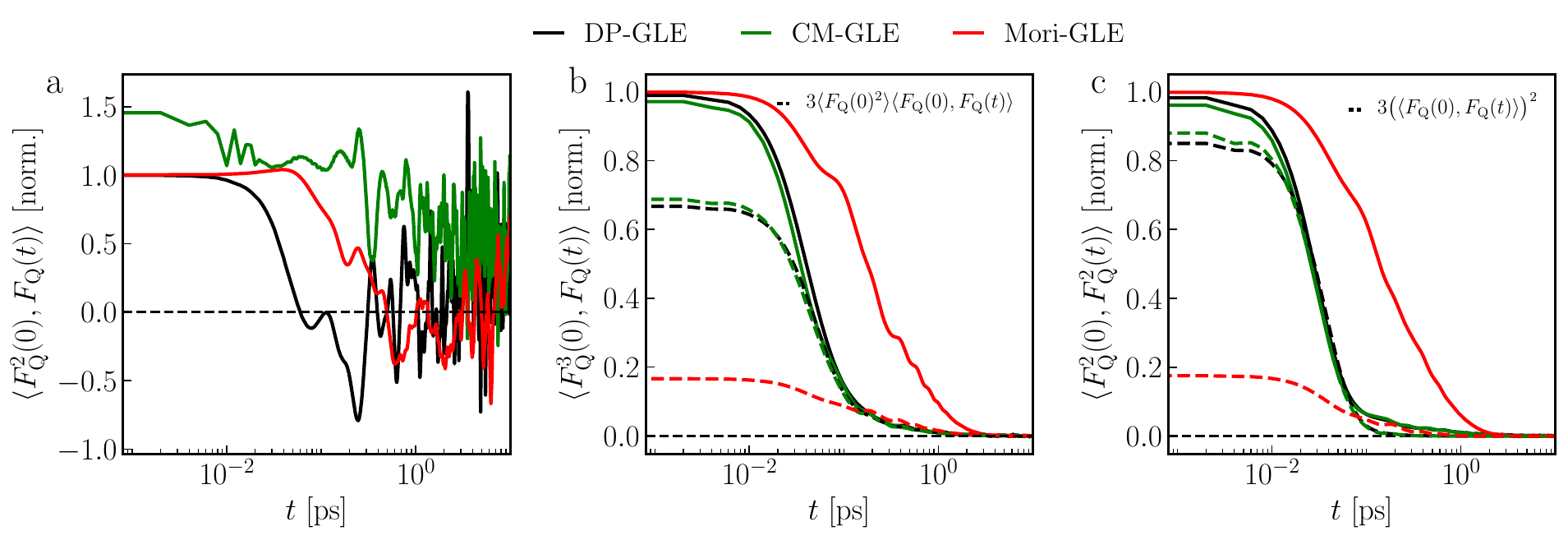}
\caption{Higher-order correlation functions of the butane dihedral angle's orthogonal force $F_\text{Q}(t)$ (Eq.~\eqref{eq:random_force} in the main text) using different GLEs (see Fig.~\ref{fig:butane_check_fdt} in the main text for details).  Note that the values are normalized by their initial values. Broken lines in (b) denote the scaled ACFs $3\langle F_\text{Q}(0)^2\rangle \langle F_\text{Q}(0), F_\text{Q}(t) \rangle$ and broken lines in (c) denote the squared ACFs $3\bigl(\langle F_\text{Q}(0), F_\text{Q}(t) \rangle\bigr)^2$.}
\label{fig:butane_fr_higher_orders}
\end{figure*}
Deviations from a Gaussian distribution, as observed for the orthogonal force of the dihedral angle in Fig.~\ref{fig:butane_check_fdt}(f) in the main text, are typically accompanied by non-vanishing higher-order correlations. In Fig.~\ref{fig:butane_fr_higher_orders}, we display higher-order correlation functions of the orthogonal force $F_\text{Q}(t)$ using the DP-GLE, CM-GLE, and Mori-GLE (see Fig.~\ref{fig:butane_check_fdt} in the main text for details). We normalized the higher-order correlations by their respective initial values. All correlation functions $\langle F_\text{Q}^2(0), F_\text{Q}(t) \rangle$ in (a) are negligible. 
The third- and fourth-order correlations from different GLEs in (b) and (c) differ from each other. Moreover, the correlation functions $\langle F_\text{Q}^3(0), F_\text{Q}(t) \rangle$ and $\langle F_\text{Q}^2(0), F_\text{Q}^2(t) \rangle$ differ from the scaled ACFs $3\langle F_\text{Q}(0)^2\rangle \langle F_\text{Q}(0), F_\text{Q}(t) \rangle$ and squared ACFs $3\bigl(\langle F_\text{Q}(0), F_\text{Q}(t) \rangle\bigr)^2$, respectively, denoted as broken lines in (b, c), referring to non-Gaussian behavior in all orthogonal-force trajectories \cite{wang2012brownian, cherstvy2019non}.

\section{Markovian Embedding Technique using the CM-GLE}
\label{app:const_mass_ME}
\begin{figure*}
\centering
\includegraphics[width=0.66
\linewidth]{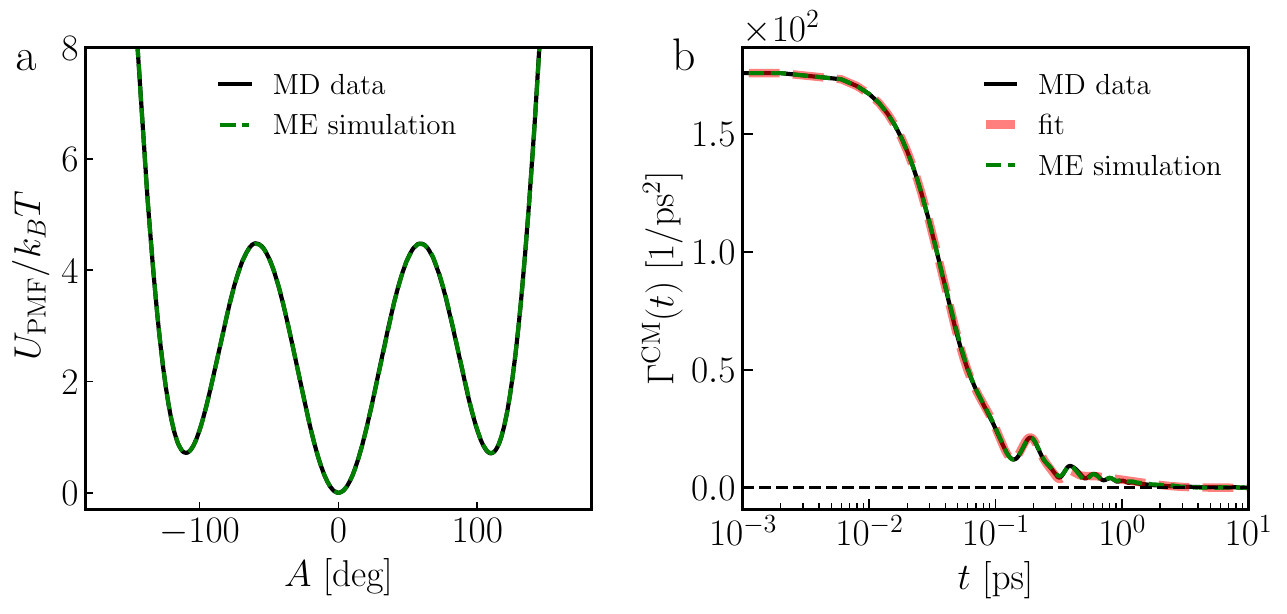}
\caption{Demonstration of the Markovian embedding (ME) simulation scheme for the butane dihedral angle. For the ME simulation (green), we parameterize the memory kernel (black) in the CM-GLE ($\Gamma^\text{CM}(t)$, taken from Fig.~\ref{fig:butane_check_fdt}(d) in the main text) using Eq.~\eqref{eq:fit_kernel} (see red broken line in (b)) and solve Eqs.~\eqref{eq:markov_embedding_1D} numerically. A simulation of $A(t)$ (green) with a length of 1$\:\mu$s perfectly reproduces the PMF $U_{\text{PMF}}$ (a) and the memory kernel $\Gamma^\text{CM}(t)$ (b), which we both extract from the simulated MD and ME trajectories $A(t)$ using the Volterra method described in supplementary material Sec.~\ref{app:MemKernExtrac}.}
\label{fig:butane_me_POC}
\end{figure*} 
We perform Markovian embedding (ME) simulations with the position-dependent mass of the observable assumed to be constant, $M(A) = M_0$. The DP-GLE in Eq.~\eqref{eq:dp_gle} then has the approximate form in Eq.~\eqref{eq:const_mass_gle} in the main text. Moreover, we approximate the extracted memory kernel from the CM-GLE in Eq.~\eqref{eq:const_mass_gle} in the main text as a sum of $m=5$ oscillatory-exponentials with memory times $\tau_j$ and periods $w_j$
\begin{eqnarray}
\label{eq:fit_kernel}
\Gamma^\text{CM}(t) = \sum_{j=1}^{m=5} k_je^{-t/\tau_{j}}\Bigl[\frac{1}{\tau_jw_j}\sin{(w_j t)}+ \cos{(w_j t)}\Bigr].
\end{eqnarray}
Using this form and assuming the orthogonal force to follow a stationary Gaussian process, the GLE in Eq.~\eqref{eq:const_mass_gle} in the main text is equivalent to the following Markovian system \cite{brunig2022timedependent, brunig2022pair}
\begin{align}
\label{eq:markov_embedding_1D}
\ddot{A}(t) =& - M_0^{-1}\frac{d}{dA} U_{\text{PMF}}(A(t)) + \sum_{j=1}^{m=5} k_j[y_j(t) - A(t)], \\ \nonumber
m^{y}_j \Ddot{y}_j(t) &= - \gamma_j \Dot{y}_j(t) + k_j[A(t) - y_j(t)] + \zeta_j(t),
\end{align}
where $m^y_j$ and $\gamma_j$ correspond to the fitting parameters $k_j$, $\tau_j$ and $w_j$ via
\begin{align}
m^y_j &= \frac{k_j}{\tau_j^{-2} + w_j^2}, \\
\gamma_j &= 2\frac{m_y^j}{\tau_j}.
\end{align}
Using fitting constants $k_j$,$\tau_j$, and $w_j$ obtained from the memory kernel data, we solve the Markovian system in Eqs.~\eqref{eq:markov_embedding_1D} numerically via a 4th-order Runge-Kutta scheme with a time step $\Delta t$ = 2 fs to generate the trajectory of $A(t)$, with a length equal to the MD data. The orthogonal forces $\zeta_j(t)$ are drawn from stationary Gaussian processes with zero mean and $\langle \zeta_j(t), \zeta_i(s)\rangle = 2  \langle \dot{A}_0^2 \rangle \gamma_j \delta_{ij} \delta(|t-s|)$, which is why, in Markovian embedding simulations, the orthogonal forces are referred to as random forces. Note that this Markovian embedding  is only equivalent to the CM-GLE in Eq.~\eqref{eq:const_mass_gle} if the relation
$\langle  F_\text{Q}^\text{CM}(0) F_\text{Q}^\text{CM}(t) \rangle = \langle \dot{A}^2_0 \rangle \Gamma^\text{CM}(t)$ holds \cite{brunig2022timedependent}. \\
\indent In Fig.~\ref{fig:butane_me_POC}, we demonstrate the robustness of the ME simulation technique for the butane dihedral angle; the memory kernel $\Gamma^\text{CM}(t)$ is taken from Fig.~\ref{fig:butane_check_fdt}(d) in the main text. Simulating a trajectory with a time step of 2 fs and a length of 1$\:\mu$s perfectly reproduces the PMF and the memory kernel, indicating that a ME simulation with the CM-GLE in Eq.~\eqref{eq:const_mass_gle} in the main text is self-consistent.

\section{Self-Consistency of the NGF Simulation Technique for Dihedral Angle Dynamics}
\label{app:ngf_poc_butane}

\begin{figure*}[hbt!]
\centering
\includegraphics[width=1
\linewidth]{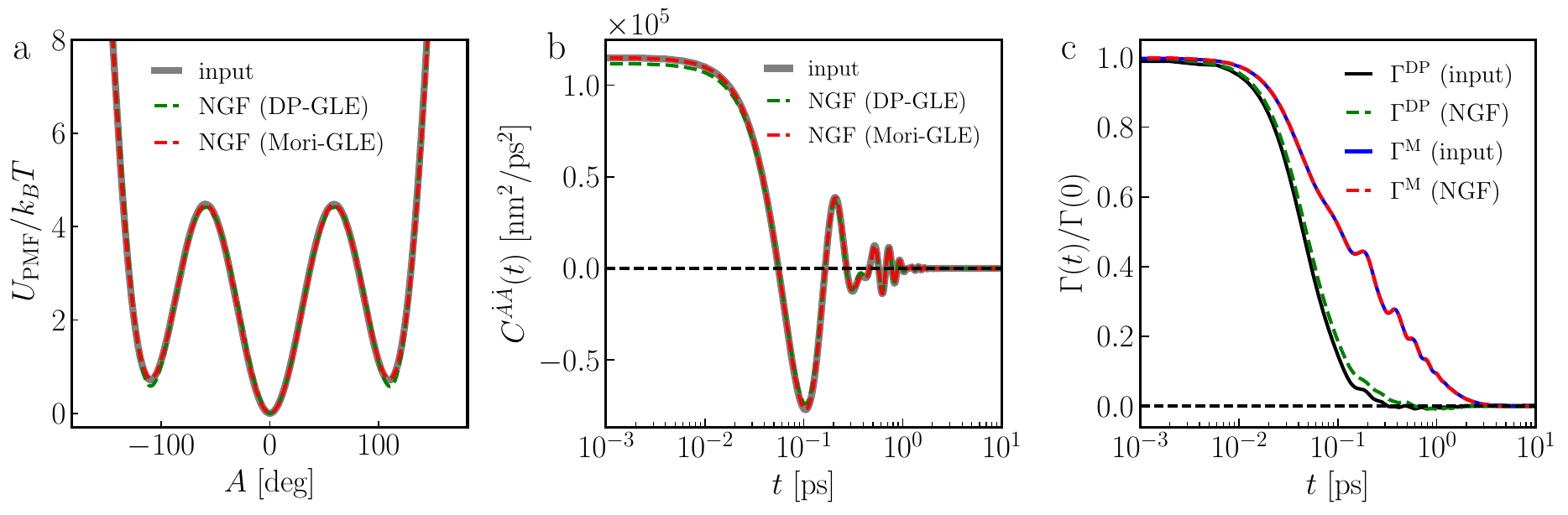}
\caption{Illustration of the self-consistency of the NGF GLE simulation (see Appendix \ref{sec:ngf}) for the dihedral angle data shown in Fig.~\ref{fig:butane_check_fdt}. With the extracted memory kernels from the DP-GLE in Eq.~\eqref{eq:const_mass_gle} (black line in (c)) and from the Mori-GLE in Eq.~\eqref{eq:gle_mori} in the main text (blue line in (c)), both extracted by the Volterra extraction scheme (supplementary material Sec.~\ref{app:MemKernExtrac}), from the trajectory $A(t)$, we compute randomly selected orthogonal-force trajectories $\bigl(F^\text{DP}_\text{Q}(t)$ or $F^\text{M}_\text{Q}(t)\bigr)$ via Eq.~\eqref{eq:random_force} for NGF simulations of $A(t)$ at arbitrary starting points. The initial conditions of the simulations are selected randomly but are not the same as for the orthogonal force computation (compare Fig.~\ref{fig:butane4} in the main text). We compare the input of the PMF $U_{\text{PMF}}$ (gray in (a)), the velocity ACF $C^{\dot{A}\dot{A}}(t)$ (gray in (b)), and the memory kernel (c, $\Gamma^\text{DP}(t)$ for the DP-GLE (black) and $\Gamma^\text{M}(t)$ for the Mori-GLE (blue)) with the extracted data from the NGF simulation (green for the DP-GLE and red for the Mori-GLE).}
\label{fig:brute_test_butane}
\end{figure*}

In Fig.~\ref{fig:brute_test_butane}, we verify the applicability of NGF simulations for the butane dihedral angle data shown in Fig.~\ref{fig:butane_check_fdt} in the main text. Following the procedure in Appendix \ref{sec:ngf}, we perform GLE simulations by solving Eq.~\eqref{eq:ODE_discret}, here exemplary for the DP-GLE in Eq.~\eqref{eq:dp_gle} and the Mori-GLE in Eq.~\eqref{eq:gle_mori} in the main text. In the simulations, we ensure that the orthogonal forces, which follow from Eq.~\eqref{eq:random_force} in the main text with the respective GLEs, are not initialized at the same randomly selected starting points of the simulations. The NGF simulations with the Mori-GLE reproduce the input velocity ACF of the data in (b, gray) and the input memory kernel in (c, blue), demonstrating self-consistency. Most importantly, even for the harmonic potential in the Mori-GLE in Eq.~\eqref{eq:gle_mori} in the main text, the correct inclusion of the orthogonal force accurately reconstructs the non-Gaussian distribution of the observable in Fig.~\ref{fig:brute_test_butane}(a, gray). The NGF simulation with the DP-GLE (green) slightly differs from the input data, most visible for the memory kernel $\Gamma^\text{DP}(t)$ in (c, black), which is presumably due to numerical instability, caused by the inclusion of the configuration-dependent mass \cite{ayaz_embedding_nl} (compare supplementary material Sec.~\ref{app:acc_brute}).

\section{GLE Parameters and NGF Simulations for Higher-Viscosity Butane Dihedral Dynamics}
\label{app:butane_params_higher_visc}

\begin{figure*}
\centering
\includegraphics[width=0.7\textwidth]{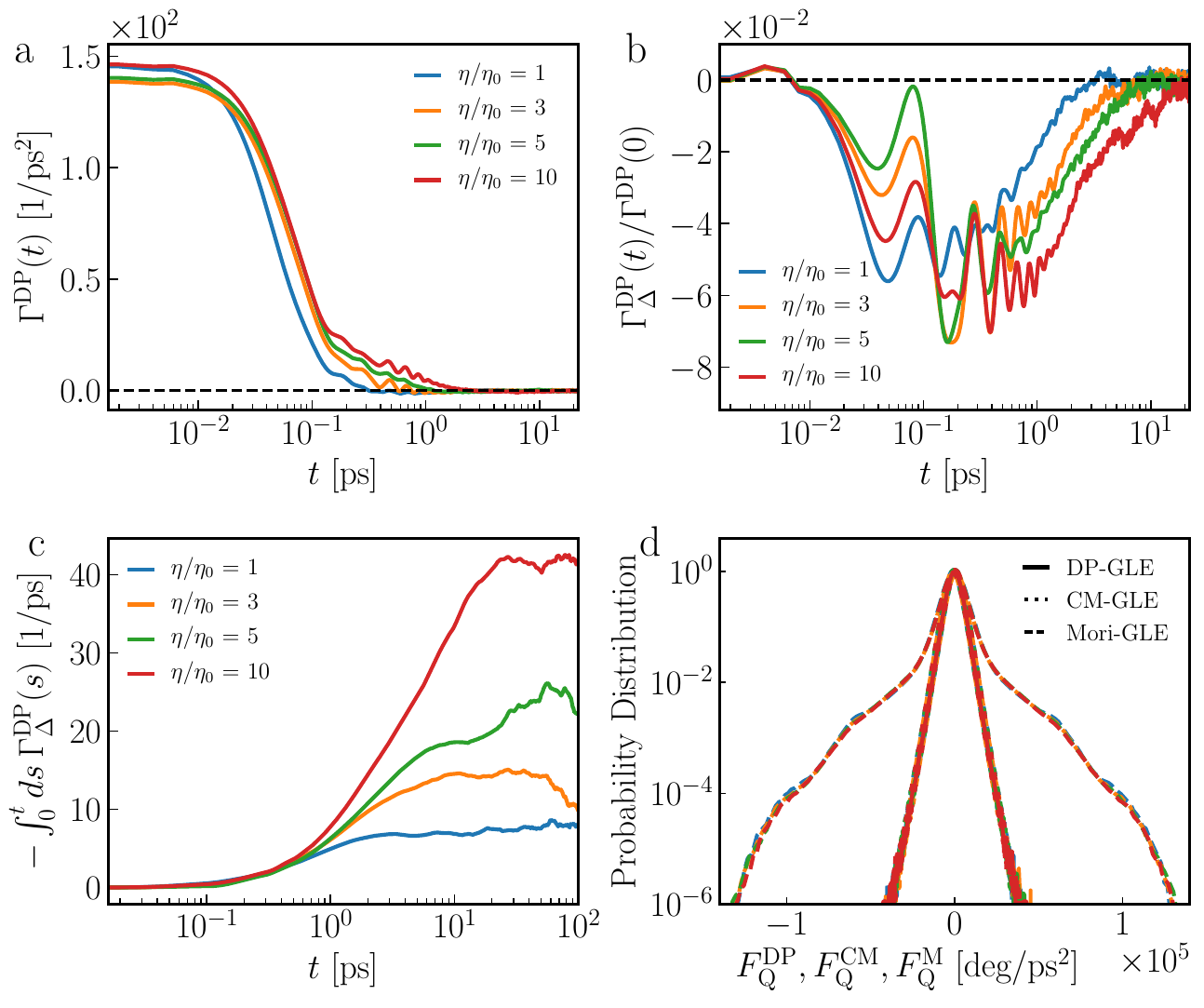}
\caption{(a - c) Analysis of $\Gamma^\text{DP}(t)$ and $\Gamma^\text{DP}_{\Delta}(t) = \Gamma^\text{DP}(t) - \Gamma^\text{DP}_\text{Q}(t)$ for the butane dihedral angle (compare Fig.~\ref{fig:butane_check_fdt}(d - f) in the main text), here for different water viscosities in the MD simulation $\eta/\eta_0$ (see Appendix \ref{sec:mds}), where $\eta_0$ is the standard viscosity. (c) displays the running integral over $\Gamma^\text{DP}_{\Delta}(t)$. (d) Orthogonal force distributions for different water viscosities, computed from the DP-GLE in Eq.~\eqref{eq:dp_gle} (solid lines), from the CM-GLE in Eq.~\eqref{eq:const_mass_gle} (dotted lines) and from the Mori-GLE in Eq.~\eqref{eq:gle_mori} (broken lines), using Eq.~\eqref{eq:random_force} in the main text; for details, see Fig.~\ref{fig:butane_check_fdt} in the main text.}
\label{fig:butane3}
\end{figure*}

\begin{figure*}
\centering
\includegraphics[width=0.4\textwidth]{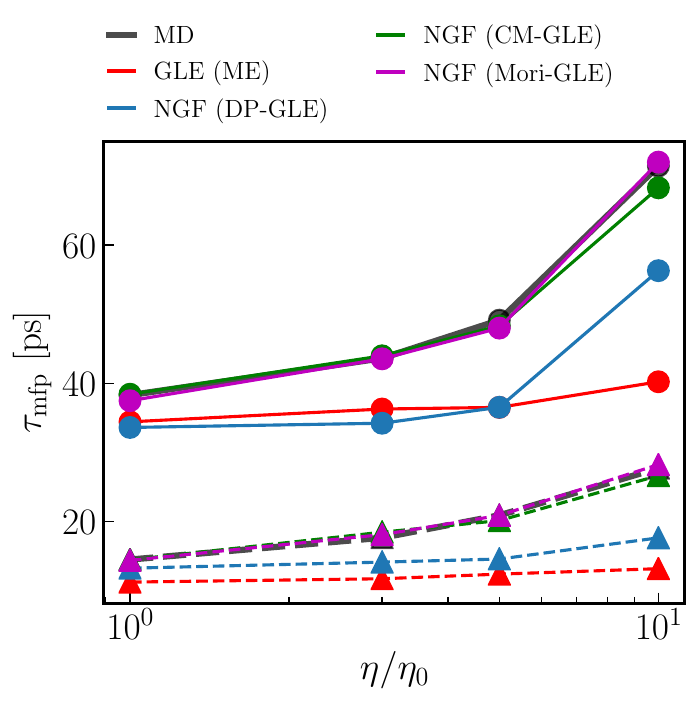}
\caption{Mean first-passage times $\tau_\text{mfp}$ for MD simulations with different water viscosities $\eta/\eta_0$ compared with GLE simulations (compare see Fig.~\ref{fig:butane4}(d) in the main text). We show the mean first-passage times going from the trans- to the cis-state (spheres) and from the cis- to the trans-state (triangles).}
\label{fig:butane4_visc}
\end{figure*}

In Fig.~\ref{fig:butane3}, we repeat the analysis of the butane dihedral angle dynamics from Fig.~\ref{fig:butane_check_fdt} in the main text, but for higher solvent viscosities $\eta/\eta_0 > 1$ in the MD simulation, where $\eta_0$ is the standard viscosity, produced via mass scaling (see Appendix \ref{sec:mds} for details). 
Fig.~\ref{fig:butane3}(b) shows no direct correlation between increasing viscosity and an enhancement of $\Gamma^\text{DP}_{\Delta}(t)$. However, notable features are observed: the decay of $\Gamma^\text{DP}_{\Delta}(t)$ for $\eta/\eta_0 = 1$, occurring between 0.1 and 1 ps, shifts to longer times as viscosity increases. Furthermore, the running integral over the difference kernel in Fig.~\ref{fig:butane3}(c) increases with higher water viscosity. Interestingly, in Fig.~\ref{fig:butane3}(d), the non-Gaussian behavior of the orthogonal force $F^\text{DP}_\text{Q}(t)$ in Eq.~\eqref{eq:dp_gle} remains unchanged while varying the viscosity. The same holds for the orthogonal forces in the CM- and Mori-GLEs.\\
\indent In Fig.~\ref{fig:butane4_visc}, we summarize the results of NGF simulations for butane in water with higher viscosities $\eta/\eta_0$. We find that the mean first-passage times for transitions between the trans-state (0 deg) and the cis-state (110 deg) predicted from Mori- and CM-GLE simulations are in very good agreement with the MD data for all viscosities. ME and DP-GLE perform poorly, which underlines our results in the main text.

\section{Simulation of Transition-Path and First-Passage Times of the Dihedral Angle}
\label{app:transition_times_butane}
\begin{figure*}
\centering
\includegraphics[width=1
\linewidth]{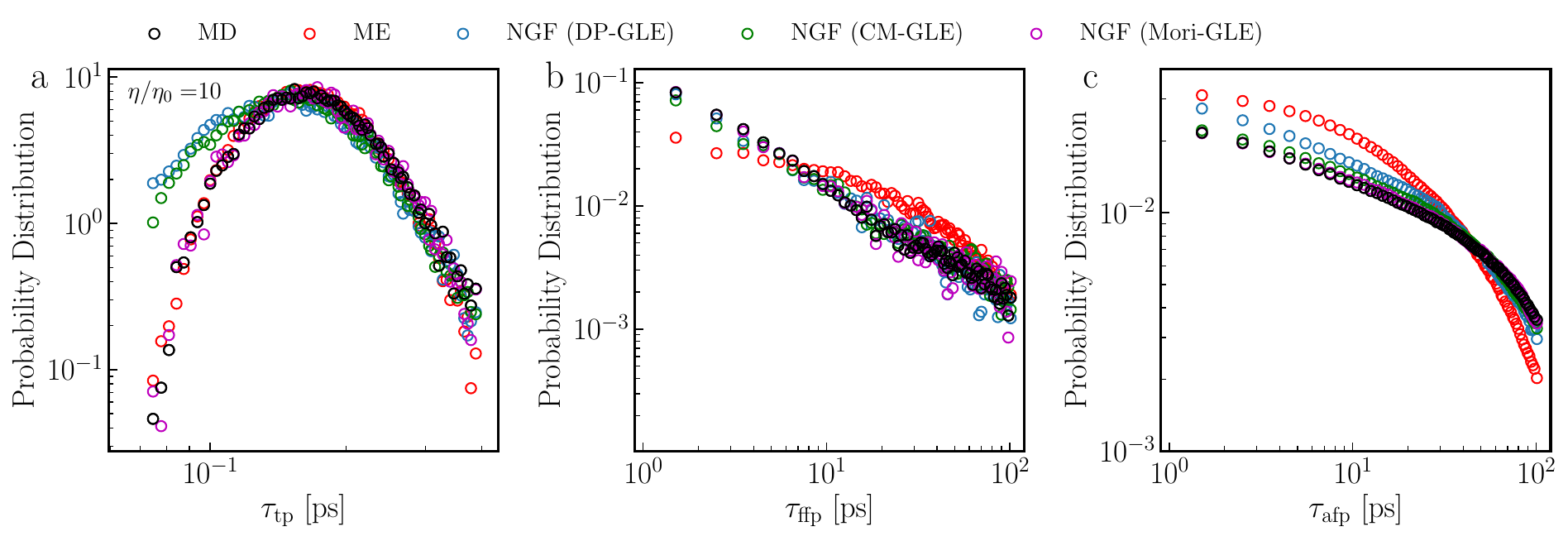}
\caption{Transition-path and first-passage time distributions for the butane dihedral angle from MD and GLE simulation data discussed in Fig.~\ref{fig:butane4} in the main text. We show the results for the heavy-water simulations ($\eta/\eta_0 = 10$) for transitions from the trans-state (0 deg) to the cis-state (110 deg). (a) Transition-path time ($\tau_\text{tp}$) distributions. (b) Distributions of the first-to-first-passage times $\tau_\text{ffp}$. (c) Distributions of the all-to-first-passage times $\tau_\text{afp}$. We accumulate the distributions using equally separated histogram bins but show the logarithmically spaced data in the figure.}
\label{fig:butane_visc_10_passage_times}
\end{figure*} 

Considering state-recrossing events is important for predicting short passage times in a GLE simulation, as recrossings influence butane isomerization \cite{dalton2024role}. 
To accurately calculate mean first-passage times, reproducing the transition-path and passage time distributions is essential. \\
\indent
In Fig.~\ref{fig:butane_visc_10_passage_times}, we show the simulated transition-path time $\tau_\text{tp}$ (a) and passage time (b, c) distributions from the MD and GLE simulation data discussed in Fig.~\ref{fig:butane4} in the main text. A transition path between the two states $A_\text{s} \to A_\text{f}$ is a path that leaves $A_\text{s}$ and reaches $A_\text{f}$ for the first time without recrossing $A_\text{s}$. Subfigure (b) summarizes the results for the first-to-first-passage times $\tau_\text{ffp}$, defined as the time between first crossing the trans-state and first crossing the cis-state, which is, therefore, strongly impacted by state-recrossing dynamics \cite{dalton2024role, zhou2024rapid}. Subfigure (c) includes all-to-first-passage events $\tau_\text{afp}$, which is the time between first crossing the cis-state and all previous crossings of the trans-state; the mean of this distribution is $\tau_{\text{mfp}}$, as discussed in the main text. Here, we choose to show the results for the heavy-water simulations ($\eta/\eta_0 = 10$), since for this viscosity, deviations between MD and GLE simulations for $\tau_{\text{mfp}}$ are found to be the most significant (compare Fig.~\ref{fig:butane4_visc}). \\
\indent The transition-path time distribution from MD data, shown in Fig.~\ref{fig:butane_visc_10_passage_times}(a), is well reproduced by all simulation techniques, except for short times for the DP- and CM-GLE methods. Significant deviations are also observed in the $\tau_\text{ffp}$- and $\tau_\text{afp}$-distributions, as shown in Fig.~\ref{fig:butane_visc_10_passage_times}(b). The ME GLE method underestimates short first-to-first-passage times and overestimates long ones, likely due to over-weighting state-recrossing events. Deviations are also evident in the all-to-first-passage time distributions in Fig.~\ref{fig:butane_visc_10_passage_times}(c), where the ME GLE method diverges substantially from the MD data. The DP-GLE and ME simulation slightly overestimate the probability of short passage times by incorporating excessive recrossing events, resulting in too fast mean first-passage time predictions, contributing to the deviations observed in Fig.~\ref{fig:butane4_visc} for $\eta/\eta_0 = 10$. Based on these findings, we conclude that the Mori-GLE simulations are the most accurate in capturing the MD dihedral dynamics, as they successfully reproduce both short- and long-term transition dynamics with high fidelity.

\section{Details on Accuracy of the NGF Simulation Technique}
\label{app:acc_brute}
\begin{figure*}
\centering
\includegraphics[width=1
\linewidth]{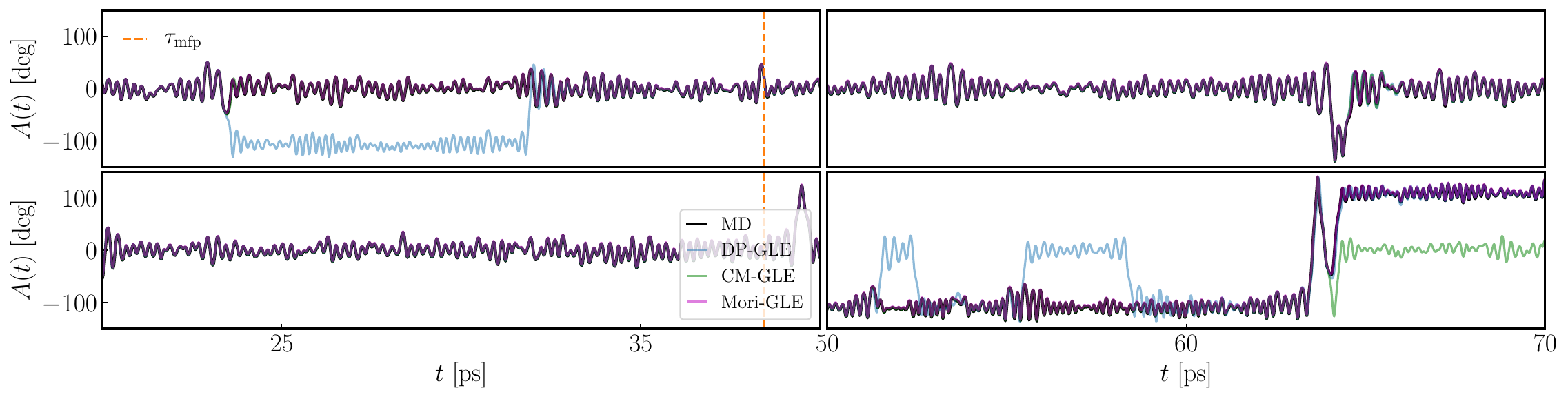}
\caption{Zoomed-in regions of the results shown in Fig.~\ref{fig:butane_brute_vf_samples} in the main text at times where the NGF simulation trajectories differ from the MD trajectories.}
\label{fig:butane_brute_vf_samples2}
\end{figure*} 

\begin{figure*}
\centering
\includegraphics[width=1
\linewidth]{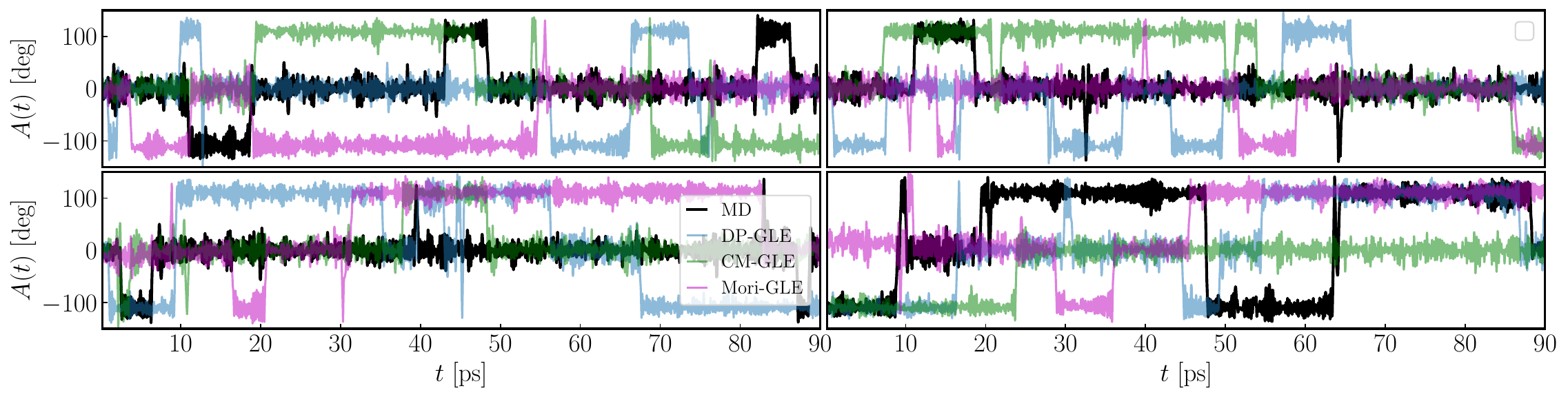}
\caption{Same analysis as in Fig.~\ref{fig:butane_brute_vf_samples} in the main text, but where the initial conditions $A_0$ and $\dot{A}_0$ are different from the ones for the computation of the orthogonal forces and for the NGF simulation (compare Fig.~\ref{fig:butane4} in the main text).}
\label{fig:butane_brute_vf_samples_diff_init_cond}
\end{figure*} 

Dephasing effects revealed in Fig.~\ref{fig:butane_brute_vf_samples} in the main text are better visible in the zoomed-in trajectories in Fig.~\ref{fig:butane_brute_vf_samples2}. The deviations of GLE simulations from the MD trajectory do not propagate, but after a certain time, GLE simulations and MD trajectories return to be aligned. 
Deviations presumably are due to numerical inaccuracies regarding that several non-linear force terms are included in the GLE simulations solving Eq.~\eqref{eq:ODE_discret} numerically (supplementary material Sec.~\ref{app:pos_dep_fr}), and contribute to the discrepancy in mean first-passage times between the MD and NGF simulation data (compare Fig.~\ref{fig:butane4} in the main text). 
\\ \indent In the case of two frozen carbon atoms (Fig.~\ref{fig:butane_check_fdt_2constr}), for high viscosities, the mean first-passage times are higher than 100 ps. Dephasing of MD and GLE trajectories consequently occurs much more frequently, leading to large deviations in the prediction of the first-passage times.\\
\indent For the results displayed in Fig.~\ref{fig:butane_brute_vf_samples_diff_init_cond}, we repeat the simulations in Fig.~\ref{fig:butane_brute_vf_samples} in the main text, but choose initial conditions $A_0$ and $\dot{A}_0$ that are different from the starting points of the orthogonal force-computation with Eq.~\eqref{eq:random_force} in the main text, as done for the results shown in Fig.~\ref{fig:butane4} in the main text. As expected, the simulated trajectories from all GLEs differ from the MD data.

\section{Position-Dependent Distributions of the Orthogonal Force}
\label{app:pos_dep_fr}

\begin{figure*}[hbt!]
\centering
\includegraphics[width=0.7\textwidth]{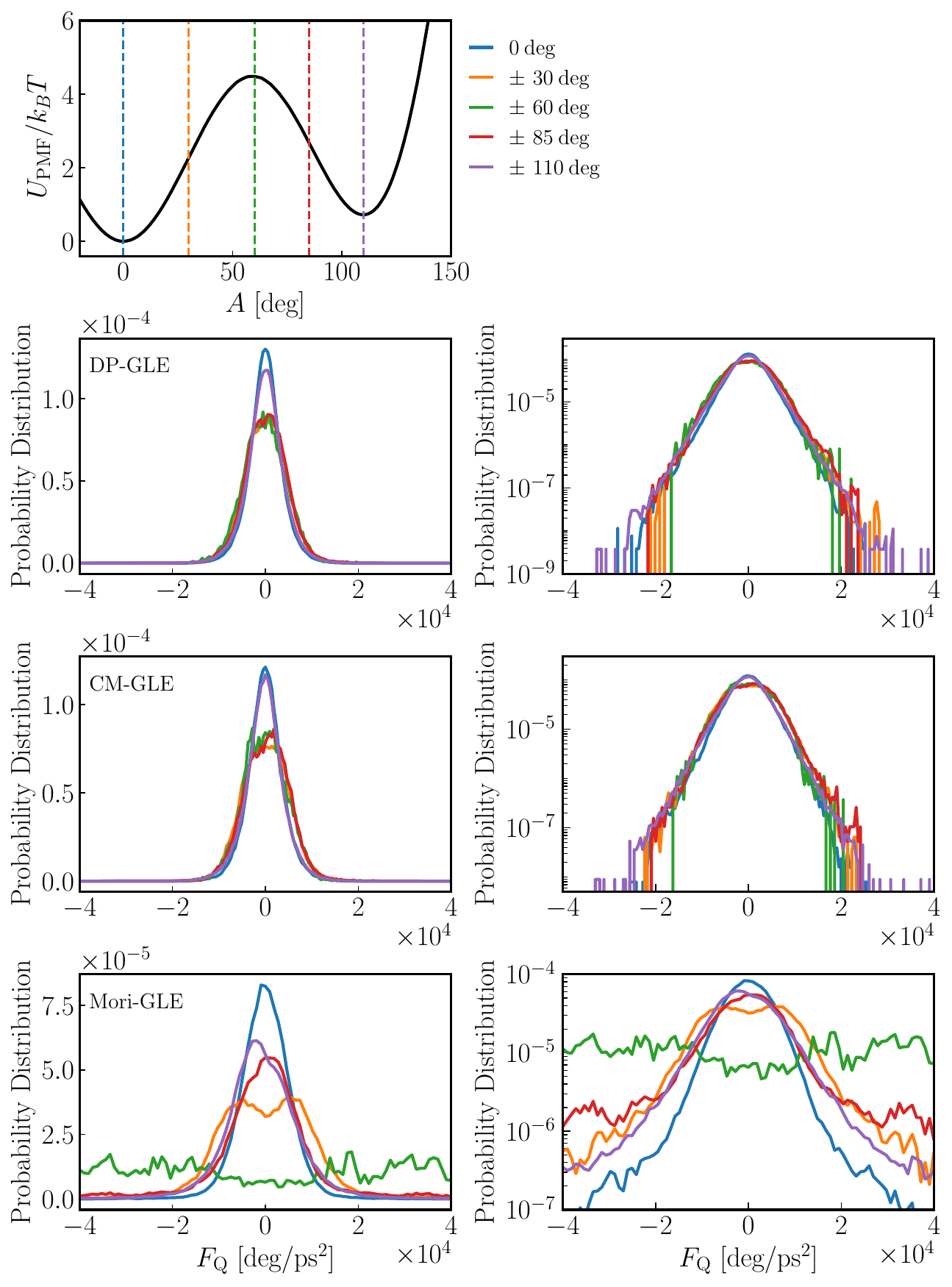}
\caption{Distributions of the butane dihedral angle's orthogonal force $F_\text{Q}(t)$ (Eq.~\eqref{eq:random_force} in the main text) in different GLEs (see Fig.~\ref{fig:butane_check_fdt} in the main text for details) in linear (left) and linear-logarithmic (right) representation, sampled at different trajectory positions. The position bins were chosen at different positions in the PMF (compare Fig.~\ref{fig:butane_check_fdt}(a) in the main text; see upper plot. }
\label{fig:cond_fr_dihedral_butane}
\end{figure*}
Fig.~\ref{fig:cond_fr_dihedral_butane} displays orthogonal force distributions for the butane dihedral angle, sampled at different ranges of $A$. The orthogonal force distributions from the three GLEs are all position-dependent, most pronounced for the Mori-GLE. As we show in supplementary material Sec.~\ref{app:acc_brute}, an NGF simulation using orthogonal forces sampled from a trajectory generated by a GLE with a non-linear deterministic force encounters numerical issues. In conjunction, the sampled orthogonal force at a given time point may not correspond to the correct distribution w.r.t. $A$, eventually leading to a mismatch between the effective force $F_\text{eff}$ at a certain position and the orthogonal force $F_\text{Q}$. This lack of coordination, together with numerical inaccuracies in solving Eq.~\eqref{eq:ODE_discret}, explains the dephasing problem discussed in supplementary material Sec.~\ref{app:acc_brute} and is particularly pronounced in the DP-GLE, where the mass $M$ is position-dependent. In contrast, the Mori-GLE, which employs a linear deterministic force, avoids this issue or, at the very least, does not introduce significant numerical inaccuracies.

\section{Test of the NGF Simulation Technique for a Model System}
\label{app:ngf_model_system}

\begin{figure}
\centering
\includegraphics[width=1
\linewidth]{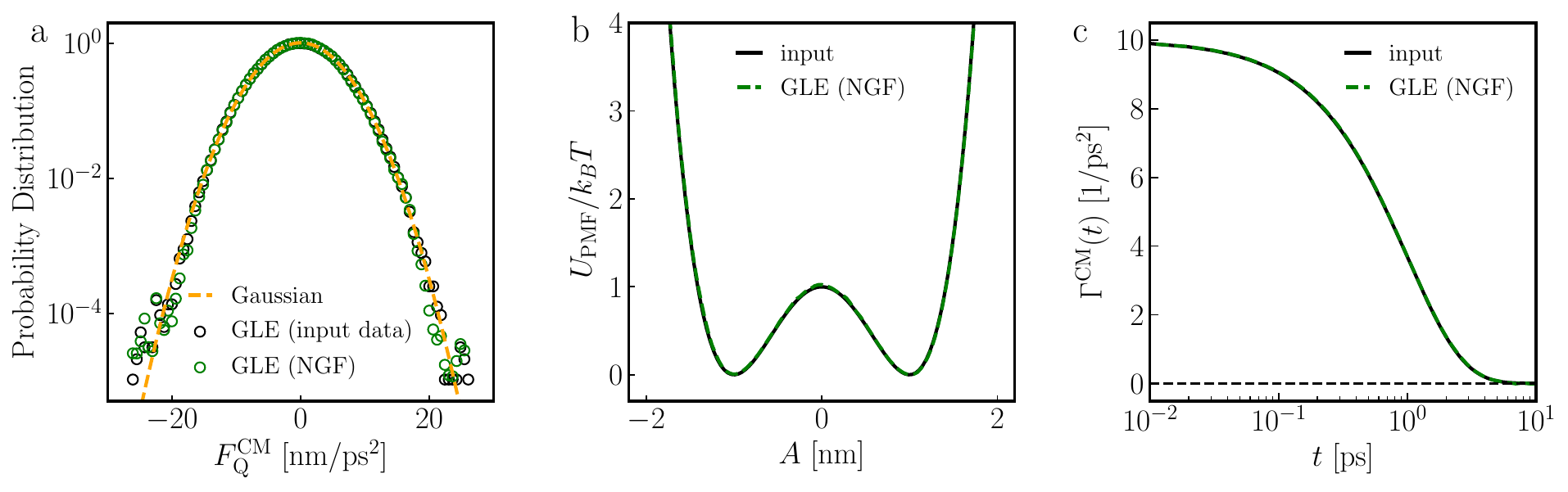}
\caption{Test of the NGF CM-GLE simulation described in Appendix \ref{sec:ngf} on a model system with single-exponential memory, i.e. $\Gamma^\text{CM}(t) = (\gamma/\tau) e^{-t/\tau}$, where $A$ is confined in a double-well potential, i.e. $U_{\text{PMF}}(A) = U_0\left((A/L)^2-1\right)^2$. We first generate a trajectory $A(t)$ using the Markovian embedding scheme with Eqs.~\eqref{eq:markov_embedding_1D_const_mass}. We then extract the memory kernel $\Gamma^\text{CM}(t)$ using the CM-GLE in Eq.~\eqref{eq:const_mass_gle} in the main text (supplementary material Sec.~\ref{app:MemKernExtrac}) and randomly select orthogonal-force trajectories $F^\text{CM}_\text{Q}(t)$ (Eq.~\eqref{eq:random_force} in the main text) for simulations of the generated trajectory $A(t)$. We compare the input of the PMF $U_{\text{PMF}}$ (b) and the memory kernel $\Gamma^\text{CM}(t)$ (c) with the extracted data from the NGF CM-GLE simulation (green), where the initial conditions of the simulation are selected randomly from the generated trajectory, but are not the same as for the orthogonal-force calculation. In (a), we demonstrate that the orthogonal forces computed from the ME (black) and NGF CM-GLE (green) trajectories follow the input Gaussian distribution with zero mean and variance $\langle \Dot{A}_0^2 \rangle\Gamma^\text{CM}(0)$ (orange broken line).}
\label{fig:brute_test_model}
\end{figure}

\begin{figure}
\centering
\includegraphics[width=0.7\linewidth]{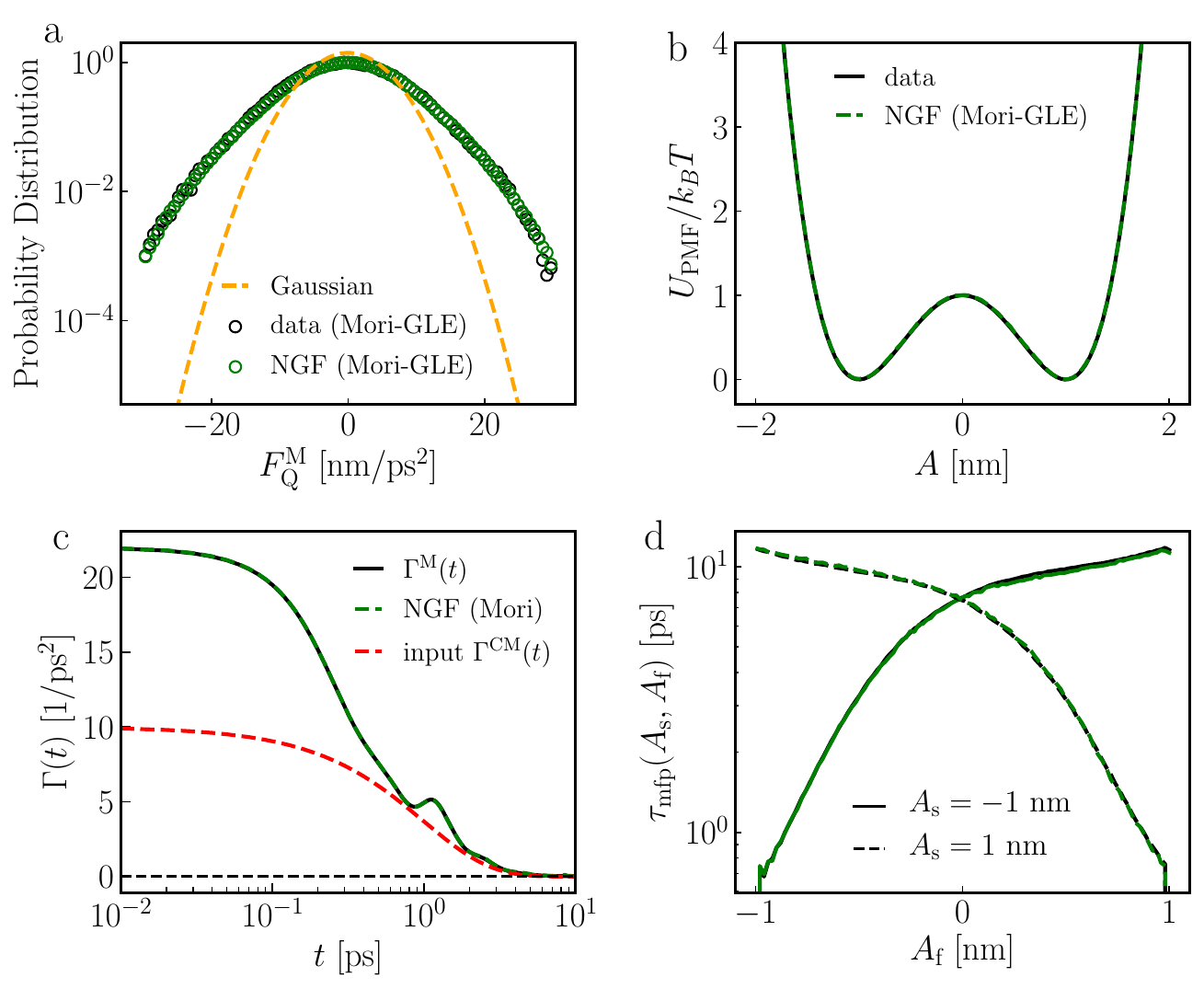}
\caption{Same analysis as shown in Fig.~\ref{fig:brute_test_model}. Contrary, we extract the memory kernel $\Gamma^\text{M}(t)$ from the Mori-GLE in Eq.~\eqref{eq:gle_mori} in the main text and randomly select orthogonal-force trajectories $F^\text{M}_\text{Q}(t)$ for Mori-GLE simulations at arbitrary starting points in the generated trajectory $A(t)$ that originates from Eqs.~\eqref{eq:markov_embedding_1D_const_mass}. In (a), we demonstrate that the orthogonal force distributions computed from the ME (black) and NGF Mori-GLE (green) trajectories coincide but do not agree with the input Gaussian distribution with zero mean and variance $\langle \Dot{A}_0^2 \rangle\Gamma^\text{CM}(0)$ (orange broken line). In (b), we compare the input PMF $U_{\text{PMF}}$ (black) with the extracted data from the NGF Mori-GLE simulation. In (c), we display the input memory kernel $\Gamma^\text{CM}(t)$ of the ME simulation (red), together with the memory kernel in the Mori-GLE $\Gamma^\text{M}(t)$ extracted from the ME data (black) and from the NGF Mori-GLE simulation (green). Subplot (d) shows the mean first-passage time $\tau_{\text{mfp}}$ of the ME and NGF Mori-GLE simulations as a function of the final position $A_\text{f}$ for different starting positions $A_\text{s}$. The color coding is the same as in (b).}
\label{fig:brute_test_model_mori}
\end{figure}

We test the NGF simulation technique introduced in Appendix \ref{sec:ngf} with data generated from a simple model system, described by the CM-GLE in Eq.~\eqref{eq:const_mass_gle} in the main text. The model system includes single-exponential memory with friction coefficient $\gamma$ and memory time $\tau$, i.e. $\Gamma^\text{CM}(t) = \frac{\gamma}{\tau}e^{-t/\tau}$, and $A$ is confined in a double-well potential, i.e. $U_{\text{PMF}}(A) = U_0\left((A/L)^2-1\right)^2$.
Accordingly, we generate a trajectory of the underlying system by the following Markovian embedding \cite{kappler2018memory, ayaz2021non}
\begin{align}
\label{eq:markov_embedding_1D_const_mass}
\ddot{A}(t) =& - M_0^{-1}\frac{d}{dA} U_{\text{PMF}}\bigl(A(t)\bigr) + \frac{\gamma}{\tau}[y(t) - A(t)], \\ \nonumber
\Dot{y}(t) =& - \tau^{-1} [y(t) - A(t)] + \zeta(t).
\end{align}
The orthogonal force $\zeta(t)$ follows a stationary Gaussian process with zero mean and $\langle \zeta(t), \zeta(s) \rangle = 2 \langle \dot{A}_0^2 \rangle \gamma^{-1} \delta(|t-s|)$. We simulate a trajectory $A(t)$ with a length of $5\cdot10^{7}$ steps via a 4th order Runge-Kutta scheme and the following parameters: $L$ = 1 nm, $U_0$ = 1 $k_BT$, $M_0$ = 1u, $\gamma$ = 10 ps$^{-1}$, $\tau$ = 1 ps, and a time resolution of $\Delta t$ = 0.01 ps. Using the memory kernel $\Gamma^\text{CM}(t)$ calculated from the simulated trajectory using the Volterra extraction scheme for the CM-GLE in Eq.~\eqref{eq:const_mass_gle} (supplementary material Sec.~\ref{app:MemKernExtrac}), we compute 50 realizations of the orthogonal force $F^\text{CM}_\text{Q}(t)$ (Eq.~\eqref{eq:random_force} in the main text) with a length of $10^6$ time steps each starting at arbitrary time points of $A(t)$ and insert them into Eq.~\eqref{eq:ODE_discret} to perform 50 NGF simulations.
Therefore, the NGF simulation's total length is the same as that of the trajectory generated by the Markovian embedding scheme. The initial conditions of the NGF simulations, $A_0$ and $\dot{A}_0$, are chosen randomly from the trajectory but are not the starting points for the orthogonal force computation.\\
\indent In Fig.~\ref{fig:brute_test_model}, we find that the simulated PMF $U_{\text{PMF}}(A)$ and memory kernel $\Gamma^\text{CM}(t)$ agree perfectly with the input functions. This indicates that the NGF simulation is numerically robust and self-consistent for the model system. Furthermore, the orthogonal forces computed from the ME and NGF trajectories follow a Gaussian distribution (orange broken line), as shown in Fig.~\ref{fig:brute_test_model}(a), which points to the accuracy of the orthogonal-force calculation with Eq.~\eqref{eq:random_force} in the main text.\\
\indent For the results shown in Fig.~\ref{fig:brute_test_model_mori}, we repeat the procedure, but perform memory kernel extraction and NGF simulation of the trajectory from Eqs.~\eqref{eq:markov_embedding_1D_const_mass} with the Mori-GLE (Eq.~\eqref{eq:gle_mori} in the main text). As expected, the orthogonal force distribution in (a) deviates from the input Gaussian form in the ME trajectory generation using Eqs.~\eqref{eq:markov_embedding_1D_const_mass}, and the memory kernel $\Gamma^\text{M}(t)$ in (c) deviates from the input memory kernel $\Gamma^\text{CM}(t)$, as the non-linearity of the double-well potential is absorbed into the memory kernel (compare red and black in (c)), as well as into the orthogonal force. Nonetheless, the NGF simulation is very accurate, as it reproduces the distribution of $A$ (b), the memory kernel (c), and the mean first-passage time (d) of the underlying system perfectly.
\section{Second-Order Expansion of Eq.~S13 at $t = 0$}
\label{app:second_order_expansion}

By differentiation of the integrals in Eq.~\eqref{eq:proof_fdts10} in supplementary material Sec.~\ref{app:proof_equiv_kernels}, and given that $\langle F_\text{Q}^\text{H}(0), \Ddot{A}(0) \rangle/\langle \Dot{A}_{0}^{2} \rangle = \Gamma_\text{L}^\text{H}(0)$ from Eq.~\eqref{eq:hybrid-kernel}, we obtain

\begin{gather}
\Delta\Gamma^{(2)}(t) = \frac{\langle F_\text{Q}^\text{H}(0), \Dot{A}(t) \partial_{A}\Gamma_\text{NL}^\text{H}\bigl(0, A(t)\bigr) \rangle}{\langle \Dot{A}_{0}^{2}\rangle} + \frac{\langle F_\text{Q}^\text{H}(0), \partial_{t}\Gamma_\text{NL}^\text{H}\bigl(0, A(t)\bigr) \rangle}{\langle \Dot{A}_{0}^{2}\rangle} + \nonumber \\
\int_{0}^{t} ds \: \frac{\langle F_\text{Q}^\text{H}(0), \partial_{t}^{2}\Gamma_\text{NL}^\text{H}\bigl(t-s, A(s)\bigr) \rangle}{\langle \Dot{A}_{0}^{2}\rangle} + \Gamma_{L}^{H}(0) \left( \int_{0}^{t} ds \: \frac{\langle \Ddot{A}_{0} ,\Gamma_\text{NL}^\text{H}\bigl(t-s, A(s)\bigr) \rangle}{\langle \Dot{A}_{0}^{2} \rangle} + \right. \nonumber \\
\left. \sum_{n \geq 2} \prod_{i = 2}^{n+1} \int_{0}^{t\delta_{i,2} + t_{i-1}(1-\delta_{i,2})} dt_{i} \frac{\langle \Ddot{A}_{0}, \Dot{A}(t - t_{2}) \rangle}{\langle \Dot{A}_{0}^{2}\rangle} \prod_{j=3}^{n} \frac{\langle \Ddot{A}_{0}, \Dot{A}(t_{j-1} - t_{j}) \rangle}{\langle \Dot{A}_{0}^{2}\rangle} \frac{\langle \Ddot{A}_{0},\Gamma_\text{NL}^\text{H}\bigl(t_{n}-t_{n+1}, A(t_{n+1})\bigr) \rangle}{\langle \Dot{A}_{0}^{2} \rangle} \right) \nonumber \\
+ \int_{0}^{t} dt_{1} \int_{0}^{t_{1}} dt_{2} \: \frac{\langle F_\text{Q}^\text{H}(0), \dddot{A}(t-t_{1}) \rangle}{\langle \Dot{A}_{0}^{2} \rangle} \frac{\langle \Ddot{A}_{0}, \Gamma_\text{NL}^\text{H}\bigl(t_{1}-t_{2}, A(t_{2})\bigr) \rangle}{\langle \Dot{A}_{0}^{2} \rangle}  \nonumber \\
+ \sum_{n \geq 2} \prod_{i = 1}^{n+1} \int_{0}^{t\delta_{i,1} + t_{i-1}(1-\delta_{i,1})} dt_{i} \frac{\langle F_\text{Q}^\text{H}(0), \dddot{A}(t-t_{1}) \rangle}{\langle \Dot{A}_{0}^{2}\rangle} \prod_{j=2}^{n} \frac{\langle \Ddot{A}_{0}, \Dot{A}(t_{j-1} - t_{j}) \rangle}{\langle \Dot{A}_{0}^{2}\rangle} \frac{\langle \Ddot{A}_{0}, \Gamma_\text{NL}^\text{H}\bigl(t_{n}-t_{n+1}, A(t_{n+1})\bigr) \rangle}{\langle \Dot{A}_{0}^{2} \rangle}. \nonumber \\
\label{eq:proof_fdts21}
\end{gather}
At time $t = 0$, all the integral terms from $0$ to $t$ on the r.h.s. in Eq.~\eqref{eq:proof_fdts21} vanish, and $\Delta\Gamma^{(2)}(0)$ is given by

\begin{eqnarray}
\Delta\Gamma^{(2)}(0) = \frac{\langle F_\text{Q}^\text{H}(0), \Dot{A}_{0} \partial_{A}\Gamma_\text{NL}^\text{H}\bigl(0, A_{0}\bigr) \rangle}{\langle \Dot{A}_{0}^{2}\rangle} + \frac{\langle F_\text{Q}^\text{H}(0), \partial_{t}\Gamma_\text{NL}^\text{H}(0, A_{0}) \rangle}{\langle \Dot{A}_{0}^{2}\rangle}.
\label{eq:proof_fdts22}
\end{eqnarray}
The correlation of $F_\text{Q}^\text{H}(0)$ with any function of $A_{0}$ vanishes, as discussed in supplementary material Sec.~\ref{app:proof_equiv_kernels} (see Eq.~\ref{eq:orthogonality_proof_4}). Thus, $\langle F_\text{Q}^\text{H}(0), \partial_{t}\Gamma_\text{NL}^\text{H}(0, A_{0}) \rangle = 0$. Moreover, from Eq.~\eqref{eq:hybrid_gle} we obtain $F_\text{Q}^\text{H}(0) = \Ddot{A}_{0} - F_\text{eff}(A_{0})$ \cite{ayaz2022generalized}, and decompose the prefactor of the remaining term on the r.h.s. in Eq.~\eqref{eq:proof_fdts22} into

\begin{eqnarray}
\langle F_\text{Q}^\text{H}(0), \Dot{A}_{0} \partial_{A}\Gamma_\text{NL}^\text{H}(0, A_{0}) \rangle = \langle \Ddot{A}_{0}, \Dot{A}_{0} \partial_{A}\Gamma_\text{NL}^\text{H}(0, A_{0}) \rangle - \langle F_\text{eff}(A_{0}), \Dot{A}_{0} \partial_{A}\Gamma_\text{NL}^\text{H}(0, A_{0}) \rangle.
\label{eq:proof_fdts23}
\end{eqnarray}
Since $A_{0}$ depends only on positional degrees of freedom, $\Dot{A}_{0}$ is a function that is odd in momenta. Moreover, we recall that the distribution  $\rho_\text{eq}(\omega)$, defined in the main text, is the Boltzmann distribution determined by $H(\omega)$. It follows that the second term of Eq.~\eqref{eq:proof_fdts23} is the integral over $\Omega$ of a term that is odd in momenta multiplied by one that is even in momenta. Therefore, the integral vanishes, i.e. $\langle F_\text{eff}(A_{0}), \Dot{A}_{0} \partial_{A}\Gamma_\text{NL}^\text{H}(0, A_{0}) \rangle = 0$. For the remaining term in the r.h.s. of Eq.~\eqref{eq:proof_fdts23}, we recall that $\Ddot{A}_{0} = \mathcal{L} \Dot{A}_{0}$ and use that $\mathcal{L}$ is a first-order partial linear anti self-adjoint differential operator to rewrite $\langle \Ddot{A}_{0}, \Dot{A}_{0} \partial_{A}\Gamma_\text{NL}^\text{H}(0, A_{0}) \rangle$ as

\begin{eqnarray}
\langle \Ddot{A}_{0}, \Dot{A}_{0} \partial_{A}\Gamma_\text{NL}^\text{H}(0, A_{0}) \rangle = - \langle \Ddot{A}_{0}, \Dot{A}_{0} \partial_{A}\Gamma_\text{NL}^\text{H}(0, A_{0}) \rangle - \langle \Dot{A}_{0}^{3}, \partial_{A}^{2}\Gamma_\text{NL}^\text{H}(0, A_{0}) \rangle,  \\
\to \langle \Ddot{A}_{0}, \Dot{A}_{0} \partial_{A}\Gamma_\text{NL}^\text{H}(0, A_{0}) \rangle = - \frac{\langle \Dot{A}_{0}^{3}, \partial_{A}^{2}\Gamma_\text{NL}^\text{H}(0, A_{0}) \rangle}{2}.
\label{eq:proof_fdts24}
\end{eqnarray}
Since $A_{0}$ depends only on positional degrees of freedom, so that $\Dot{A}_{0}$ is odd in momenta, and $\rho_\text{eq}(\omega)$ is even in momenta, Eq.~\eqref{eq:proof_fdts24} is the integral over $\Omega$ of an odd quantity in momenta, and thus vanishes, i.e. $\langle \Ddot{A}_{0}, \Dot{A}_{0} \partial_{A}\Gamma_\text{NL}^\text{H}(0, A_{0}) \rangle = 0$. We, therefore, have proven that the second derivative of $\Delta\Gamma$ at $t=0$  in Eq.~\eqref{eq:proof_fdts22} vanishes

\begin{eqnarray}
\Delta\Gamma^{(2)}(0) = 0.
\label{eq:proof_fdts25}
\end{eqnarray}

\section{Derivation of Eq.~S23}
\label{app:deriv_master_rel}

After multiplying both Eqs.~\eqref{eq:hybrid_gle} and \eqref{eq:dp_gle} in the main text with $\Dot{A}_0$ and performing an ensemble average, we obtain
\begin{align}
\label{eq:Volterra1}
\langle \Dot{A}_0,\ddot{A}(t) \rangle = - \langle \Dot{A}_0,F_{\text{eff}}\bigl(A(t)\bigr) \rangle - \int_0^t ds\:\Gamma^\text{DP}_\text{Q}(t-s)\langle \Dot{A}_0, \Dot{A}(s) \rangle - \int_0^t ds\:\Gamma^\text{DP}_{\Delta}(t-s)\langle \Dot{A}_0, \Dot{A}(s) \rangle, \\
\label{eq:Volterra2}
\langle \Dot{A}_0,\ddot{A}(t) \rangle = - \langle \Dot{A}_0,F_{\text{eff}}\bigl(A(t)\bigr) \rangle - \int_0^t ds\:\Gamma^\text{H}_\text{L}(t-s)\langle \Dot{A}_0, \Dot{A}(s) \rangle + \int_0^t ds\:\langle \Dot{A}_0,\Gamma^\text{H}_\text{NL}\bigl(A(s), t-s\bigr) \rangle,
\end{align}
where we employed $\langle \dot{A}_0, F^\text{DP}_\text{Q}(t) \rangle = 0$ and $\langle \dot{A}_0, F^\text{H}_\text{Q}(t) \rangle = 0$.
Using $\Gamma^\text{DP}_\text{Q} = \Gamma^\text{H}_\text{L}$ (supplementary material Sec.~\ref{app:proof_equiv_kernels}) and equating both expressions, we arrive at the relation in Eq.~\eqref{eq:master_relation}, which connects the DP- and the hybrid GLE kernel functions
\begin{equation}
\label{eq:master_relation2}
- \int_0^t ds\:\Gamma^\text{DP}_{\Delta}(t-s)\langle \Dot{A}_0, \Dot{A}(s) \rangle = \int_0^t ds \langle \Dot{A}_0,\Gamma^\text{H}_\text{NL}\bigl(A(s), t-s\bigr) \rangle.
\end{equation}

\section{Position-Dependent Memory-Friction Kernels for Butane}
\label{app:non_linear_kernels}
\begin{figure*}[hbt!]
\centering
\includegraphics[width=0.7
\linewidth]{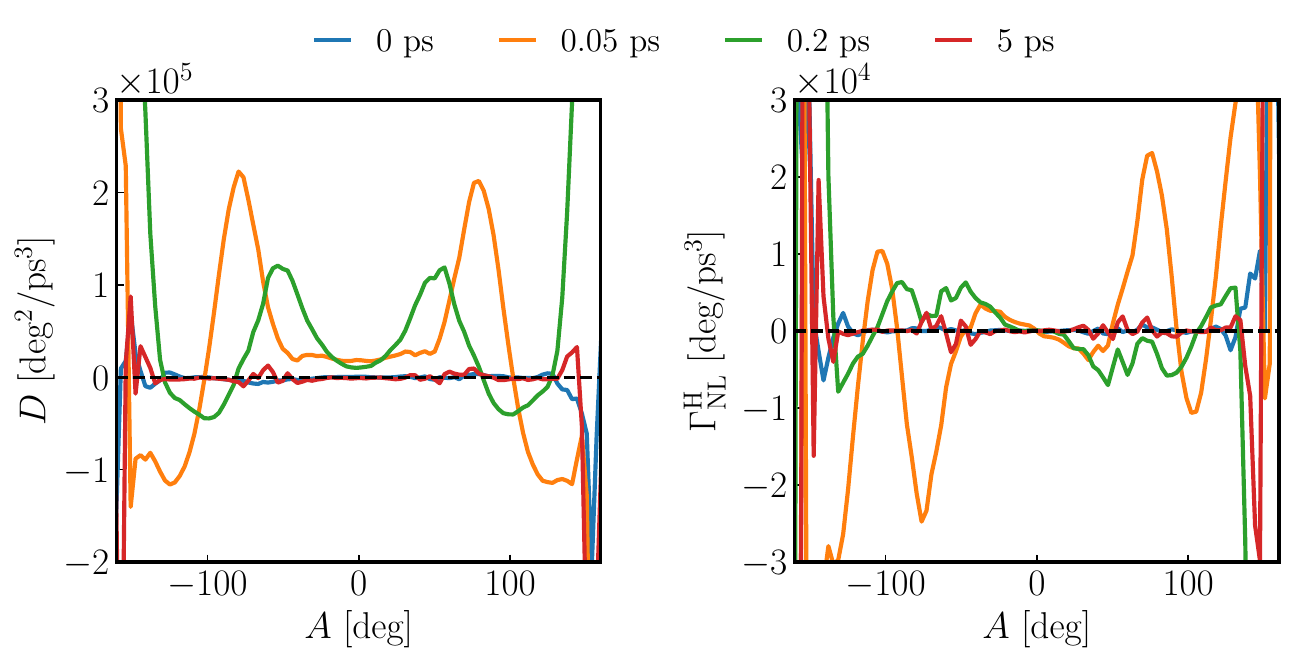}
\caption{Butane dynamics $D$-function (Eq.~\eqref{eq:dfunc} in the main text) and the non-linear memory kernel $\Gamma^\text{H}_\text{NL}(A(t),s)$ (Eq.~\eqref{eq:kernel_nl} in the main text) for the (non-frozen) butane dihedral angle data shown in Fig.~\ref{fig:butane_check_fdt} in the main text, extracted with the forward propagation scheme explained in Ref.~\onlinecite{ayaz2022generalized}. We show both quantities as a function of the position $A$ and at different times $t$.}
\label{fig:butane_hybrid}
\end{figure*} 
\begin{figure*}[hbt!]
\centering
\includegraphics[width=0.7
\linewidth]{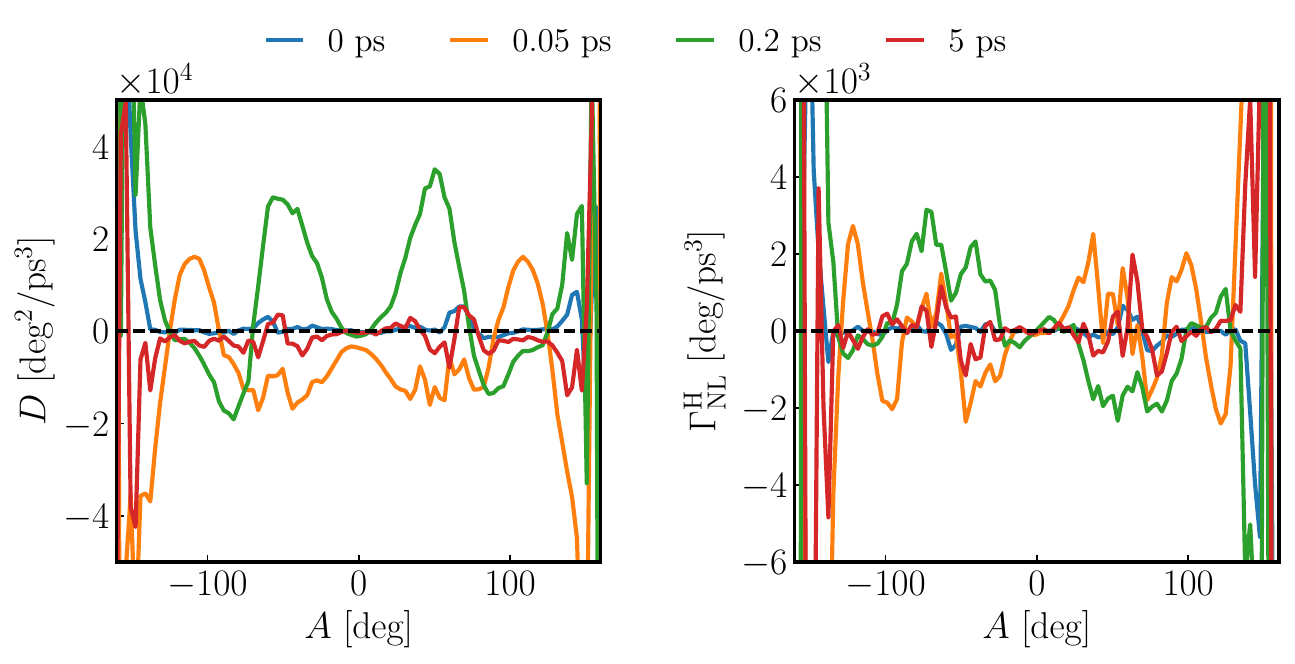}
\caption{Same results as in Fig.~\ref{fig:butane_hybrid} but where two inner carbon atoms in butane are frozen in space (compare supplementary material Sec.~\ref{app:results_2constr}).}
\label{fig:butane_hybrid_2constr}
\end{figure*} 

Fig.~\ref{fig:butane_hybrid} and Fig.~\ref{fig:butane_hybrid_2constr} display extracted parameters from the hybrid GLE in Eq.~\eqref{eq:hybrid_gle} using the forward propagation method developed in Ref.~\onlinecite{ayaz2022generalized}, for the dihedral angle of butane where the carbon atoms are not fixed (data we discuss in the main text) and where the inner two carbon atoms are frozen (see supplementary material Sec.~\ref{app:results_2constr}), respectively. Note that for the frozen scenario, we assume the position-dependent mass to be constant, i.e. $M(A) = M_0$, which reduces numerical inaccuracies. The functional shape of the $D$ and  $\Gamma^\text{NL}$-functions for the free butane molecule in Fig.~\ref{fig:butane_hybrid} have already been discussed in Ref.~\onlinecite{ayaz2022generalized}. Most strikingly, the absolute values for different times $t$ in the frozen scenario are significantly smaller. 
$\Gamma^\text{H}_\text{NL}$ at $t=0.2$ ps for the free case has its maximum around 8 $\cdot\:10^3$ deg/ps$^3$, and is subject to noise. The values in the frozen case are always below this value, which is why they are more affected by statistical noise.

\section{Expanding Position-Dependent Memory Kernels by Legendre-Polynomials}
\label{app:legendre_non_linear_kernels}
\begin{figure*}[hbt!]
\centering
\includegraphics[width=0.4
\linewidth]{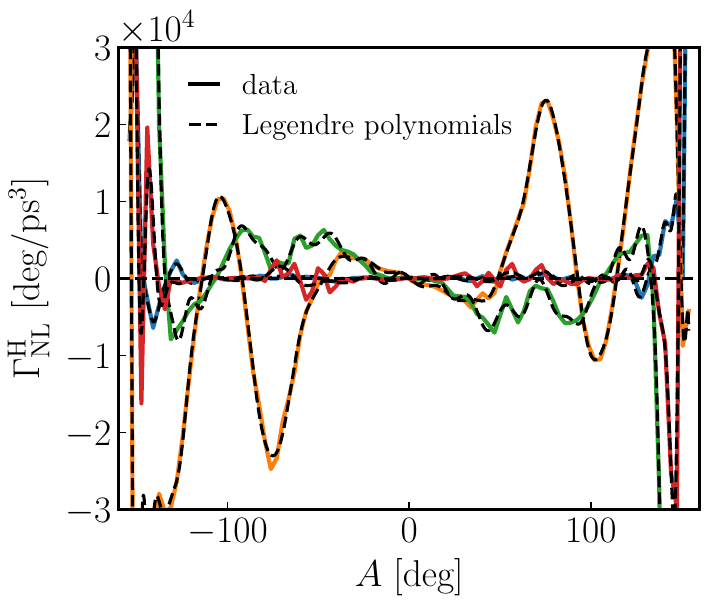}
\caption{Non-linear memory kernel $\Gamma^\text{H}_\text{NL}(A(t),s)$ from the MD simulation of a non-frozen butane  (same color coding as in Fig.~\ref{fig:butane_hybrid}), compared with the interpolation of the data (black broken lines) via Legendre polynomial expansion according to Eq.~\eqref{eq:legendre}.}
\label{fig:butane_hybrid_legendre}
\end{figure*}

Evaluating the non-linear memory kernel $\Gamma^\text{H}_\text{NL}(A,t)$ in Eq.~\eqref{eq:master_relation}, using extracted data as shown in supplementary material Sec.~\ref{app:non_linear_kernels}, requires a continuous representation of the discrete data for $\Gamma^\text{H}_\text{NL}$. Instead of using cubic splines, which is the method we used for the PMF and the position-dependent mass in the memory kernel extraction (see supplementary material Sec.~\ref{app:MemKernExtrac}) and is problematic for two-dimensional data sets subject to numerical inaccuracies (compare supplementary material Sec.~\ref{app:non_linear_kernels}), we expand the two-dimensional function $\Gamma^\text{H}_\text{NL}(A,t)$ by a sum of Legendre polynomials according to
\begin{equation}
\label{eq:legendre}
\Gamma^\text{H}_\text{NL}(A,t) = \sum_{l=0}^N K_l(t)P_l(A),
\end{equation}
where $P_l$ is the $l$-th Legendre polynomial and
\begin{equation}
K_l(t) = \frac{2l + 1}{l} \int_{-1}^1 dA\:\Gamma^\text{H}_\text{NL}(A,t)P_l(A).
\end{equation}
Here, we choose $N=50$ components. In Fig.~\ref{fig:butane_hybrid_legendre}, we present the interpolated data from Fig.~\ref{fig:butane_hybrid}.

\end{document}